\documentclass[twocolumn]{aastex63}
\usepackage{amssymb}
\usepackage{amsmath}
\usepackage{amsfonts}
\usepackage{color}
\usepackage{graphicx}
\usepackage[varg]{txfonts}
\usepackage{natbib}
\usepackage{hyperref}
\usepackage{dcolumn}
\usepackage{bm}
\usepackage{lineno}
\received{XXX}
\revised{YYY}
\accepted{ZZZ}
\submitjournal{ApJ}
\shorttitle{Binary-merger progenitors of SN~1987A}
\shortauthors{V. P. Utrobin et al.}
\begin{document}
%
\title{Supernova 1987A: 3D Mixing and light curves for explosion models based
       on binary-merger progenitors}

\correspondingauthor{V.~P.~Utrobin}
\email{utrobin@itep.ru}

\author{V.~P.~Utrobin}
\affiliation{Max-Planck-Institut f\"ur Astrophysik,
   Karl-Schwarzschild-Str. 1, 85748 Garching, Germany}
\affiliation{NRC ``Kurchatov Institute'' --
   Institute for Theoretical and Experimental Physics,
   B.~Cheremushkinskaya St. 25, 117218 Moscow, Russia}
\affiliation{Institute of Astronomy, Russian Academy of Sciences,
   Pyatnitskaya St. 48, 119017 Moscow, Russia}

\author{A.~Wongwathanarat}
\affiliation{Max-Planck-Institut f\"ur Astrophysik,
   Karl-Schwarzschild-Str. 1, 85748 Garching, Germany}

\author{H.-Th.~Janka}
\affiliation{Max-Planck-Institut f\"ur Astrophysik,
   Karl-Schwarzschild-Str. 1, 85748 Garching, Germany}

\author{E.~M\"uller}
\affiliation{Max-Planck-Institut f\"ur Astrophysik,
   Karl-Schwarzschild-Str. 1, 85748 Garching, Germany}

\author{T.~Ertl}
\affiliation{Max-Planck-Institut f\"ur Astrophysik,
   Karl-Schwarzschild-Str. 1, 85748 Garching, Germany}

\author{A.~Menon}
\affiliation{Anton Pannekoek Institute of Astronomy, University of Amsterdam,
   Science Park 904, 1098 XH Amsterdam, Netherlands}

\author{A.~Heger}
\affiliation{School of Physics and Astronomy, Monash University, VIC 3800,
   Australia}
\affiliation{ARC Centre of Excellence for Gravitational Wave Discovery (OzGrav),
   Melbourne, Australia}
\affiliation{ARC Centre of Excellence for Astrophysics in Three Dimensions
   (ASTRO-3D), Australia}
\affiliation{Joint Institute for Nuclear Astrophysics, National Superconducting
   Cyclotron Laboratory, Michigan State University, East Lansing,
   MI-48824-1321, USA}

\begin{abstract}
Six binary-merger progenitors of Supernova~1987A (SN~1987A) with properties
   close to those of the blue supergiant Sanduleak\,$-69^{\circ}202$ are
   exploded by neutrino heating and evolved until long after shock
   breakout in three dimensions (3D), and continued for light-curve
   calculations in spherical symmetry.
Our results confirm previous findings for single-star progenitors:
   (1)~3D neutrino-driven explosions with SN~1987A-like energies synthesize
   $^{56}$Ni masses consistent with the radioactive light-curve tail;
   (2)~hydrodynamic models mix hydrogen inward to minimum velocities below
   40\,km\,s$^{-1}$ compatible with spectral observations of SN~1987A; and
   (3)~for given explosion energy the efficiency of outward radioactive
   $^{56}$Ni mixing depends mainly on high growth factors of Rayleigh-Taylor
   instabilities at the (C+O)/He and He/H composition interfaces and a weak
   interaction of fast plumes with the reverse shock occurring below the
   He/H interface.
All binary-merger models possess presupernova radii matching the
   photometric radius of Sanduleak\,$-69^{\circ}202$ and a structure of
   the outer layers allowing them to reproduce the observed initial
   luminosity peak in the first $\sim$7 days.
Models that mix about 0.5\,$M_{\sun}$ of
   hydrogen into the He-shell and exhibit strong outward mixing of
   $^{56}$Ni with maximum velocities exceeding the 3000\,km\,s$^{-1}$
   observed for the bulk of ejected $^{56}$Ni have light-curve shapes
   in good agreement with the dome of the SN~1987A light curve.
A comparative analysis of the best representatives of our 3D neutrino-driven
   explosion models of SN~1987A based on single-star and binary-merger
   progenitors reveals that only one binary model fulfills all observational
   constraints, except one.
\end{abstract}


\section{Introduction}
\label{sec:intro}
On February 23, 1987, the outburst of the Type II plateau supernova (SN) 1987A
   was discovered by Ian Shelton in the Large Magellanic Cloud (LMC)
   \citep[see][]{KMS_87}.
This SN became a uniquely peculiar event and one of the most thoroughly
   studied objects observed from radio wavelengths to gamma-rays.
Accurate positional measurement of the SN showed that its position
   coincided with that of star 1 from the Sanduleak\,$-69^{\circ}202$ system
   and a careful examination of the measured $UBVRI$ magnitudes indicated that
   this Sanduleak star was a B3 Ia supergiant \citep[cf.][]{WLJS_87}.
These data suggest that the blue supergiant (BSG) star was the progenitor of
   SN~1987A and permit both single-star and binary evolution scenarios of
   Sanduleak\,$-69^{\circ}202$.%
   \footnote{Henceforward all references to Sanduleak\,$-69^{\circ}202$
   only refer to star 1.}

The fact that SN~1987A was the explosion of the BSG progenitor rather than
   a red supergiant (RSG) as expected for ordinary Type IIP SNe, triggered
   construction of a large variety of evolutionary models to explain the
   observed properties of Sanduleak\,$-69^{\circ}202$.
In the single-star scenario the relative compactness of the BSG progenitor
   was achieved either by a metal-deficient composition similar to that of
   the LMC \citep{TW_87, HHTW_87, Arn_87}, a modification of convective mixing
   through rotation-induced meridional circulation during stellar evolution
   \citep{WHT_88}, a restricted semiconvective diffusion and low metallicity
   \citep{WPE_88, Wei_89, Lan_91}, or both mass loss and convective mixing
   \citep{SNK_88}.

All of these possibilities were equally promising until the ESO ground-based
   New Technology Telescope \citep{WWB_90} and the NASA/ESA Hubble Space
   Telescope \citep[HST,][]{JAB_91} revealed an intricate triple-ring structure
   around SN~1987A.
The existence of this triple-ring structure imposes additional severe
   constraints on the evolution of the pre-SN, putting the
   single-star scenario into doubt and favoring a binary evolution scenario.
The evolution of stars in an interacting binary is so rich in possibilities
   that it permits not only the accretion of a substantial amount of matter
   from the secondary component onto the pre-SN, but also a complete merger
   of the two stars.
Both abundant accretion of matter from the companion star \citep{PJ_89} and
   the merger of the companion star with the primary RSG \citep{CS_89, HM_89,
   PJR_90, PMI_07, MH_17, UTUY_18} can explain the observed properties of
   the SN~1987A progenitor.

\citet{Pod_92} devised five observational and theoretical tests
   (the blue color of the progenitor, the ring surrounding it, the chemical
   anomalies of the progenitor, the characteristics of the SN explosion,
   and general consistency with the theory of massive stars) and confronted
   the available evolutionary models for the progenitor of SN 1987A with
   them.
\citeauthor{Pod_92} concluded that it is most likely that only binary
   scenario models (accretion and merger models) are able to fit all above
   constraints.

It is noteworthy that the chemical anomalies of the progenitor included
   a high nitrogen abundance that was revealed in the circumstellar matter
   of SN~1987A by an analysis of ultraviolet lines \citep{FCG_89} and
   the overabundance of barium in the ejecta that was estimated by
   a non-local thermodynamic equilibrium (non-LTE) modeling of optical spectra
   of SN~1987A \citep{MLB_92}.
The latter anomaly was revisited and not confirmed when time-dependent effects
   were taken into account in addition to the non-LTE treatment of spectra
   \citep{UC_05, DH_08}, thus softening this constraint.

Over the past quarter of a century, the characteristics of the SN explosion
   extracted from photometric and spectral observations of SN~1987A have
   become more and more detailed, especially regarding mixing of radioactive
   $^{56}$Ni and hydrogen in the ejected envelope \citep[see][for details]
   {UWJM_15}.
\citet{CHELH_94} analyzed the infrared emission lines of [\ion{Ni}{2}] and
   [\ion{Fe}{2}] in the nebular phase and found that the bulk of radioactive
   $^{56}$Ni was moving with a maximum velocity of $\sim$3000\,km\,s$^{-1}$.
Additionally, at a higher radial velocity of about $+3900$\,km\,s$^{-1}$
   a unique feature in the infrared emission lines of [\ion{Fe}{2}] was found
   and interpreted as a fast-moving iron clump \citep{HCE_90}.
Using the effect of an occultation of this clump by the photosphere on days
   29 and 41, \citet{UCA_95} estimated its transversal velocity, which together
   with the radial velocity gives an absolute velocity of 4700\,km\,s$^{-1}$,
   and identified this high-velocity feature with a fast $^{56}$Ni clump
   whose luminosity corresponds to a mass of $\sim$10$^{-3}\,M_{\sun}$.
The analysis of the profiles of hydrogen emission lines in SN~1987A during
   the nebular phase showed that hydrogen is mixed down into the innermost
   ejecta to velocities of $\le 700$\,km\,s$^{-1}$ \citep{Chu_91, KF_98,
   MJS_12}.
The reconstruction of a three-dimensional (3D) view of H$\alpha$ emission
   \citep{LFS_16} and molecular hydrogen \citep{LSF_19} revealed that the
   lowest velocities of hydrogen are observed at 450\,km\,s$^{-1}$ and
   400\,km\,s$^{-1}$, respectively.
\citet{KF_98} also constrained quantitatively the mass of hydrogen-rich matter
   which expanded with velocities less than 2000\,km\,s$^{-1}$ to about
   2.2\,$M_{\sun}$.

Systematic optical and infrared photometric observations of the radioactive
   tail of SN~1987A carried out by different groups with different instruments
   permitted them to construct the bolometric light curve and, consequently,
   to estimate the total mass of ejected $^{56}$Ni -- one of the important
   properties of SN explosions.
\citet{DDA_88} obtained a $^{56}$Ni mass of 0.085\,$M_{\sun}$ at the Mount
   Stromlo and Siding Spring Observatories, \citet{SHM_88} got 0.071\,$M_{\sun}$
   at the Cerro Tololo Inter-American Observatory, and \citet{CWM_89} found
   0.078\,$M_{\sun}$ at the Sutherland Observatory of the South African
   Astronomical Observatory.
All these masses were obtained with the same distance modulus for
   the LMC of $m-M=18.5$\,mag, but different reddening $E(B-V)$.
Note that \citet{SB_90} found a range of 0.055\,$M_{\sun}$ to 0.090\,$M_{\sun}$
   for an admissible mass of radioactive $^{56}$Ni using the extreme values of
   the distance modulus to the LMC (18.3\,mag and 18.6\,mag) and the reddening
   (0.15\,mag and 0.20\,mag).
More reliable estimates of the $^{56}$Ni mass with all essential sources of
   uncertainties taken into account are ($0.069\pm0.003$)\,$M_{\sun}$
   \citep{BPS_91, McC_93} and ($0.071\pm0.003$)\,$M_{\sun}$ \citep{STM_14}.
   
One of the central characteristics of an SN explosion is the mass of oxygen
   in the ejecta.
Assuming the most favorable conditions for powering the luminosity of the
   forbidden oxygen doublet [\ion{O}{1}] $\lambda\lambda 6300, 6364\,\AA$,
   \citet{McC_93} estimated that a lower limit of the mass of oxygen is about
   0.3\,$M_{\sun}$.
\citet{LM_92} calculated the evolution of the intensity and the profile of the
   nebular [\ion{O}{1}] $\lambda\lambda 6300, 6364\,\AA$ doublet using
   a three-zone model with clumps of oxygen and fitted the observational data
   with an oxygen mass of 1.3\,$M_{\sun}$.
Identifying the fluctuations in the profile of the oxygen doublet [\ion{O}{1}]
   $\lambda\lambda 6300, 6364\,\AA$ with statistical fluctuations of
   the oxygen clump number, \citet{Chu_94} found the net oxygen mass to be 
   in the range of 1.2\,$M_{\sun}$ to 1.5\,$M_{\sun}$.
\citet{CCKC_97} analyzed the $2000-8000\,\AA$ spectrum of SN~1987A, taken with
   the HST eight years after the explosion, and concluded that the luminosity
   of the [\ion{O}{1}] $\lambda\lambda 6300, 6364\,\AA$ doublet is consistent
   with an oxygen mass in the range of 1.5\,$M_{\sun}$--2.0\,$M_{\sun}$.
Applying spectral synthesis non-LTE modeling to the nebular phase spectra,
   \citet{KF_98} found that the mass of oxygen gas moving with velocities
   less than 2000\,km\,s$^{-1}$ is about 1.9\,$M_{\sun}$.
In turn, \citet{JSS_15} used a more elaborate spectral synthesis modeling
   and estimated an oxygen mass as low as 0.7\,$M_{\sun}$, taking the existing
   uncertainties into account.
The mass of oxygen in SN~1987A is thus poorly constrained and ranges from
   0.7\,$M_{\sun}$ to 2.0\,$M_{\sun}$.

Since SN ejecta expand freely in the nebular phase, when the matter
   velocity is directly proportional to the radius, it is possible to combine
   the spectral information and the images along the line of sight to the
   observer to infer the 3D distribution of the ejecta for the
   observed emission lines.
\citet{KLF_10} studied the morphology of the ejecta using images and
   spectra for the emission lines of [\ion{Si}{1}]+[\ion{Fe}{2}]
   ($\lambda1.64\,\mu\mbox{m}$) and \ion{He}{1} ($\lambda2.058\,\mu\mbox{m}$)
   from the integral field spectroscopy on the Very Large Telescope
   (VLT) in Chile, which were obtained in October and November 2005.
\citet{LFK_13} presented spectral and imaging observations obtained with
   the HST between 1994 and 2011 and with the VLT in 2011 at optical and
   near-infrared wavelengths, particularly in the emission lines of H$\alpha$
   and [\ion{Si}{1}]+[\ion{Fe}{2}].
Both \citeauthor{KLF_10} and \citeauthor{LFK_13} stated that all these
   observations show that the ejecta morphology is highly non-spherical.
In the case of the [\ion{Si}{1}]+[\ion{Fe}{2}] and \ion{He}{1} lines,
   the innermost regions of the ejecta are seen and the observed emission
   represents well the density distribution of these elements.
\citet{KLF_10} approximated the exact 3D shape of the ejecta by a non-spherical
   distribution and found that an elongated triaxial ellipsoid fits the
   observations most accurately.

The early hard X-ray and gamma-ray observations of SN~1987A are a powerful tool
   to test 3D hydrodynamic simulations of neutrino-driven explosions.
These observations were carried out with the Roentgen Observatory, the Solar
   Maximum Mission, the X-ray astronomy satellite {\it Ginga}, and
   balloon-borne experiments.
\citet{ALM_19} computed the hard X-ray and gamma-ray emission based on
   3D neutrino-driven explosion simulations of \citet{WJM_13, WMJ_15, WJM_17}
   and compared the emergent emission with the corresponding spectra,
   continuum light curves, and line fluxes of SN~1987A.
\citet{JWJ_20} reproduced the gamma-ray decay lines and the UVOIR bolometric
   light curve of SN~1987A up to 600\,days by calculating the total gamma-ray
   deposition for these 3D simulations and found that an ejecta mass around
   14\,$M_{\sun}$ is favored for an explosion energy of
   $1.5\times10^{51}$\,erg.

In our previous papers we studied SN~1987A in the framework of the
   neutrino-driven explosion mechanism \citep{UWJM_15, UWJ_19} using
   available pre-SN models obtained in the scenario of single-star evolution
   \citep{SN_90, WPE_88, WHWL_97, SEWBJ_16, UWJ_19}.
The corresponding 3D explosion simulations with an approximate, parameterized
   neutrino engine, when it is tuned to yield the explosion energy observed in
   SN~1987A, and subsequent hydrodynamic light-curve modeling can explain only
   some basic observational data.
None of the computed models fits all observational constraints, namely,
   the total $^{56}$Ni mass ejected during the explosion, the amounts of
   outward $^{56}$Ni mixing and inward hydrogen mixing, the mass of hydrogen
   mixed to velocities lower than 2000\,km\,s$^{-1}$, and the oxygen mass
   in the SN ejecta.
It is noteworthy that only one model yields a maximum velocity of the bulk of
   $^{56}$Ni consistent with spectral observations of SN~1987A, and only
   this model is able to reproduce the dome of the light curve in reasonable
   agreement with the observed light curve. 
However, the corresponding progenitor model cannot fulfill the observational
   requirements of the location of the BSG Sanduleak\,$-69^{\circ}202$ star
   in the Hertzsprung-Russell diagram (HRD).

This failure in satisfying all basic observational constraints by an adequate
   progenitor model in the single-star evolution scenario and progress
   in explaining the formation of the mysterious triple-ring
   system in the binary-merger evolution scenario \citep{PMI_07, MP_09}
   motivated us to analyze binary-merger progenitor models in the framework of
   the neutrino-driven explosion mechanism as the next step in our study of
   SN~1987A.
\citet{MH_17} presented the results of a systematic study of binary-merger
   models for the progenitor of SN~1987A, which were evolved until the pre-SN
   stage.
The majority of these models are compact and hot BSG stars, of which six are
   located in the region of the observed properties of
   Sanduleak\,$-69^{\circ}202$ in the HRD.
Moreover, these evolutionary models reproduce the observed enrichment of
   helium and nitrogen in the triple-ring nebula, which is characterized by
   number ratios of $\mathrm{He/H} = 0.14 \pm 0.06$ \citep{FMP_11},
   $\mathrm{N/C} = 5 \pm 2$ and $\mathrm{N/O} = 1.1 \pm 0.4$ \citep{LF_96}.
An additional impulse to address the binary-merger progenitors comes from
   \citet{MUH_19}, who computed bolometric light curves using these
   binary-merger progenitors after exploding them by a piston model in
   spherical symmetry and applying artificial ``boxcar'' mixing of chemical
   species.
They obtained a much better agreement of the dome of the light
   curves with the observations of SN~1987A than that reported by
   \citet{UWJ_19} for the hydrodynamic models based on the single-star
   progenitor models.

Our present study of 3D explosion simulations of SN~1987A in the framework of
   neutrino-driven explosions based on binary-merger progenitors provides
   an opportunity to compare them with those based on the single-star
   progenitor models in \citet{UWJ_19}.
\citet{ONF_20} have recently performed 3D hydrodynamic simulations of mixing
   in SN~1987A triggering the explosion by injecting energy near the Fe/Si
   composition interface with parameterized aspherical perturbations 
   of the SN shock to mimic the effect of non-radial instabilities.
To reproduce the observed morphology of the ejecta, \citet{ONF_20}
   investigated two BSG progenitors: one model being the result of 
   a single-star evolution and another model based on a slow merger
   scenario developed by \citet{UTUY_18}.

Our paper is structured as follows.
In Section~\ref{sec:modmeth-psn} we briefly describe the binary-merger
   progenitor models and in Section~\ref{sec:modmeth-numrcs} we summarize
   our numerical approach to the 3D simulations of the neutrino-driven
   explosion and to the hydrodynamic light-curve modeling.
Section~\ref{sec:results} presents and analyzes the results obtained with
   the binary-merger pre-SN models and compares them with SN~1987A
   observations.
Section~\ref{sec:prmass} formulates possible requirements for the SN~1987A
   progenitor that are imposed by the hydrodynamic light-curve modeling.
In Section~\ref{sec:ssvsbm} we confront the hydrodynamic models based on
   the single-star and binary-merger pre-SN models with a set of the available
   observational constraints of SN~1987A.
Finally, in Section~\ref{sec:conclsns} we summarize our results.

\begin{deluxetable*}{ l c c c c c c c c c c c c c c c}
\tabletypesize{\small}
\tablewidth{0pt}
\tablecaption{Binary-merger presupernova models for blue supergiants%
\label{tab:presnm}}
\tablehead{
\colhead{Model} & \colhead{$M_1$}
                & \colhead{$M_2$}
                & \colhead{$f_\mathrm{c}$}
                & \colhead{$M_\mathrm{CO}^{\,\mathrm{core}}$}
                & \colhead{$M_\mathrm{He}^{\,\mathrm{core}}$}
                & \colhead{$M_\mathrm{He}^{\,\mathrm{shell}}$}
                & \colhead{$M_\mathrm{pSN}$}
                & \colhead{$\Delta R_\mathrm{He}^{\,\mathrm{shell}}$}
                & \colhead{$R_\mathrm{pSN}$}
                & \colhead{$X_\mathrm{surf}$}
                & \colhead{$Y_\mathrm{surf}$}
                & \colhead{$Z_\mathrm{surf}$}
                & \colhead{He/H$_\mathrm{surf}$}
                & \colhead{N/C$_\mathrm{surf}$}
                & \colhead{N/O$_\mathrm{surf}$} \\
\colhead{} & \multicolumn{2}{c}{$(M_{\sun})$}
           & \colhead{$(\%)$}
           & \multicolumn{4}{c}{$(M_{\sun})$}
           & \multicolumn{2}{c}{$(R_{\sun})$}
           & \multicolumn{2}{c}{}
           & \colhead{$(10^{-2})$}
           & \multicolumn{3}{c}{}
}
\startdata
 M15-7b & 15 & 7 & 17.5 & 2.48 & 2.90 & 0.42 & 21.06 & 0.117 & 37.0 & 0.6498 & 0.3447 & 0.5542 & 0.134 & 7.11 & 1.30 \\
 M15-8b & 15 & 8 & 17.5 & 2.50 & 2.95 & 0.46 & 22.05 & 0.143 & 31.8 & 0.6529 & 0.3416 & 0.5542 & 0.132 & 6.34 & 1.26 \\
 M16-4a & 16 & 4 & \phantom{e}3.3 & 3.02 & 4.10 & 1.08 & 19.00 & 0.275 & 34.9 & 0.6458 & 0.3487 & 0.5536 & 0.136 & 7.87 & 1.40 \\
 M16-7b & 16 & 7 & 16.6 & 2.81 & 3.41 & 0.60 & 21.98 & 0.182 & 37.3 & 0.6398 & 0.3547 & 0.5536 & 0.140 & 7.55 & 1.38 \\
 M17-7a & 17 & 7 & 15.6 & 3.29 & 4.25 & 0.96 & 22.82 & 0.182 & 34.5 & 0.6540 & 0.3404 & 0.5538 & 0.131 & 6.44 & 1.31 \\
 M17-8a & 17 & 8 & 15.6 & 3.29 & 4.23 & 0.94 & 23.81 & 0.221 & 33.4 & 0.6574 & 0.3371 & 0.5539 & 0.129 & 6.07 & 1.27 \\
\enddata
\tablecomments{%
The columns give the name of the pre-SN model;
   the primary mass, $M_1$;
   the secondary mass, $M_2$;
   the fraction of the He core of the primary that was dredged up,
   $f_\mathrm{c}$;
   the CO-core mass, $M_\mathrm{CO}^{\,\mathrm{core}}$;
   the He-core mass, $M_\mathrm{He}^{\,\mathrm{core}}$;
   the He-shell mass, $M_\mathrm{He}^{\,\mathrm{shell}}$;
   the pre-SN mass, $M_\mathrm{pSN}$; 
   the width of the He shell, $\Delta R_\mathrm{He}^{\,\mathrm{shell}}$;
   the pre-SN radius, $R_\mathrm{pSN}$;
   the mass fractions of hydrogen, $X_\mathrm{surf}$,
      helium, $Y_\mathrm{surf}$, heavy elements, $Z_\mathrm{surf}$,
   the number ratios of helium to hydrogen, He/H$_\mathrm{surf}$,
      nitrogen to carbon, N/C$_\mathrm{surf}$, and nitrogen to oxygen,
      N/O$_\mathrm{surf}$,
   in the hydrogen-rich envelope at the stage of core collapse.
}
\vspace{-3.5em}
\end{deluxetable*}
%
\section{Model overview and numerical approach}
\label{sec:modmeth}
%
We consider six pre-SN models obtained in the scenario of binary-merger
   evolution and use these models as the initial data in our 3D simulations.
As in our previous papers, we follow the numerical approach presented in
   \citet{UWJM_15}.
It consists of the three steps: the 3D neutrino-driven explosion simulation
   until about one day after core collapse, the subsequent mapping of the
   3D explosion model to 1D geometry, and the radiation-hydrodynamic modeling
   of the light curve.
We briefly review each step in the corresponding subsection.

\subsection{Presupernova models}
\label{sec:modmeth-psn}
%
\citet{MH_17} carried out a stellar evolution study of binary-merger models
   for the progenitor of SN~1987A, based on the scenario proposed by
   \citet{PJH_92, PMI_07}.
In this scenario a main-sequence secondary star merges with a primary RSG
   at the end of core helium burning when helium is depleted down to
   a mass fraction of $1\%$ in the CO rich convective core,\footnote%
     {We define the helium-core mass as the mass enclosed by the shell
     where the mass fraction of hydrogen $X$ drops below a value of $X=0.01$
     when moving inward from the surface of a star.
     The CO-core mass is determined in the same way, but the mass fraction
     of helium $Y$ having a value of $Y=0.01$.}
   surrounded by a helium layer and an extended, expanding hydrogen envelope
   driving the merger.
During the resulting common envelope phase a fraction of the hydrogen-rich
   secondary mass penetrates the helium core of the primary, whereby
   an equivalent mass fraction, $f_\mathrm{c}$, of the He core is dredged up
   and mixed uniformly in the envelope.
\citeauthor{MH_17} explored an extended parameter space: primary masses of
   15\,$M_{\sun}$, 16\,$M_{\sun}$, and 17\,$M_{\sun}$; secondary masses of
   2\,$M_{\sun}$, 3\,$M_{\sun}$,..., 8\,$M_{\sun}$; and different depths
   up to which the secondary star penetrates into the He core of the primary
   star during the common-envelope phase.
After the common-envelope phase, the merged star is evolved until just before
   iron-core collapse.
Six of the 84 pre-SN models match the color-luminosity properties of the BSG
   Sanduleak\,$-69^{\circ}202$ star.

\begin{figure*}[t]
\centering
   \includegraphics[width=0.48\hsize, clip, trim=18 153 67 99]{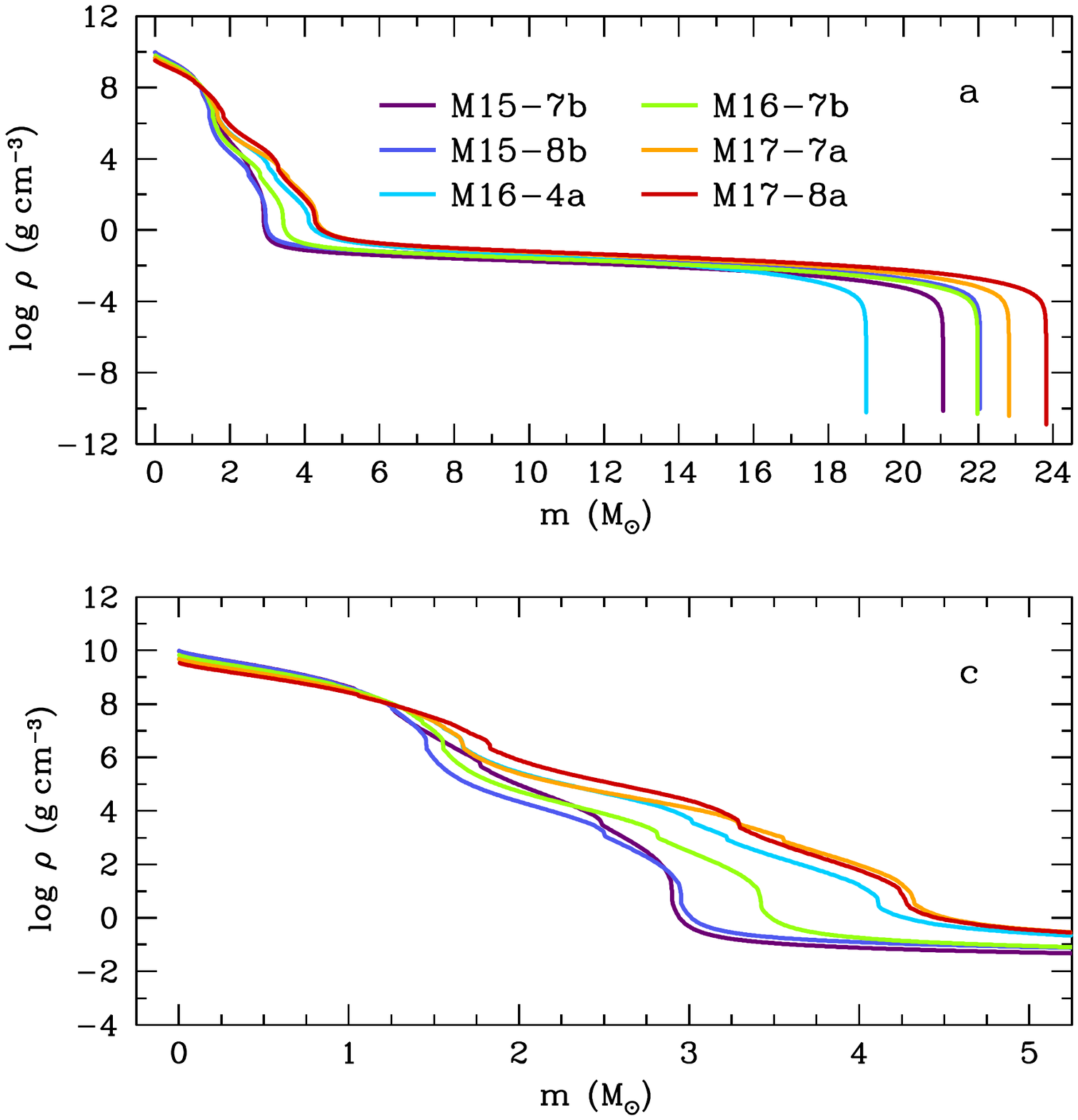}
   \hspace{0.5cm}
   \includegraphics[width=0.48\hsize, clip, trim=18 153 67 99]{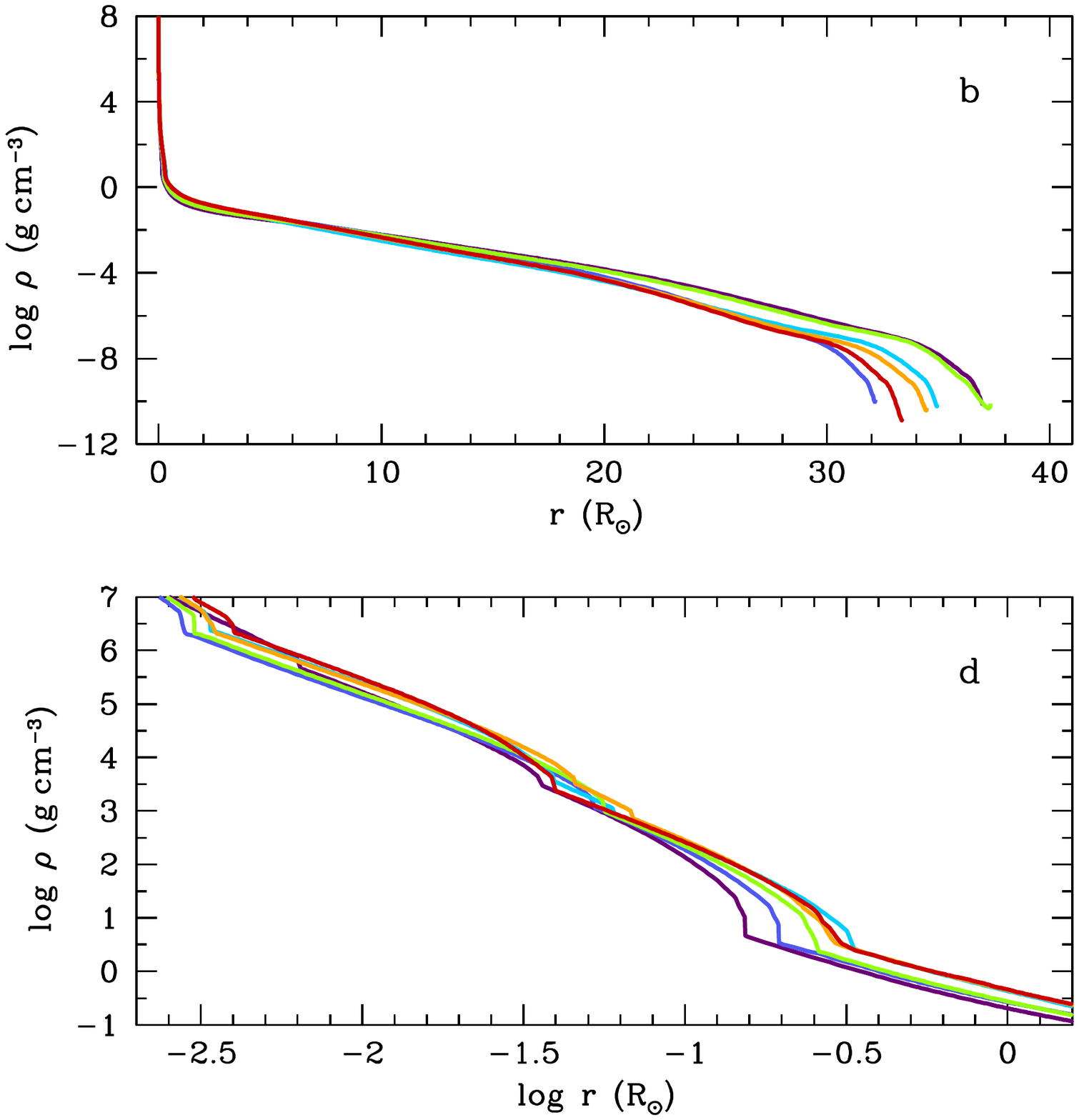}\\
   \caption{%
   Density distributions as functions of interior mass for the whole star
      (a) and the inner region of 5\,$M_{\sun}$ (c), and as functions of
      radius for the whole star (b) and the inner region of 1\,$R_{\sun}$ (d)
      in the pre-SN models M15-7b, M15-8b, M16-4a, M16-7b, M17-7a, and M17-8a
      given in Table~\ref{tab:presnm}.
   }
   \label{fig:denmr}
\end{figure*}

\begin{figure*}[t]
\centering
   \includegraphics[width=0.33\hsize, clip, trim=37 163 323 213]{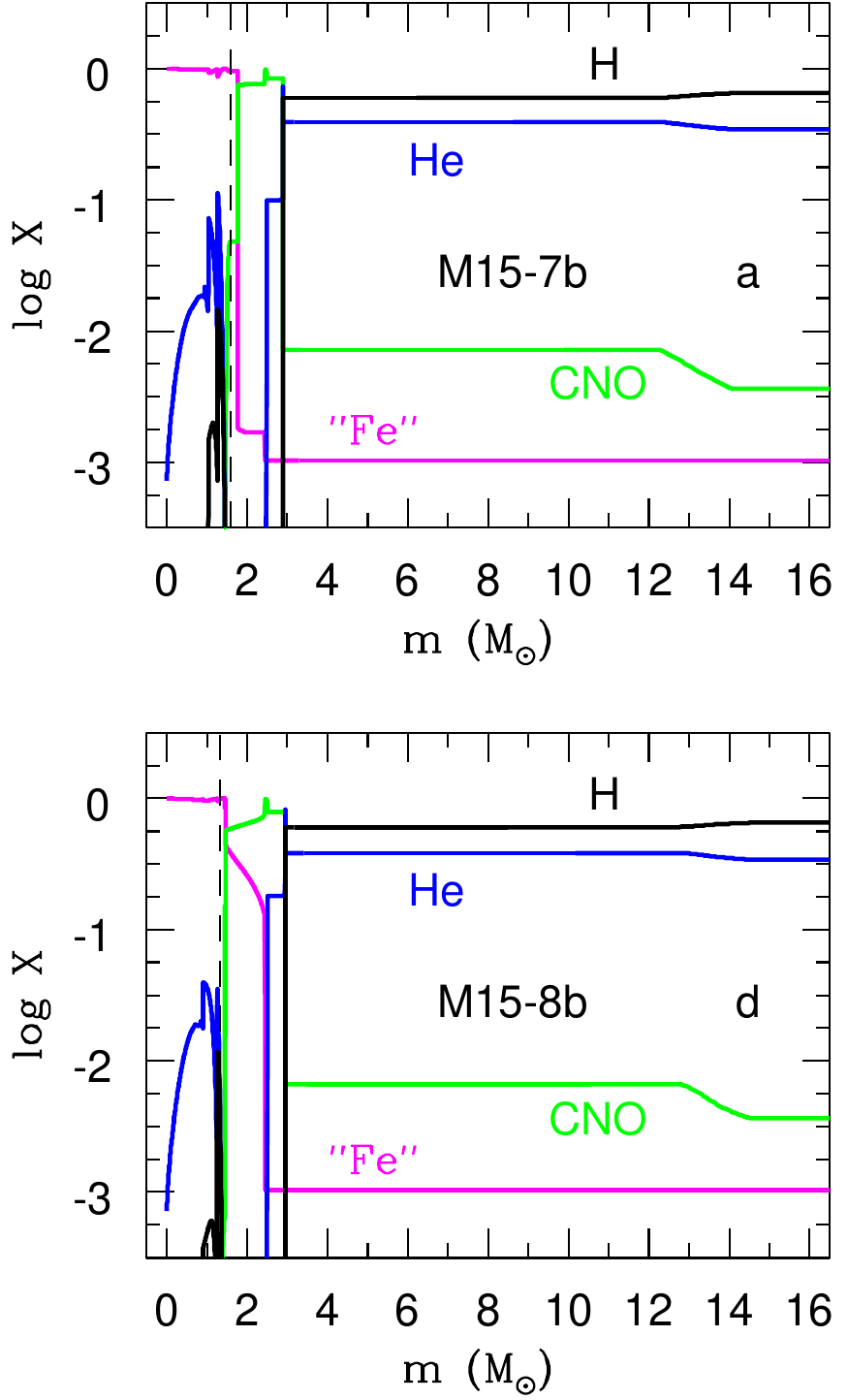}
   \includegraphics[width=0.33\hsize, clip, trim=37 163 323 213]{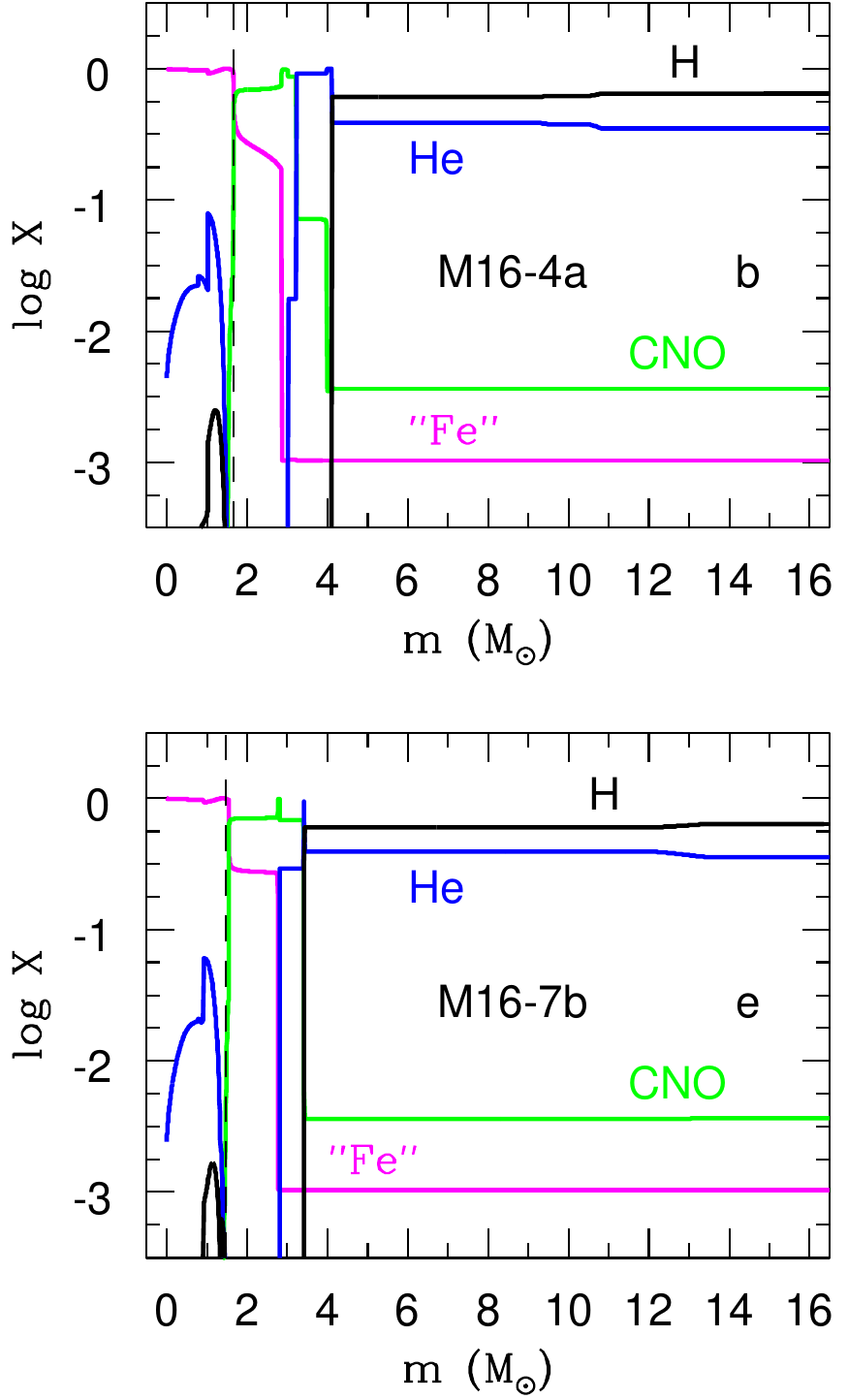}
   \includegraphics[width=0.33\hsize, clip, trim=37 163 323 213]{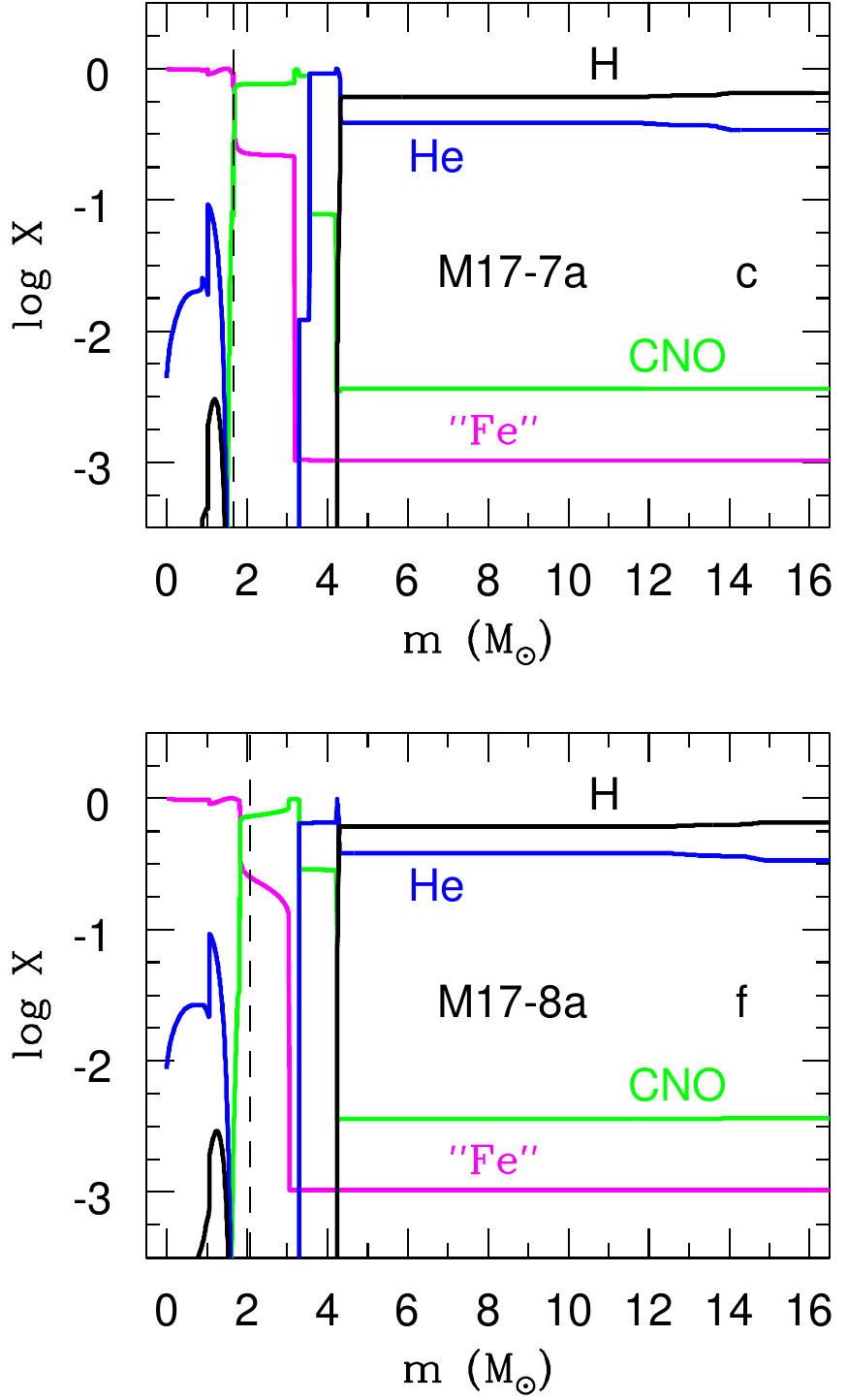}\\
   \caption{%
   Mass fractions of hydrogen (black line), helium
      (blue line), CNO group elements (green line),
      and ``iron-group'' elements containing the heavy elements starting
      from silicon (magenta line) in the pre-SN models
      M15-7b (a), M16-4a (b), M17-7a (c), M15-8b (d), M16-7b (e), and
      M17-8a (f) (Table~\ref{tab:presnm}).
   The vertical black dashed lines mark the location of the final mass cut
      in the corresponding six reference 3D explosion models M15-7b-3,
      M15-8b-1, M16-4a-1, M16-7b-2, M17-7a-2, and M17-8a-4
      (Table~\ref{tab:hydmod}).
   }
   \label{fig:chcom}
\end{figure*}

We have investigated these pre-SN models M15-7b, M15-8b, M16-4a, M16-7b, M17-7a,
   and M17-8a, whose basic properties are listed in Table~\ref{tab:presnm},
   and whose structure and chemical composition are illustrated by
   Figures~\ref{fig:denmr} and \ref{fig:chcom}.
These binary-merger models having primary stars of 15\,$M_{\sun}$ to
   17\,$M_{\sun}$ ($M_1$), secondaries from 4\,$M_{\sun}$ to
   8\,$M_{\sun}$ ($M_2$), and a fraction of the He core of the primary that
   was dredged up of $3.3\%$ to $17.5\%$ evolve to
   compact blue pre-SN models with masses from 19.00\,$M_{\sun}$ to
   23.81\,$M_{\sun}$ ($M_\mathrm{pSN}$) and radii of 31.8\,$R_{\sun}$ to
   37.3\,$R_{\sun}$ ($R_\mathrm{pSN}$).

Because some fraction of the He-shell mass of the helium core\footnote%
     {Note that \citet{MH_17} restrict mixing to be inside the helium shell
     in their models, quantified by the parameter $f_\mathrm{sh}$ in
     their work.}
   of the primary star is dredged up during the merger, all resultant pre-SN
   models have helium-core masses ($M_\mathrm{He}^{\,\mathrm{core}}$) lower
   than those of the initial primary stars, while their CO-core masses
   ($M_\mathrm{CO}^{\,\mathrm{core}}$) remain practically unchanged.
As a result, these pre-SN models have much higher mass ratios of the CO core
   to the helium core in the range of 0.737--0.855 (Table~\ref{tab:presnm})
   compared to the progenitor models that are evolved in the single-star
   scenario to a BSG configuration with this mass ratio being in the range of
   0.392--0.414 \citep{UWJ_19}.

The helium-core masses of 2.90\,$M_{\sun}$ to 4.25\,$M_{\sun}$
   (Table~\ref{tab:presnm}) are much less than 
   the 6\,$M_{\sun}$ suggested for Sanduleak\,$-69^{\circ}202$
   as a single star \citep{SNK_88, WPE_88}.
This should favor stronger outward
   mixing of radioactive $^{56}$Ni into the hydrogen-rich envelope as
   observed in SN~1987A, because 3D neutrino-driven explosion
   simulations with BSG models obtained in the scenario of single-star
   evolution exhibit $^{56}$Ni mixing whose strength scales very roughly
   inversely with the helium-core mass \citep{UWJ_19}.
In turn, the unchanged CO-core masses of 2.48\,$M_{\sun}$ to 3.29\,$M_{\sun}$
   inherited from moderately massive primary stars (Table~\ref{tab:presnm})
   should allow for a better match of the oxygen mass observed in
   the SN~1987A ejecta.

The pre-SN models have different helium core masses in a relatively narrow
   range of 2.90\,$M_{\sun}$ to 4.25\,$M_{\sun}$ (Table~\ref{tab:presnm},
   Figures~\ref{fig:denmr}(a) and (c)), and high ratios of the CO-core mass
   to the helium-core mass in the range of 0.737--0.855 (Table~\ref{tab:presnm}).
These facts imply that the structure of the helium core in these models should
   be quite different from that in the single-star progenitor models.
Indeed, the difference becomes evident from an inspection of
   Figure~\ref{fig:denmr}(d), Figure~1(d) in \citet{UWJM_15}, and Figure~1(d)
   in \citet{UWJ_19}, and we extend our previous work on analyzing the
   sensitivity of the amount of outward $^{56}$Ni mixing and inward hydrogen
   mixing to the structure of the helium core and the He/H composition
   interface.

\begin{deluxetable*}{ l c c c c c c c c c c c c c c c }
\tabletypesize{\small}
\tablewidth{0pt}
\tablecaption{Basic properties of the 3D explosion models%
\label{tab:hydmod}}
\tablehead{
\colhead{Model} & \colhead{$M_\mathrm{NS}$}
                & \colhead{$M_\mathrm{ej}$}
                & \colhead{$E_\mathrm{exp}$}
       & \colhead{$M_\mathrm{Ni}^{\,\mathrm{min}}$}
       & \colhead{$M_\mathrm{Ni}^{\,\mathrm{max}}$}
       & \colhead{$M_\mathrm{Ni}^{\,\mathrm{i}}$}
       & \colhead{$M_\mathrm{Ni}^{\,\mathrm{f}}$}
       & \colhead{$v_\mathrm{Ni}^{\,\mathrm{bulk}}$}
       & \colhead{$\langle v \rangle_\mathrm{Ni}^\mathrm{tail}$}
       & \colhead{$v_\mathrm{H}^{\,\mathrm{mix}}$}
       & \colhead{$\delta M_\mathrm{H}^\mathrm{mix}$}
       & \colhead{$\Delta M_\mathrm{H}^\mathrm{2000}$}
       & \colhead{$M_\mathrm{O}$}
       & \colhead{$t_\mathrm{map}$}
       & \colhead{$t_\mathrm{SB}$} \\
\colhead{} & \multicolumn{2}{c}{$(M_{\sun})$}
       & \colhead{(B)}
       & \multicolumn{4}{c}{$(10^{-2}\,M_{\sun})$}
       & \multicolumn{3}{c}{(km\,s$^{-1}$)}
       & \multicolumn{3}{c}{$(M_{\sun})$}
       & \multicolumn{2}{c}{($10^{3}$\,s)}
}
\startdata
M15-7b-1 & 1.57 & 19.48 & 1.404 & 3.86 & 14.23 & 7.37 & 7.34 & 3018 & 3185 & 29 & 0.48 & 3.21 & 0.83 & 86.39 & 5.32 \\
M15-7b-2 & 1.55 & 19.49 & 1.428 & 4.91 & 14.96 & 7.34 & 7.31 & 3406 & 3607 & 28 & 0.48 & 3.07 & 0.84 & 86.40 & 5.34 \\
M15-7b-3 & 1.58 & 19.46 & 1.432 & 4.79 & 14.64 & 7.31 & 7.28 & 2980 & 3193 & 29 & 0.49 & 3.10 & 0.83 & 86.39 & 5.20 \\
M15-7b-4 & 1.49 & 19.56 & 1.780 & 4.88 & 17.08 & 7.33 & 7.31 & 3344 & 3553 & 28 & 0.50 & 2.20 & 0.86 & 86.40 & 4.74 \\
\noalign{\smallskip}
\hline
\noalign{\smallskip}
M15-8b-1 & 1.32 & 20.73 & 1.567 & 2.67 & 12.07 & 7.42 & 7.40 & 1829 & 1949 & 27 & 0.52 & 3.14 & 0.96 & 86.41 & 4.54 \\
M15-8b-2 & 1.38 & 20.66 & 1.116 & 2.32 &  8.63 & 7.37 & 7.32 & 1439 & 1535 & 27 & 0.44 & 5.50 & 0.96 & 86.40 & 5.29 \\
\noalign{\smallskip}
\hline
\noalign{\smallskip}
M16-4a-1 & 1.66 & 17.34 & 1.562 & 4.84 & 14.99 & 7.50 & 7.41 & 2436 & 2732 & 29 & 0.55 & 2.13 & 0.97 & 86.41 & 4.26 \\
M16-4a-2 & 1.87 & 17.13 & 1.068 & 3.79 & 10.25 & 7.74 & 7.28 & 2237 & 2445 & 32 & 0.43 & 3.55 & 0.88 & 86.39 & 5.02 \\
\noalign{\smallskip}
\hline
\noalign{\smallskip}
M16-7b-1 & 1.53 & 20.44 & 1.168 & 2.83 & 10.00 & 7.35 & 7.26 & 1639 & 1759 & 29 & 0.51 & 5.01 & 1.13 & 86.41 & 5.87 \\
M16-7b-2 & 1.46 & 20.51 & 1.412 & 2.97 & 11.47 & 7.39 & 7.34 & 1714 & 1838 & 28 & 0.56 & 3.80 & 1.16 & 86.40 & 5.35 \\
\noalign{\smallskip}
\hline
\noalign{\smallskip}
M17-7a-1 & 1.68 & 21.13 & 1.516 & 4.28 & 13.91 & 7.48 & 7.36 & 1807 & 1910 & 30 & 0.59 & 3.75 & 1.28 & 86.42 & 4.65 \\
M17-7a-2 & 1.67 & 21.14 & 1.559 & 4.60 & 14.10 & 7.52 & 7.41 & 1713 & 1888 & 29 & 0.56 & 3.60 & 1.29 & 86.41 & 4.55 \\
\noalign{\smallskip}
\hline
\noalign{\smallskip}
M17-8a-3 & 2.22 & 21.58 & 1.075 & 7.68 & 11.67 & 7.69 & 7.22 & 2461 & 2653 & 32 & 0.51 & 6.41 & 0.90 & 86.39 & 5.27 \\
M17-8a-4 & 2.06 & 21.75 & 1.216 & 6.52 & 12.67 & 7.51 & 7.25 & 2411 & 2520 & 32 & 0.53 & 5.62 & 1.02 & 86.41 & 5.02 \\
\enddata
\tablecomments{%
The 3D explosion models are based on the corresponding pre-SN models
   of Table~\ref{tab:presnm}.
$M_\mathrm{NS}$ is the baryonic mass of the neutron star at the end of the 3D
   simulations;
   $M_\mathrm{ej}$ the ejecta mass;
   $E_\mathrm{exp}$ the explosion energy;
   $M_\mathrm{Ni}^{\,\mathrm{min}}$ the mass of radioactive $^{56}$Ni produced
      directly by our $\alpha$-chain reaction network;
   $M_\mathrm{Ni}^{\,\mathrm{max}}$ the aggregate mass of directly produced
      $^{56}$Ni and tracer nucleus;
   $M^{i}_{\mathrm{Ni}}$ the initial $^{56}$Ni mass at the onset of the light
      curve modeling;
   $M^{f}_{\mathrm{Ni}}$ the $^{56}$Ni mass ejected at day 150;
   $v_\mathrm{Ni}^{\,\mathrm{bulk}}$ the maximum velocity of the bulk mass of
      $^{56}$Ni;
   $\langle v \rangle_\mathrm{Ni}^\mathrm{tail}$ the mean velocity of the fast
      moving $^{56}$Ni tail;
   $v_\mathrm{H}^{\,\mathrm{mix}}$ the minimum velocity of hydrogen mixed
      into the He shell, specified at the level where the mass fraction of
      hydrogen $X$ drops to a value of $X=0.01$;
   $\delta M_\mathrm{H}^\mathrm{mix}$ the mass of hydrogen mixed into
      the He shell;
   $\Delta M_\mathrm{H}^\mathrm{2000}$ the mass of hydrogen confined to the
      inner layers ejected with velocities less than 2000\,km\,s$^{-1}$; and
   $M_\mathrm{O}$ the total mass of oxygen in the ejecta.
$t_\mathrm{map}$ is the time at which the 3D simulations are mapped to
   a spherically symmetric grid.
$t_\mathrm{SB}$ is the epoch of shock breakout in the 1D simulations.
}
\vspace{-3.5em}
\end{deluxetable*}
The initial, on the main sequence, chemical composition of the primary and
   secondary stars was taken to be representative of the subsolar metallicity
   of the LMC.
To produce such a metallicity, the present-day solar chemical composition
   with the mass fractions of hydrogen $X = 0.7381$, helium $Y = 0.2485$,
   and metals $Z = 0.0134$ \citep{AGSS_09} was scaled by reducing the total
   abundance of heavy elements by about a factor of 2.5 compared to solar.
The chemical composition of the LMC becomes $X = 0.739$, $Y = 0.255$,
   and $Z = 0.0055$.
During the binary-merger evolution, the chemical composition at the surface
   of the primary
   is enriched with helium and the ashes of CNO-burning dredged up from the
   core to the envelope, yielding high number ratios of helium to hydrogen
   (0.129--0.140), nitrogen to carbon (6.07--7.87), and nitrogen to oxygen
   (1.26--1.40) (Table~\ref{tab:presnm}), which are comparable with
   the observational data \citep{FMP_11, LF_96}.
The resultant chemical compositions of the pre-SN models M15-7b, M16-4a, M17-7a,
   M15-8b, M16-7b, and M17-8a are shown in Figure~\ref{fig:chcom}. 

\subsection{Numerical methods}
\label{sec:modmeth-numrcs}
%
For completeness, we briefly summarize the numerical methods used in our work
   in the following, basically repeating the information that we
   already provided in our previous paper on SN~1987A \citep{UWJ_19},
   because our methodical approach here is identical with the one applied 
   there.

Our 3D neutrino-driven explosion simulations begin shortly after the stellar
   core has collapsed and a newly formed SN shock wave has propagated to a mass
   coordinate of approximately 1.25\,$M_{\sun}$ inside the iron core.
The evolution during core collapse and core bounce until about 10\,ms after
   bounce is computed in spherical symmetry (because non-radial hydrodynamic
   instabilities are not expected to grow until this time).
These 1D post-bounce data are then mapped onto a 3D grid.
The subsequent 3D calculations are carried out with the time-explicit
   finite-volume Eulerian multifluid hydrodynamics code {\sc Prometheus}
   \citep{FAM_91, MFA_91a, MFA_91b}.
Details of the physics modules implemented into the {\sc Prometheus} code
   and our numerical setup have been described in \citet{WJM_13} for
   neutrino-driven explosion simulations and in \citet{WMJ_15} for
   simulations of the late-time evolution from approximately 1.3\,s after core
   bounce onward.
Nevertheless, we briefly summarize the input physics and numerical methods
   employed by the {\sc Prometheus} code as follows.

The {\sc Prometheus} code uses a dimensionally split version of the
   piecewise parabolic method \citep{CW_84} to solve the multidimensional
   hydrodynamic equations.
A fast and efficient Riemann solver for real gases \citep{CG_85} is used
   to compute numerical fluxes at cell boundaries. 
Inside grid cells, where a strong shock wave is present, we recompute the 
   inter-cell fluxes using an approximate Riemann solver \citep{Liou_96} 
   to prevent numerical artifacts known as the odd-even decoupling 
   \citep{Quirk_94}.
The Yin-Yang overlapping grid \citep{KS_04}, implemented into {\sc Prometheus}
   as in \citet{WHM_10}, is employed for efficient spatial discretization of
   the computational domain.
Newtonian self-gravity is taken into account by solving Poisson's equation
   in its integral form, using an expansion into spherical harmonics
   \citep{MS_95}.
In addition, a general relativistic correction of the monopole term of the
   gravitational potential is applied during the explosion simulations
   following \citet{SKJM_06} and \citet{AJS_07}.

To model the explosive nucleosynthesis approximately, a small $\alpha$-chain
   reaction network, similar to the network described in \citet{KPSJM_03},
   is solved.
In order to unambiguously determine the inward mixing of hydrogen, free
   protons, which are produced when neutrino-heated matter freezes out
   from nuclear statistical equilibrium, are distinguished from hydrogen
   originating from the hydrogen-rich stellar envelope by tagging them as
   different species in our multicomponent treatment of the stellar plasma.

The revival of the stalled SN shock and the explosion are triggered by imposing
   a suitable value of the neutrino luminosities at an inner radial grid
   boundary located at an enclosed mass of 1.1\,$M_{\sun}$, well inside the
   neutrinosphere.
Outside this boundary, which shrinks with time to mimic the contraction of
   the proto-neutron star, we model neutrino-matter interactions by solving
   the neutrino radiation transport equation in a ``ray-by-ray'' manner and
   in the gray approximation as described in \citet{SKJM_06}.
The explosion energy of the model is determined by the imposed isotropic
   neutrino luminosity, whose temporal evolution we prescribe as well, and
   by the accretion luminosity that results from the progenitor-dependent mass
   accretion rate and the gravitational potential of the contracting neutron
   star.
 
Our 3D calculations are terminated at approximately one day after the onset of
   the explosion, i.e. long after the SN shock has swept through the entire
   progenitor star and has broken out from the stellar surface
   (about 1--2 hours after core bounce; see Table~\ref{tab:hydmod}).
We use a two-step procedure of mapping at two different mapping epochs:
   an early-time mapping epoch well prior to the phase of shock breakout and
   a late-time mapping epoch when the 3D explosion simulations are terminated.
The early-time mapping moment is chosen to adequately treat the shock breakout,
   because this breakout is a complex process including the effects of both
   hydrodynamics and radiative transfer and has to be modeled with a radiation
   hydrodynamics code.
To this end, we compute angle-averaged profiles of hydrodynamic quantities
   and chemical abundances of the 3D flow and interpolate these profiles
   onto the Lagrangian (mass) grid used in the 1D simulations.
The resulting data are the initial conditions for the hydrodynamic modeling
   of the SN outburst.
The further time evolution of the SN outburst is modeled in one dimension.
At the early-time mapping epoch the hydrodynamic flow is far from
   homologous expansion.
This implies that outward mixing of radioactive
   $^{56}$Ni and inward mixing of hydrogen-rich matter in the ejecta will
   continue until complete homology is reached.
To capture this mixing, we map angle-averaged profiles of
   chemical abundances of the 3D flow onto the Lagrangian grid at the
   terminal time of the 3D simulations, when the ejecta expand almost
   homologously.

After mapping the 3D simulations data to a 1D grid, the evolution of the SN
   outburst is modeled with the time-implicit Lagrangian radiation hydrodynamics
   code {\sc Crab} \citep{Utr_04, Utr_07}.
It integrates the set of spherically symmetric hydrodynamic equations including
   self-gravity and a radiation transfer equation in the gray approximation
   \citep[e.g.,][]{MM_84}.
The whole SN ejecta can be divided into two regions: the inner optically thick
   core and the outer semitransparent and optically thin layers.
In the inner opaque core, the radiation transport is reduced to the diffusion of
   equilibrium radiation in the approximation of radiative heat conduction,
   and the assumption of LTE is valid.
In the semitransparent and transparent layers, the time-dependent radiative
   transfer equation, written in a comoving frame of reference to an accuracy
   of order $v/c$ ($v$ is the fluid velocity, and $c$ is the speed of light),
   is solved as a system of equations for the zeroth and first angular moments
   of the nonequilibrium radiation intensity.
To close this system of two moment equations, a variable Eddington factor
   is directly calculated with the scattering terms included explicitly
   in the source function.
The total set of equations is discretized spatially using the method of
   lines \citep[e.g.,][]{HNW_93, HW_96}.
The resultant system of ordinary differential equations is integrated
   by the implicit method of \citet{Gea_71} with an automatic choice of both
   the time integration step and the order of accuracy of the method.

The energy deposition of gamma-rays from the decay chain $^{56}$Ni
   $\to ^{56}$Co $\to ^{56}$Fe is determined by solving the gamma-ray
   transport, while the corresponding positrons are assumed to deposit their
   kinetic energy locally.
The ionization balance for an ideal gas in a nonequilibrium radiation field
   and under the influence of non-thermal processes provides the level
   populations to calculate the equation of state, the mean opacities, and
   the thermal emission coefficient.
It includes the elements H, He, C, N, O, Ne, Na, Mg, Si, S, Ar, Ca, Fe, and
   the negative hydrogen ion H$^{-}$.
In addition, in expanding SN ejecta with a velocity gradient, the contribution
   of spectral lines to the opacity is evaluated by the generalized formula
   of \citet{CAK_75}.
We refer to \citet{UWJM_15} and references therein for more details of
   the numerical setup.

\section{Results}
\label{sec:results}
%
We carried out fourteen 3D neutrino-driven explosion simulations with the
   six binary-merger pre-SN models M15-7b, M15-8b, M16-4a, M16-7b, M17-7a,
   and M17-8a (Table~\ref{tab:presnm}) as initial data.
Table~\ref{tab:hydmod} lists basic properties of these 3D hydrodynamic models
   that were extracted at the end of the simulations.
We define the explosion energy, $E_\mathrm{exp}$, as the sum of the total
   (i.e., internal plus kinetic plus gravitational) energy of all grid cells
   at the early-time mapping epoch.
Throughout this paper, we employ the energy unit
   $1\,\mathrm{bethe}=1\,\mathrm{B}=10^{51}$\,erg.
To characterize mixing of radioactive $^{56}$Ni in radial velocity space,
   we divide the $^{56}$Ni-rich ejecta into two components: a slow-moving
   bulk of $^{56}$Ni containing $99\%$ of the total $^{56}$Ni mass
   (except for places in which it is specially defined), and a fast-moving
   $^{56}$Ni tail containing the remaining $1\%$.
This division is motivated by the observational evidence for a fast $^{56}$Ni
   clump of $\sim$10$^{-3}\,M_{\sun}$ in SN~1987A \citep{UCA_95}.


\subsection{Production of $^{56}$Ni in neutrino-driven simulations}
\label{sec:results-niprod}
%
Along with the explosion energy, the total amount of radioactive $^{56}$Ni
   plays a crucial role in powering the dome-like maximum of the observed
   light curve of SN~1987A.
To study the production of $^{56}$Ni during our 3D SN simulations,
   we solved a small $\alpha$-chain reaction network from helium through
   $^{56}$Ni and are therefore unable to determine the mass fraction
   of $^{56}$Ni in neutron-rich matter very accurately.
The so-called ``tracer'' nucleus collects iron-group and trans-iron species
   that are formed in neutron-rich ($Y_e < 0.49$) ejecta
   \citep{KPSJM_03, WJM_13}.
A possible overall production of radioactive $^{56}$Ni falls in between the
   minimum and maximum values: the mass of $^{56}$Ni produced directly by
   our $\alpha$-chain reaction network, $M_\mathrm{Ni}^{\,\mathrm{min}}$,
   and the aggregate mass of directly produced $^{56}$Ni plus tracer nucleus,
   $M_\mathrm{Ni}^{\,\mathrm{max}}$.
3D neutrino-driven explosion simulations of SN~1987A based on the single-star
   pre-SNe showed that the $^{56}$Ni production is proportional to
   the explosion energy with the correlation being valid for both the
   minimum and maximum values \citep{UWJM_15, UWJ_19}.
It should be noted that in those simulations the values of the explosion
   energy are clearly different to guarantee this correlation.
In turn, all of the 3D neutrino-driven explosion simulations based on the
   binary-merger pre-SNe confirm the results of \citet{UWJM_15, UWJ_19}
   for the single-star pre-SNe except for two pairs of hydrodynamic models
   (M15-7b-2, M15-7b-3) and (M17-8a-3, M17-8a-4) (Table~\ref{tab:hydmod}).
These exceptions are worth to be discussed in the context of uncertainties
   of the production of $^{56}$Ni \citep[see][for details]{EWS_20} and
   properties of the explosion mechanism itself.

Models M15-7b-2 and M15-7b-3 which explode with nearly the same energies of
   1.428\,B and 1.432\,B, respectively, produce nearly the same mass of
   radioactive $^{56}$Ni plus neutron-rich tracer nucleus, but slightly violate
   the above $^{56}$Ni production--explosion energy correlation for both
   the minimum and maximum amounts of $^{56}$Ni (Table~\ref{tab:hydmod}).
This small difference in the $^{56}$Ni production between models M15-7b-2
   and M15-7b-3 can arise from a number of reasons. 
First, \citet{EWS_20} argued that the mass of nucleosynthesis products that
   scales with the explosion energy should include all other elements that
   are present in neutrino-heated ejecta (like $\alpha$-particles, free
   neutrons, and protons) in addition to radioactive $^{56}$Ni and the tracer
   nucleus.
In other words, we have to consider at least the maximum amount of $^{56}$Ni
   as a measure of the explosion energy rather than the minimum amount, but
   depending on the entropy of the ejecta significant amounts of alpha
   particles and free nucleons may also be ejected.
Second, such a small scattering in the $^{56}$Ni production can result from
   a simple stochasticity of 3D hydrodynamic instabilities and turbulence.
Third, it seems that model M15-7b-3 makes more fallback and therefore some
   of the initially expelled material in the neutrino-heated ejecta falls back,
   more than in model M15-7b-2, because the mass of the neutron star in
   model M15-7b-3 is larger than in model M15-7b-2.
Note that in model M15-7b-3 the maximum velocity of the bulk mass of $^{56}$Ni
   of 2980\,km\,s$^{-1}$ is lower than the value of 3406\,km\,s$^{-1}$ in model
   M15-7b-2, favoring more fallback of iron-group material in model M15-7b-3.
So, we may state that the violation of the $^{56}$Ni production--explosion
   energy correlation in the case of models M15-7b-2 and M15-7b-3 is a subtle
   effect that depends on small differences in details of the explosion
   dynamics.

Models M17-8a-3 and M17-8a-4, in turn, show perfect scaling of the mass of
   directly produced $^{56}$Ni plus tracer nucleus,
   $M_\mathrm{Ni}^{\,\mathrm{max}}$, with the explosion energy
   (Table~\ref{tab:hydmod}), which is in agreement with the $^{56}$Ni
   production--explosion energy correlation.
Another fact consistent with this correlation is the mass of the neutron
   star in model M17-8a-4, which is lower than in model M17-8a-3 because
   more neutrino-heated matter is ejected and contributes to
   the mass of directly produced $^{56}$Ni plus tracer nucleus instead of
   being added to the neutron star.
The inverse correlation of the explosion energy with the mass of directly
   produced $^{56}$Ni, is explained by the fact that model
   M17-8a-4 produces more tracer matter than model M17-8a-3 at the expense
   of the directly produced $^{56}$Ni.

\subsection{Mixing in 3D explosion simulations}
\label{sec:results-3Dmix}
%
The development of neutrino-driven explosions after core bounce and the general
   picture of concomitant turbulent mixing in 3D simulations are described and
   studied in detail by \citet{WMJ_15}.
Here we only recall that after the SN shock wave is launched by
   the delayed neutrino-driven mechanism, supported by convective overturn
   and large-scale aspherical shock oscillations caused by the standing
   accretion shock instability (SASI), the further evolution of the explosion
   depends strongly on the density profile of the progenitor.
The shock decelerates when it encounters a density profile that falls off
   less steeply than $\rho \sim r^{-3}$, and it accelerates for density
   profiles that are steeper \citep{Sed_59}.
At the locations of the Si/O, (C+O)/He, and He/H composition interfaces
   (Figure~\ref{fig:chcom}), the value of $\rho r^3$ varies nonmonotonically
   with radius such that the shock velocity increases when the shock approaches
   a composition interface and decreases after the shock has crossed
   the interface.
A deceleration of the shock causes a density inversion in the post-shock flow,
   which means that a dense shell forms.
Such shells at the locations of the composition interfaces are subject to
   Rayleigh-Taylor (RT) instabilities because they are characterized by density
   and pressure gradients of opposite signs \citep{Che_76}.

\begin{figure*}[t]
\centering
   \includegraphics[width=0.9\hsize, clip, trim=50 154 47 117]{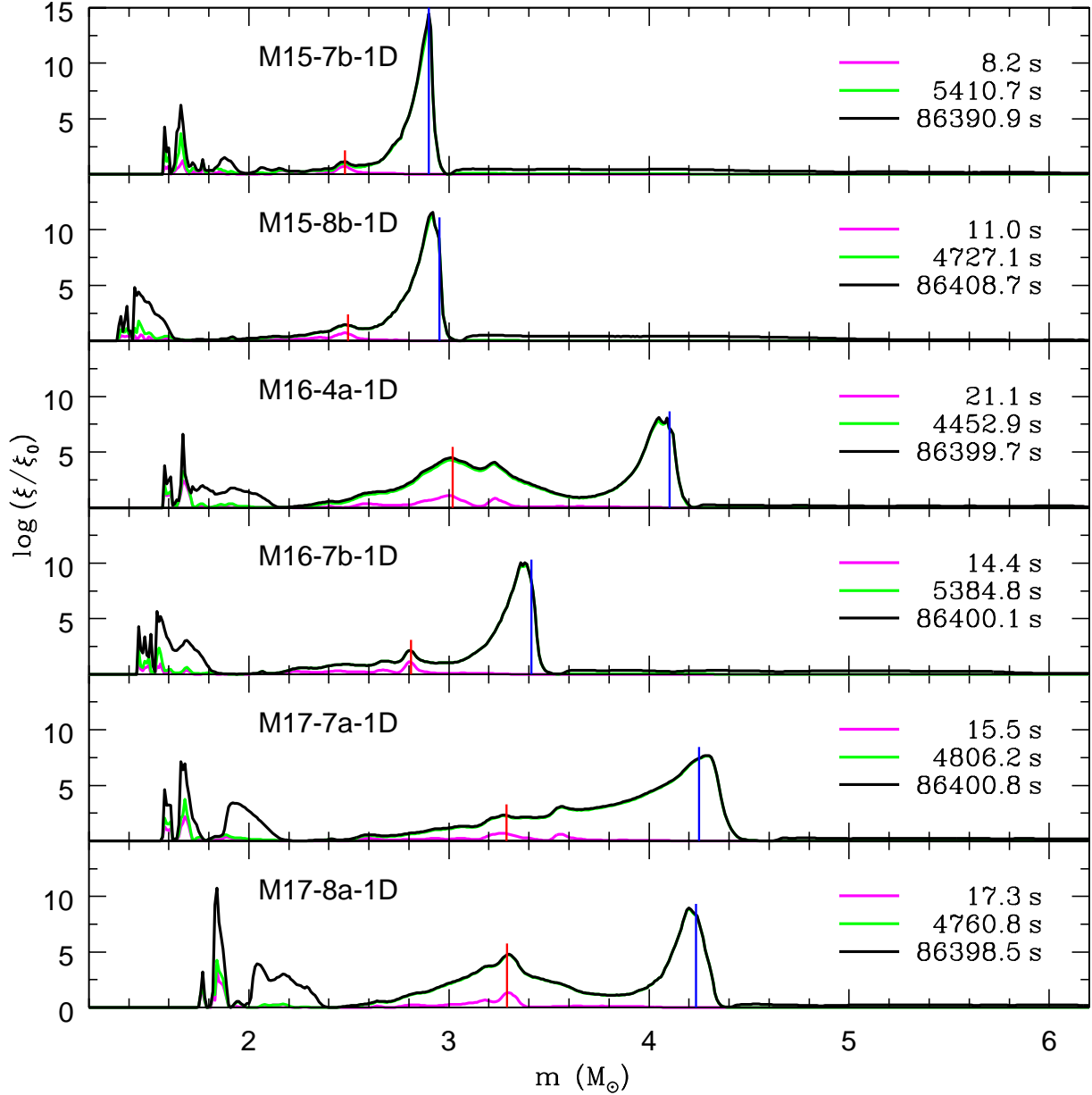}
   \caption{%
   Time-integrated RT growth factors vs. enclosed mass for the 1D explosion
      models based on the corresponding pre-SN models M15-7b, M15-8b, M16-4a,
      M16-7b, M17-7a, and M17-8a at the given times.
   The magenta, green, and black lines show the growth factors at the time
      when the SN shock crosses the He/H composition interface, at the time
      of shock breakout, and at about one day after the onset of the explosion,
      respectively.
   The vertical red and blue lines denote the mass coordinates of the
      (C+O)/He and He/H composition interfaces, respectively.
   These interfaces are RT unstable after the passage of the SN shock.
   }
   \label{fig:rtgrowth}
\end{figure*}
To compare the relative strength of the growth of RT instabilities in
   different progenitor models, we performed an additional set of
   1D neutrino-driven explosion simulations for all progenitor stars
   given in Table~\ref{tab:presnm}, using the same modeling approach
   as in our 3D simulations.
The 1D models in this set develop approximately the same explosion
   energy of 1.4\,B.
To qualitatively analyze the RT instabilities found in our 3D
   simulations, we computed linear RT growth rates as a function of
   enclosed mass for the 1D models and integrated these rates over a
   period of time much longer than the time until shock breakout
   (roughly for a period of one day; see Figure~\ref{fig:rtgrowth}).
However, because RT instabilities will quickly enter the nonlinear
   regime in 3D simulations, the results of linear perturbation theory
   can only provide qualitative information on the relative strength
   of the expected growth of RT instabilities in different layers of
   the progenitor star.
Nevertheless, the results turned out to be useful for a qualitative
   understanding of differences in the efficiency of mixing of
   $^{56}$Ni in different single-star progenitors \citet{UWJ_19},
   and were useful when analyzing the results of 3D simulations.

\begin{figure*}
\centering
   \includegraphics[width=0.49\hsize, clip, trim=52 155 22 295]{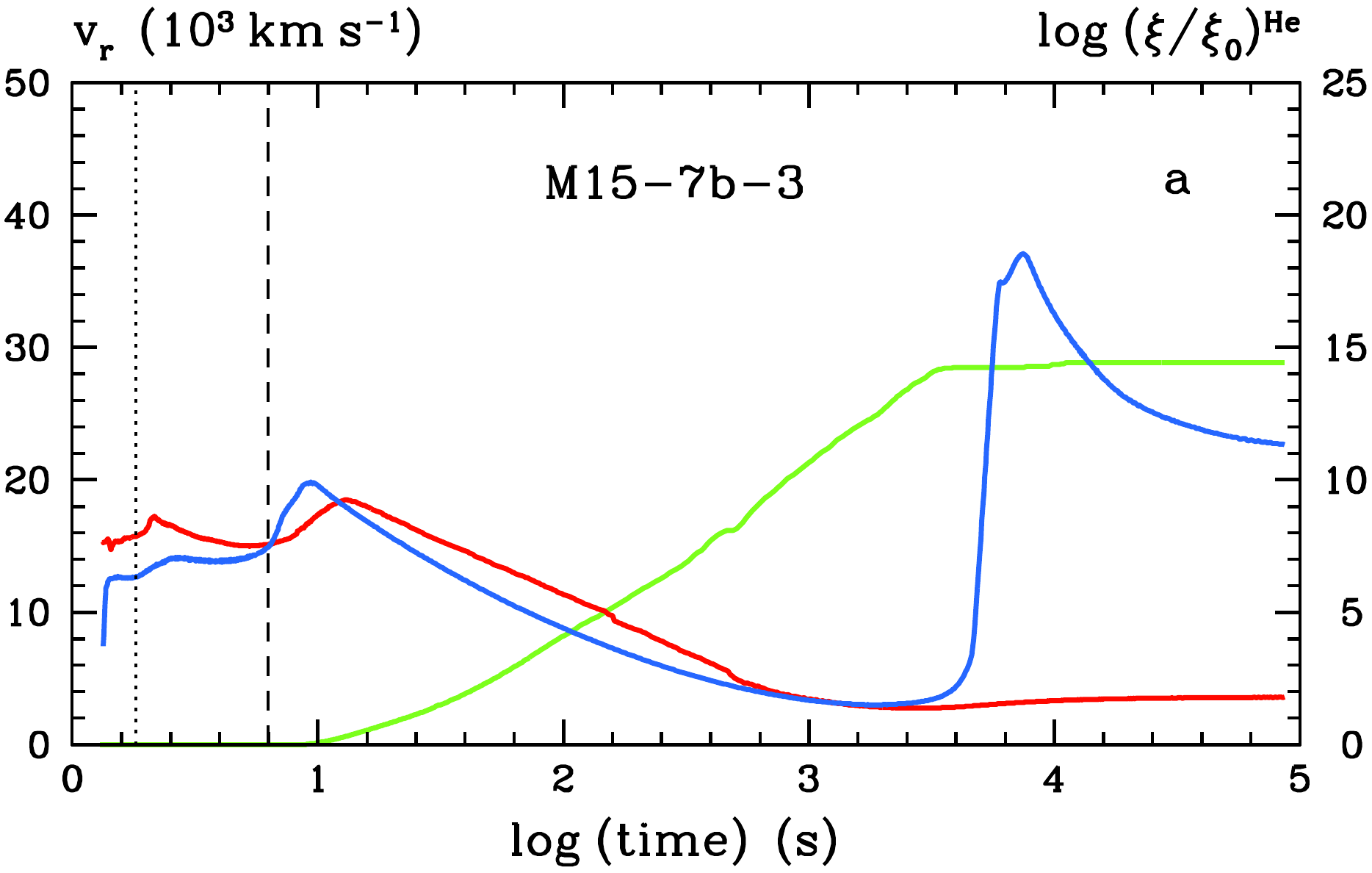}
   \hspace{0.05cm}
   \includegraphics[width=0.49\hsize, clip, trim=52 155 22 295]{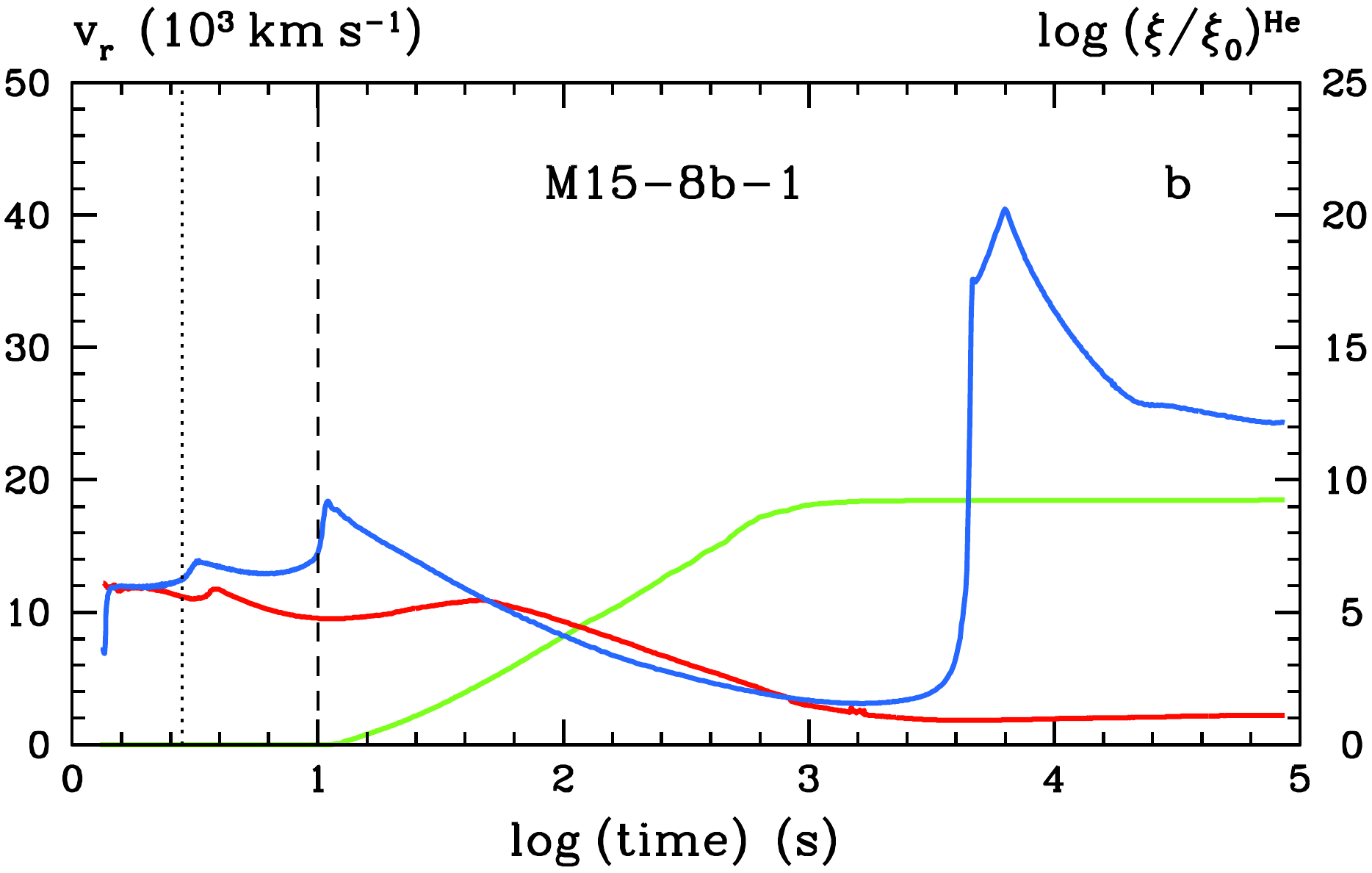}\\
   \vspace{0.5cm}
   \includegraphics[width=0.49\hsize, clip, trim=52 155 22 295]{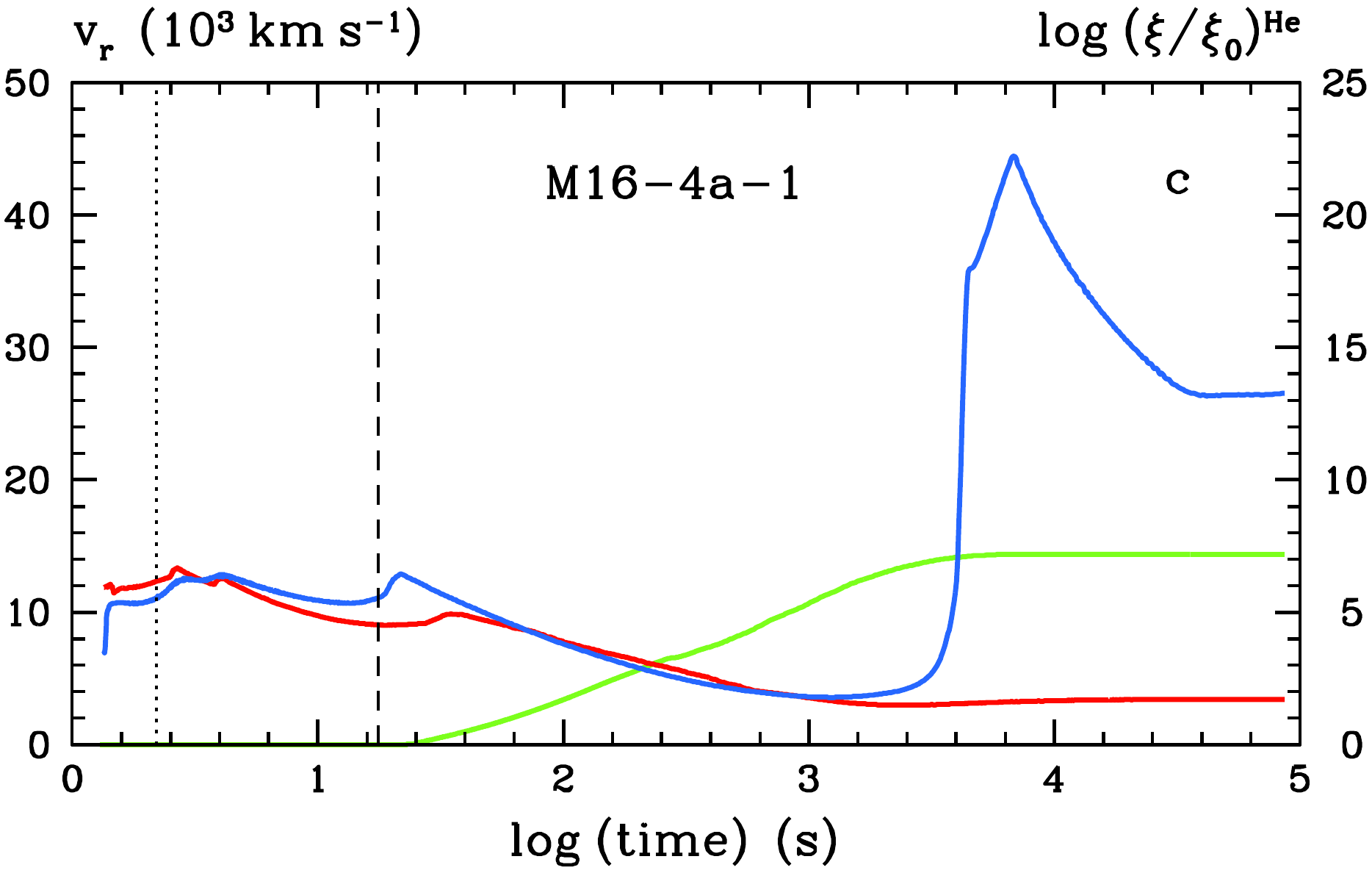}
   \hspace{0.05cm}
   \includegraphics[width=0.49\hsize, clip, trim=52 155 22 295]{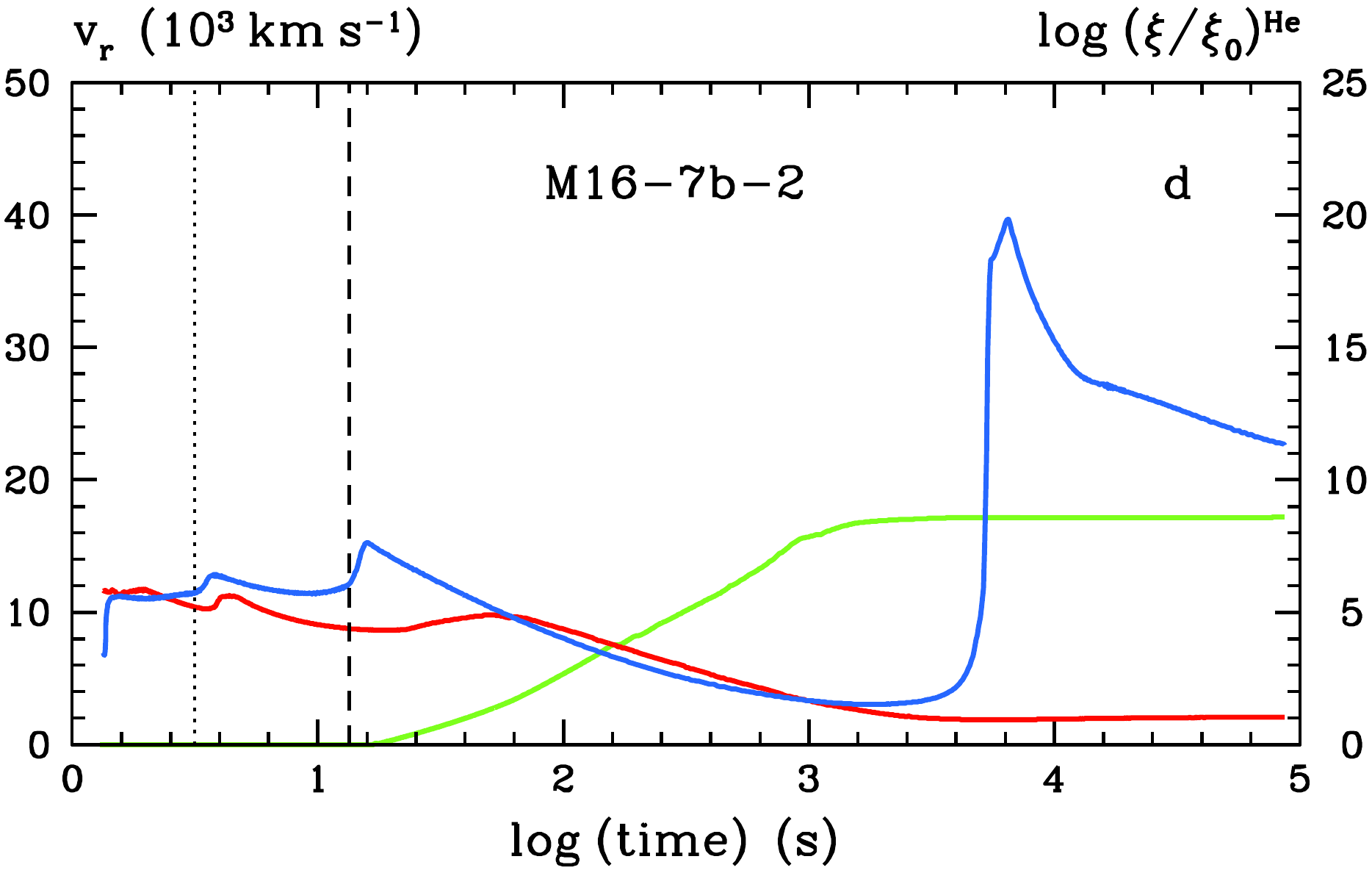}\\
   \vspace{0.5cm}
   \includegraphics[width=0.49\hsize, clip, trim=52 155 22 295]{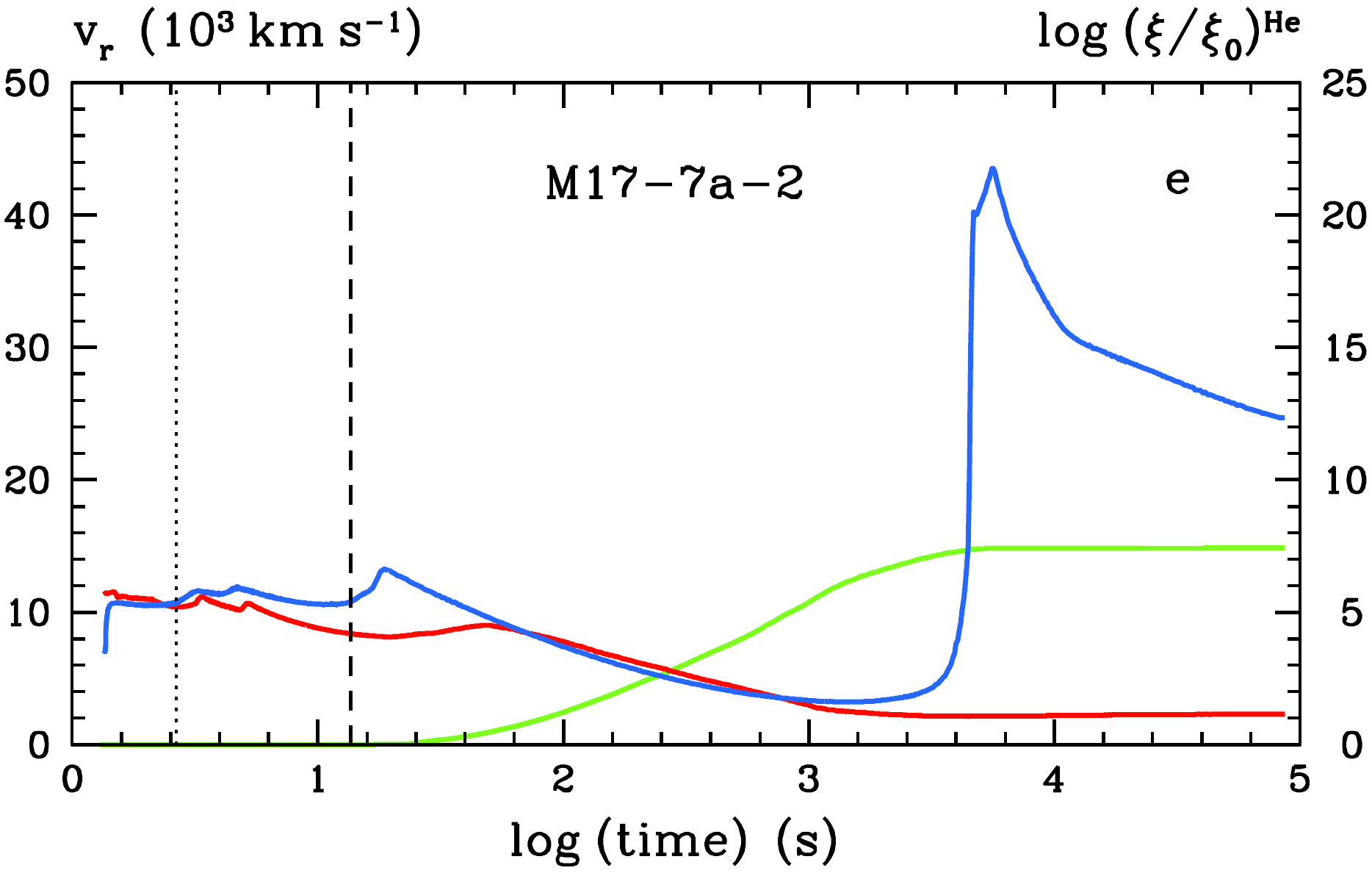}
   \hspace{0.05cm}
   \includegraphics[width=0.49\hsize, clip, trim=52 155 22 295]{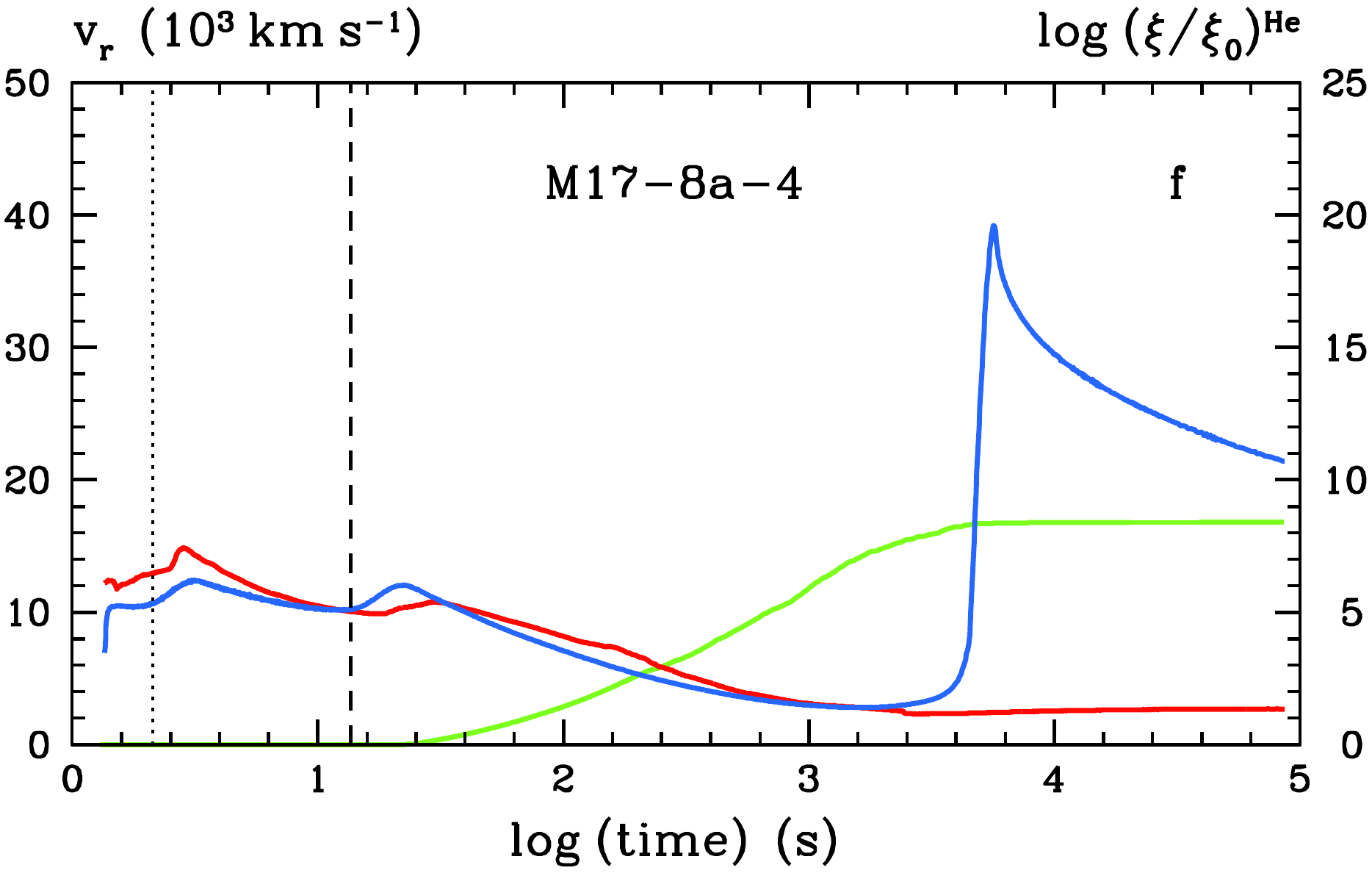}\\
   \caption{%
   Time evolution of the angle-averaged radial velocity of the SN shock
      (blue lines) and the maximum radial velocity on the surface
      where the mass fraction of $^{56}$Ni plus the neutron-rich tracer
      nucleus equals $3\%$ (red lines) for our six reference 3D explosion
      models M15-7b-3, M15-8b-1, M16-4a-1, M16-7b-2, M17-7a-2, and M17-8a-4,
      and the time-integrated RT growth factor at the He/H composition
      interface (green lines) for the same models simulated in 1D.
   The vertical dotted and dashed lines mark the times when the shock crosses
      the (C+O)/He and He/H composition interfaces, respectively.
   The former times are identical to those of the snapshots shown in columns
      one and three of Figure~\ref{fig:3D_models}.
   }
   \label{fig:vnivsh}
\end{figure*}

Here we focus on the time evolution of the time-integrated RT growth factor
   at the He/H composition interface (Figure~\ref{fig:vnivsh})
   where its value reaches a maximum in our binary-merger progenitors
   (Figure~\ref{fig:rtgrowth}).
As \citet{UWJ_19} we consider a simple phenomenological approach to
   capture multidimensional effects of RT mixing at a composition
   interface and to describe the evolution of the nickel velocity:
\begin{equation}
   {d v \over d t} = \beta \sigma_{\mathrm{RT}} v_{0} \; ,
\label{eq:model}
\end{equation}
where $v$ is the maximum velocity of the bulk mass of radioactive
   $^{56}$Ni, $\beta$ is an empirical buoyancy coefficient,
   $\sigma_{\mathrm{RT}}$ is the linear RT growth rate
\begin{equation}
   \sigma_{\mathrm{RT}} = {1 \over \rho} \sqrt{-{\partial P \over \partial r}
      {\partial \rho \over \partial r}} \; ,
\label{eq:sigmaRT}
\end{equation}
   and $v_{0}$ is the initial value of the radial velocity of $^{56}$Ni.
The solution of Equation~\ref{eq:model} is given by
\begin{equation}
   v(t) = \beta v_{0} \int_{0}^t \sigma_{\mathrm{RT}}(\tau) d\tau
        = \beta v_{0} \ln\left( \xi(t)/\xi_0 \right) \; ,
 \label{eq:rtsol}
\end{equation}
  where
\begin{equation}
   {\xi \over \xi_0}(t) = \exp\left(\int_{0}^t \sigma_{\mathrm{RT}}(\tau)\,d\tau
      \right)
\label{eq:rel_xi}
\end{equation}
   is the time-integrated RT growth factor at time $t$, and $\xi_0$ is
   the amplitude of the initial perturbation at a given Lagrangian
   mass coordinate.
Thus, the velocity growth factor $v(t)/v_0$ at time $t$ is proportional to
   the logarithm of the time-integrated RT growth factor at that time.
According to Figure~\ref{fig:vnivsh}, the growth factor $\xi(t)/\xi_0$
   increases at the He/H composition interface from a value of unity
   shortly after the SN shock crosses this interface to its maximum
   (i.e., final) value at the time of shock breakout.
During this time interval the growth factor approximately exhibits a
   power law dependence
\begin{equation}
   {\xi \over \xi_0}(t) \approx A t^{\alpha} \; ,
\label{eq:appr_xi}
\end{equation}
   as indicated by the linear slope of the green lines in
   Figure~\ref{fig:vnivsh}, and where $A$ is a constant and $\alpha$ is
   the value of the slope.
This approximation implies a simple inverse time dependence of the
   linear RT growth rate
\begin{equation}
   \sigma_{\mathrm{RT}} \approx {\alpha \over t} \; ,
\label{eq:appr_sigmaRT}
\end{equation}
   and hence the occurrence of a local maximum of $\sigma_{\mathrm{RT}}$
   at a time between shortly after the SN shock crosses the He/H
   interface and the onset of power law growth.

Before we proceed further, let us recall the origin of the binary-merger
   progenitors for SN~1987A.
The main element of the merger model of \citet{MH_17} is merging of
   a main-sequence secondary star with a RSG primary star in the
   common-envelope phase when the secondary eventually penetrates the
   helium core of the primary.
The helium core of the resulting merged star inherits the basic property of
   the RSG primary star evolved as a single star: its density profile is
   characterized by a low time-integrated RT growth factor
   near the (C+O)/He composition interface and a very large one at the He/H
   interface, since the density gradient at the He/H interface is much steeper
   than at the (C+O)/He interface (Figures~\ref{fig:denmr} and
   \ref{fig:rtgrowth}).
In contrast, the single-star BSG progenitors for SN~1987A demonstrate
   opposite behavior: a large time-integrated RT growth factor near the
   (C+O)/He composition interface and a low one at the He/H interface
   \citep{UWJ_19}.%
   \footnote{The only exception to this is model N20, which originates from
   a RSG progenitor rather than a BSG star with respect to the structure of
   the helium core \citep{NH_88, SNK_88}.}

This fundamental difference between BSG and RSG progenitors concerning the RT
   growth factors at the (C+O)/He and He/H interfaces was previously pointed
   out by \citet{WMJ_15}, and the general properties of radial mixing in the
   ejecta induced by the 3D neutrino-driven explosions were studied.
It was shown that the extent of radial mixing, i.e., the minimum velocity of
   hydrogen-rich matter and the maximum velocity of $^{56}$Ni, depends not only
   on the initial asphericity and explosion energy, but also on the density
   profiles and widths of the C+O core and the helium shell, and on the density
   gradient at the He/H composition interface.
A dominant growth rate at the He/H interface in the RSG explosions compared
   to the BSG cases gives rise to a more strongly RT-unstable layer after
   the passage of the SN shock, which then develops into more extended
   inward mixing of hydrogen into the helium core than in the BSG explosions.
On the other hand, a dominant growth rate at the (C+O)/He interface in the BSG
   explosions compared to the RSG ones results in a relatively strong outward
   mixing of $^{56}$Ni in velocity space, which turns out to be comparable to
   that in the RSG ejecta, where the extended fast $^{56}$Ni-rich fingers have
   more time to grow, because the reverse shock at the He/H interface develops
   at a much larger radius and thus much later than in the BSG progenitors.

The above findings concerning the inward mixing of hydrogen and the outward
   mixing of radioactive $^{56}$Ni are confirmed by the 3D neutrino-driven
   explosion simulations based on the single-star progenitors for SN~1987A
   \citep[BSG pre-SN models except for model N20 in][]{UWJ_19} and the
   binary-merger progenitors (pre-SN models with the RSG-like helium cores,
   see Figures~\ref{fig:denmr} and \ref{fig:rtgrowth}).
Explosions of the single-star BSG progenitors with SN~1987A-like energies
   show that hydrogen-rich matter mixed into the helium core expands with
   velocities from 270\,km\,s$^{-1}$ to 1850\,km\,s$^{-1}$, and its mass
   does not exceed 0.15\,$M_{\sun}$ \citep{UWJ_19}.
In turn, all hydrodynamic models based on the binary-merger progenitors
   demonstrate that hydrogen is mixed more deeply into the helium core down
   to velocities lower than 40\,km\,s$^{-1}$, and its mass constitutes about
   0.5\,$M_{\sun}$ (Table~\ref{tab:hydmod}).
As to the outward mixing of $^{56}$Ni, our favorite hydrodynamic models B15-2,
   based on the single-star progenitor \citep{UWJ_19}, and M15-7b-3, based
   on the binary-merger progenitor (Table~\ref{tab:hydmod}), have nearly
   the same maximum velocities of the bulk of $^{56}$Ni consistent with
   the observed value of about 3000\,km\,s$^{-1}$.

\subsection{3D morphology of explosion models}
\label{sec:results-3Dmorph}
%
\begin{figure*}
\centering
   \includegraphics[width=0.23\hsize, clip, trim=60 60 60 60]{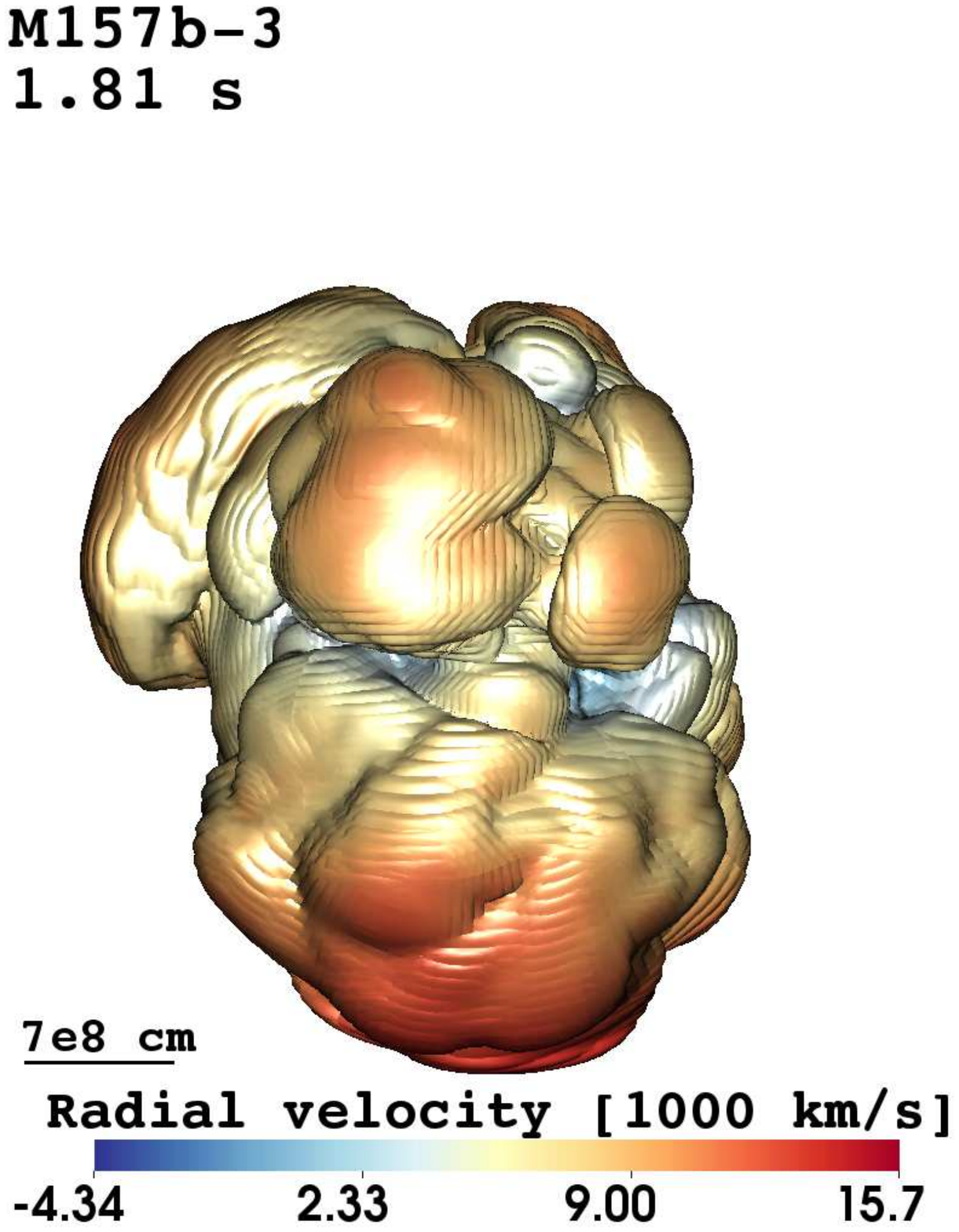}
   \includegraphics[width=0.23\hsize, clip, trim=60 60 60 60]{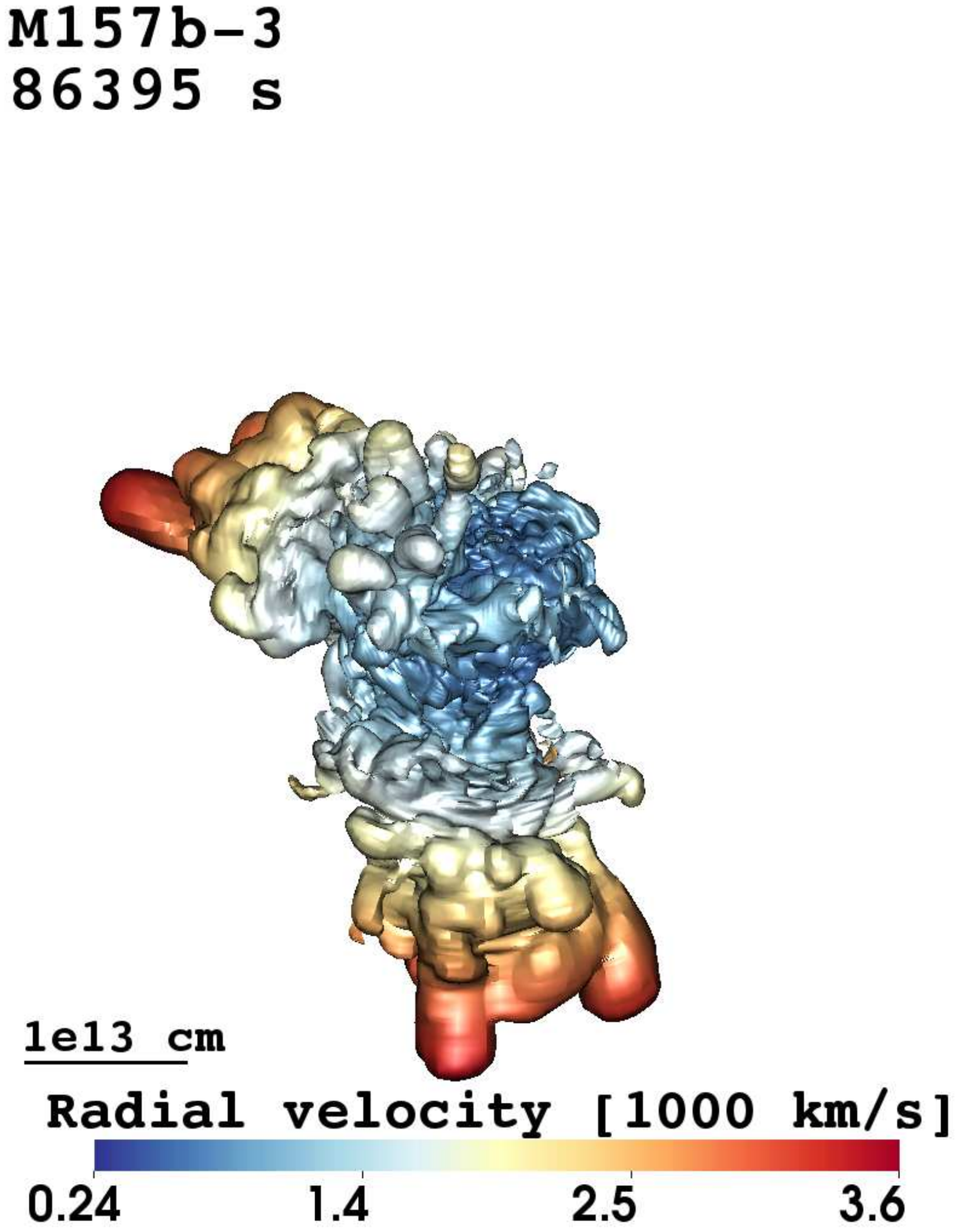}
   \hspace{0.5cm}
   \includegraphics[width=0.23\hsize, clip, trim=60 60 60 60]{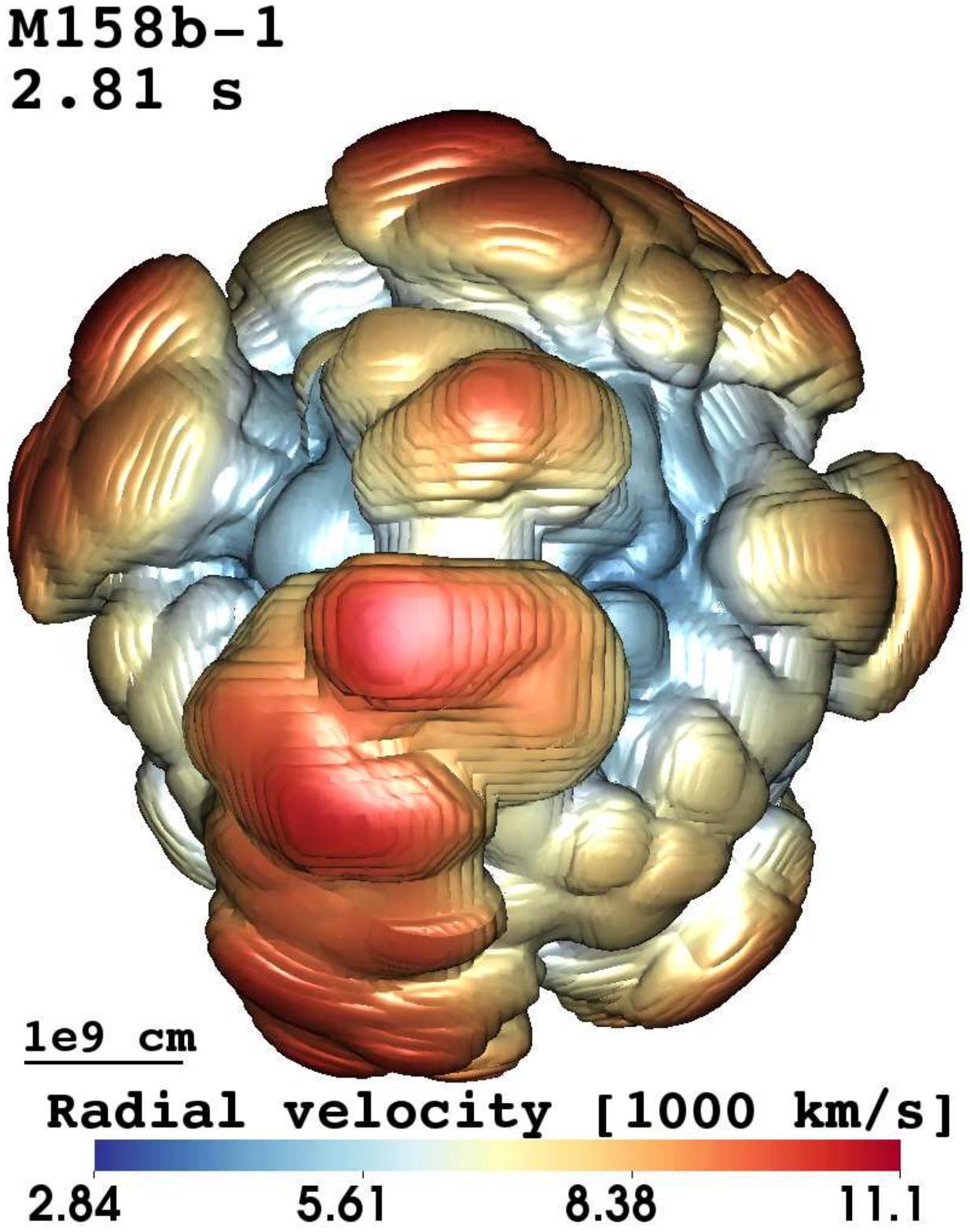}
   \includegraphics[width=0.23\hsize, clip, trim=60 60 60 60]{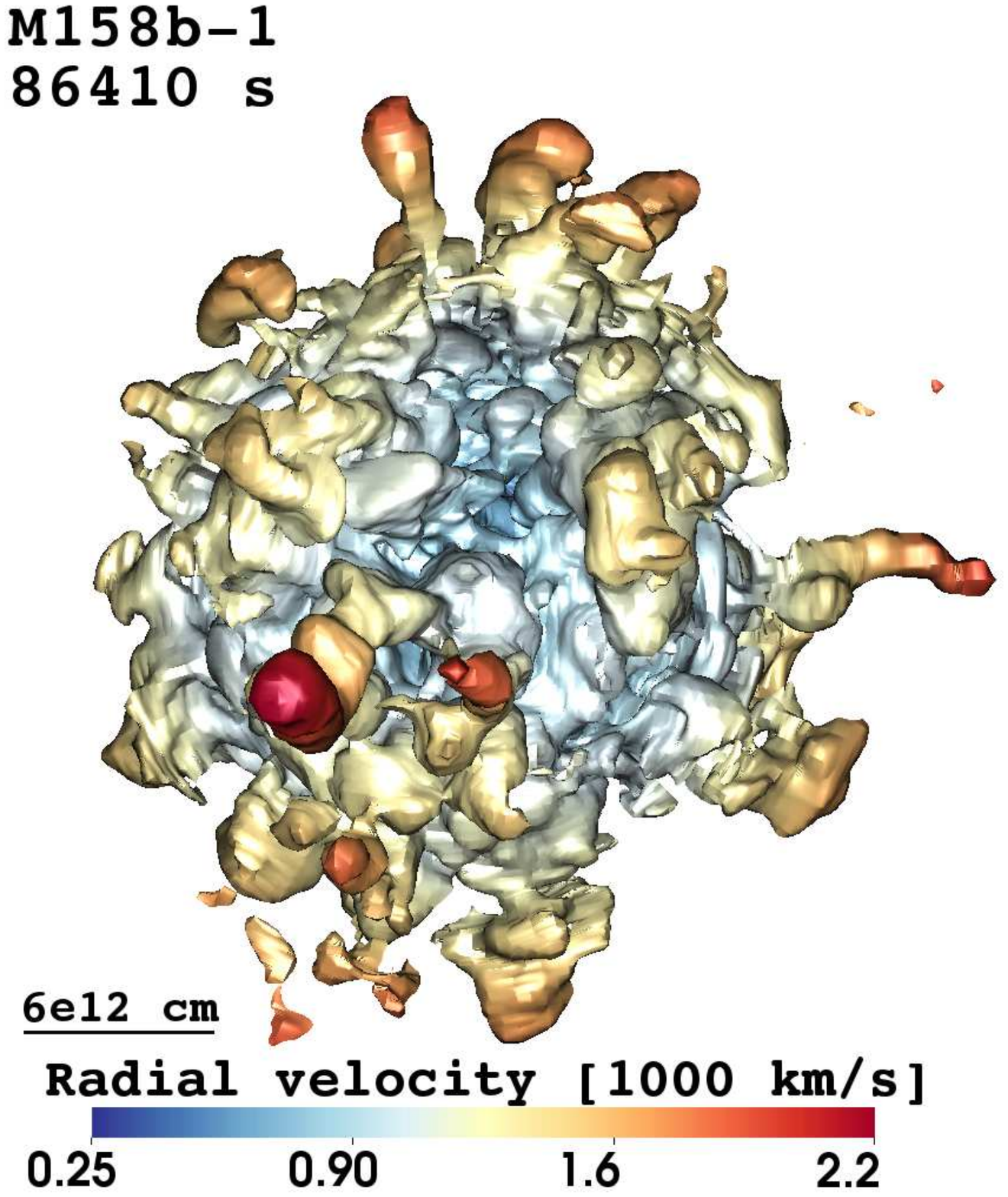}\\
   \vspace{0.25cm}
   \includegraphics[width=0.23\hsize, clip, trim=60 60 60 60]{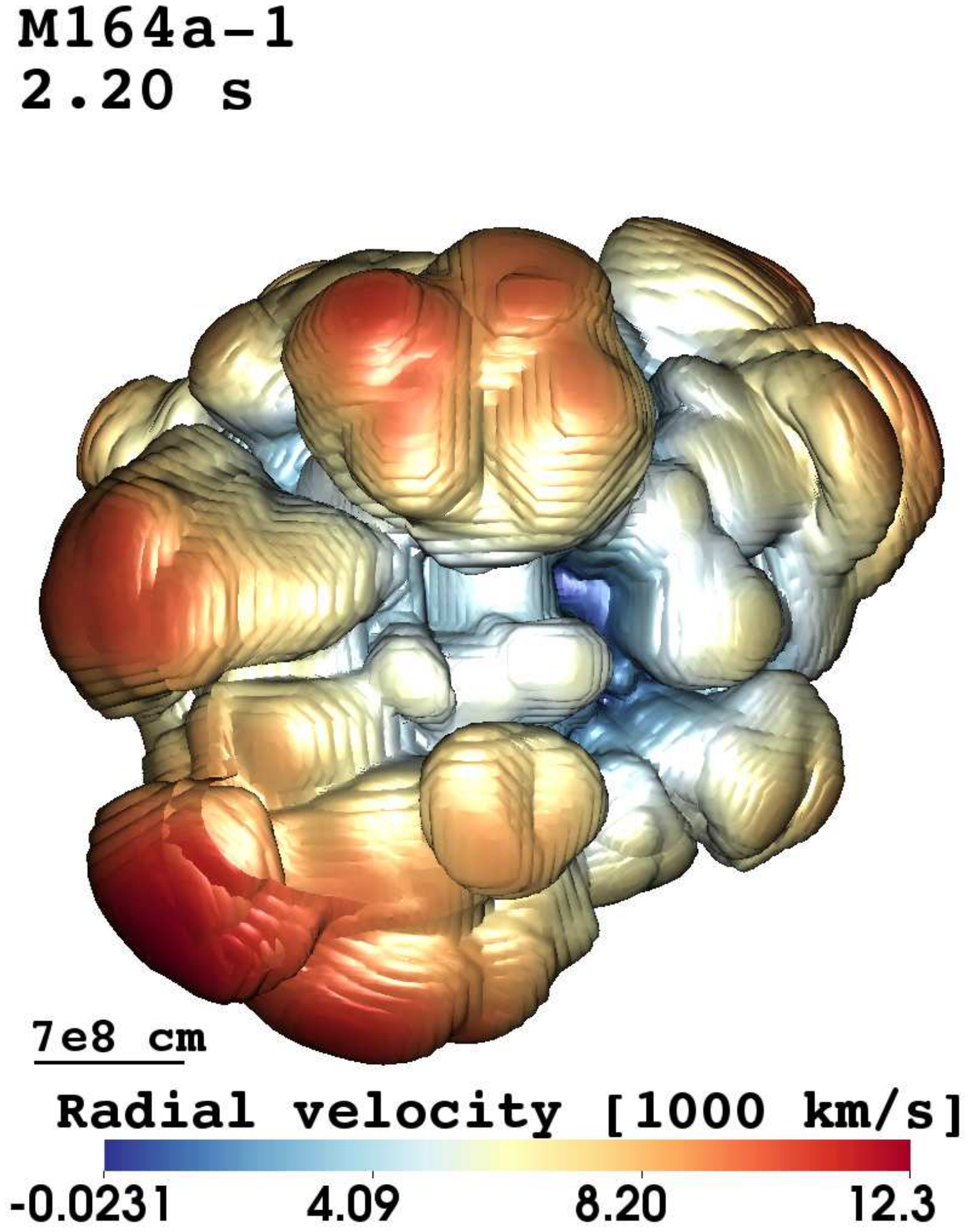}
   \includegraphics[width=0.23\hsize, clip, trim=60 60 60 60]{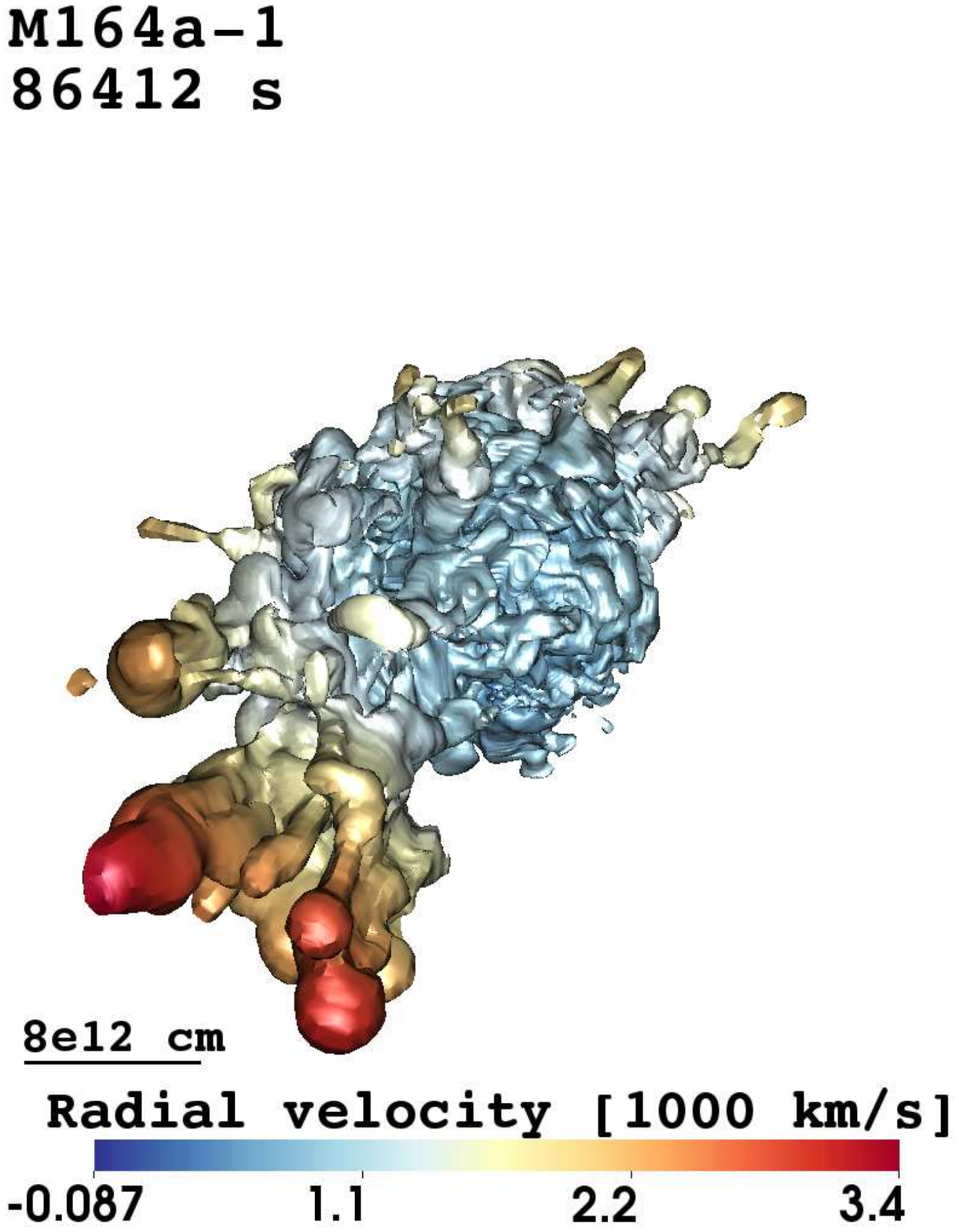}
   \hspace{0.5cm}
   \includegraphics[width=0.23\hsize, clip, trim=60 60 60 60]{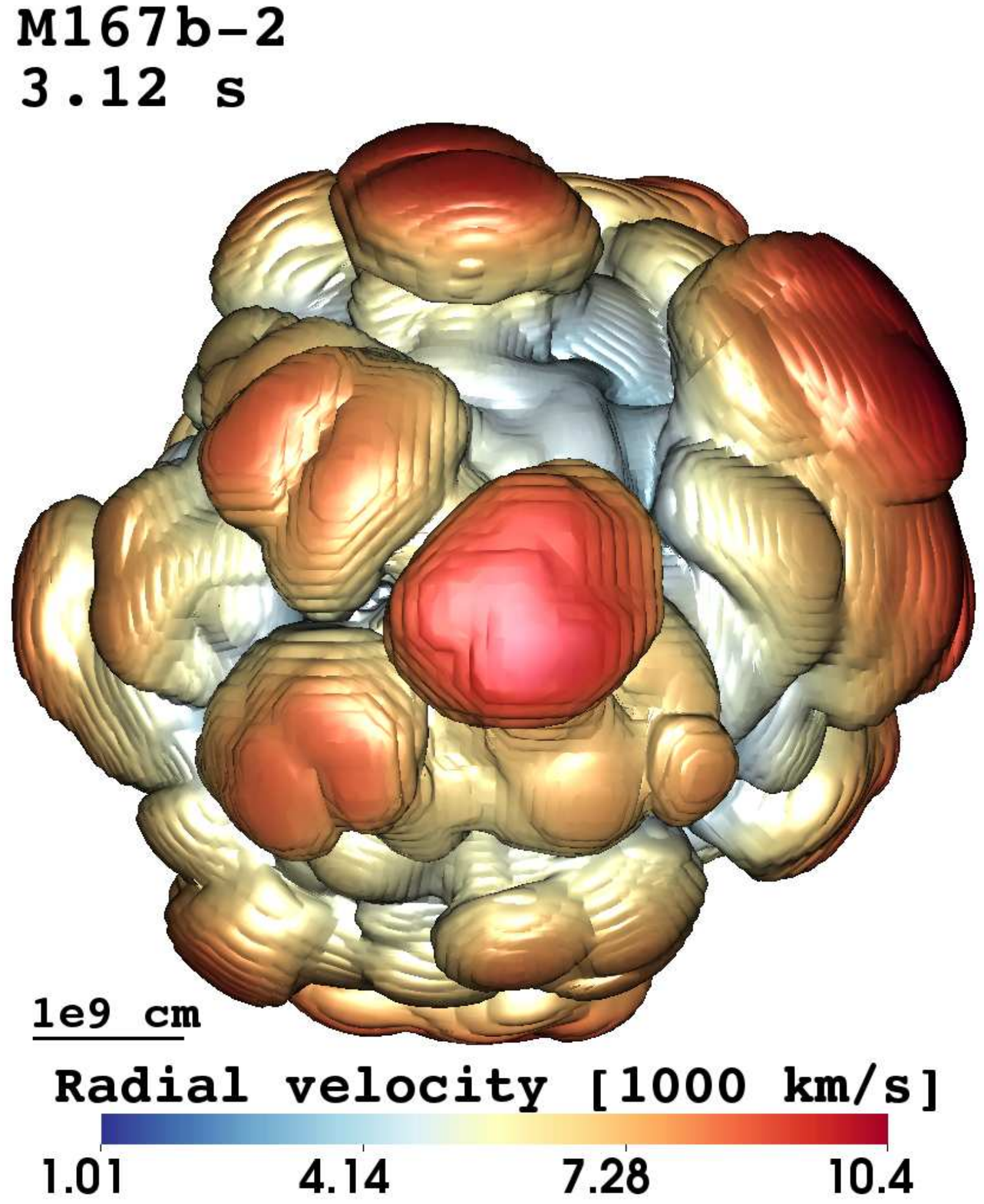}
   \includegraphics[width=0.23\hsize, clip, trim=60 60 60 60]{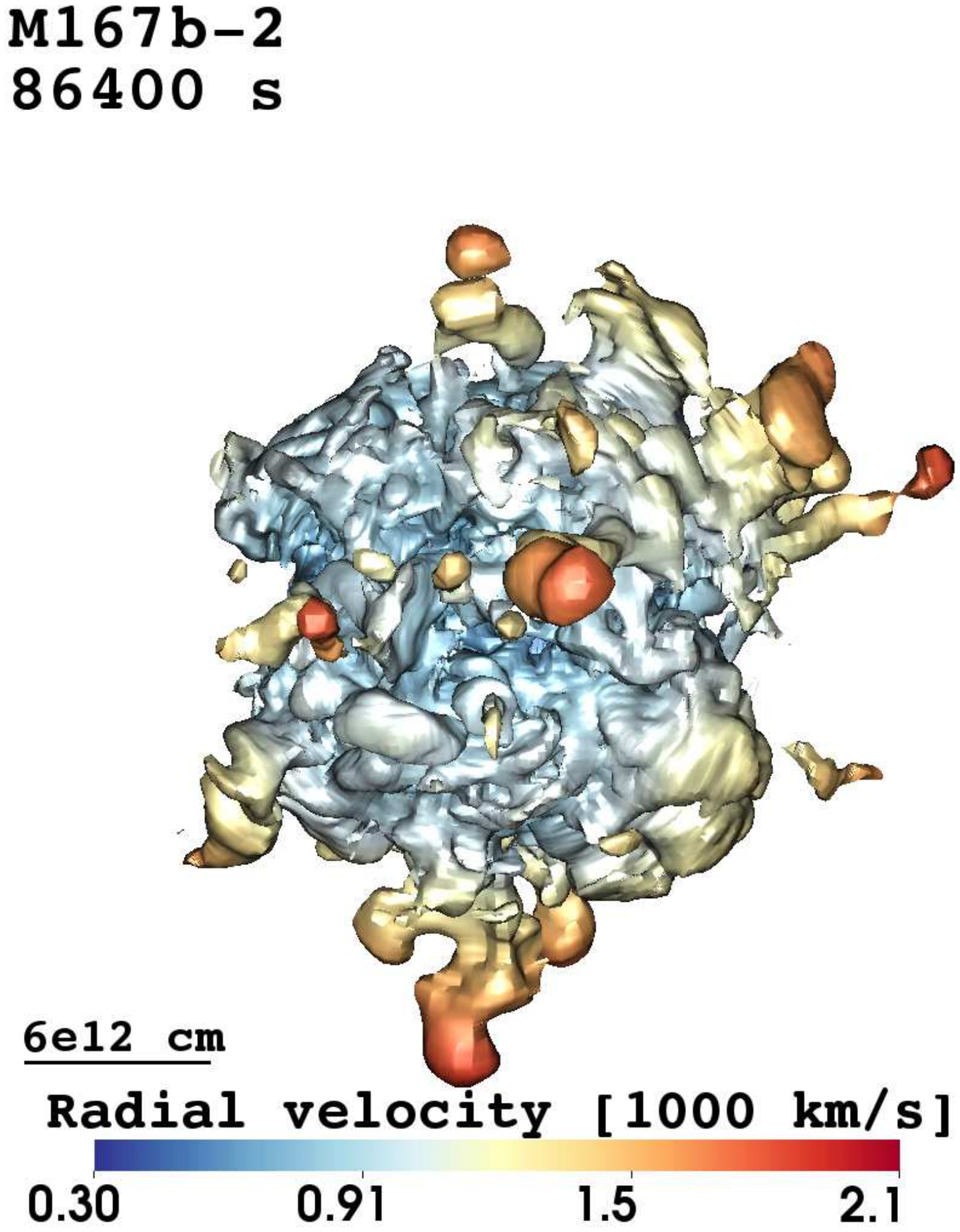}\\
   \vspace{0.25cm}
   \includegraphics[width=0.23\hsize, clip, trim=60 60 60 60]{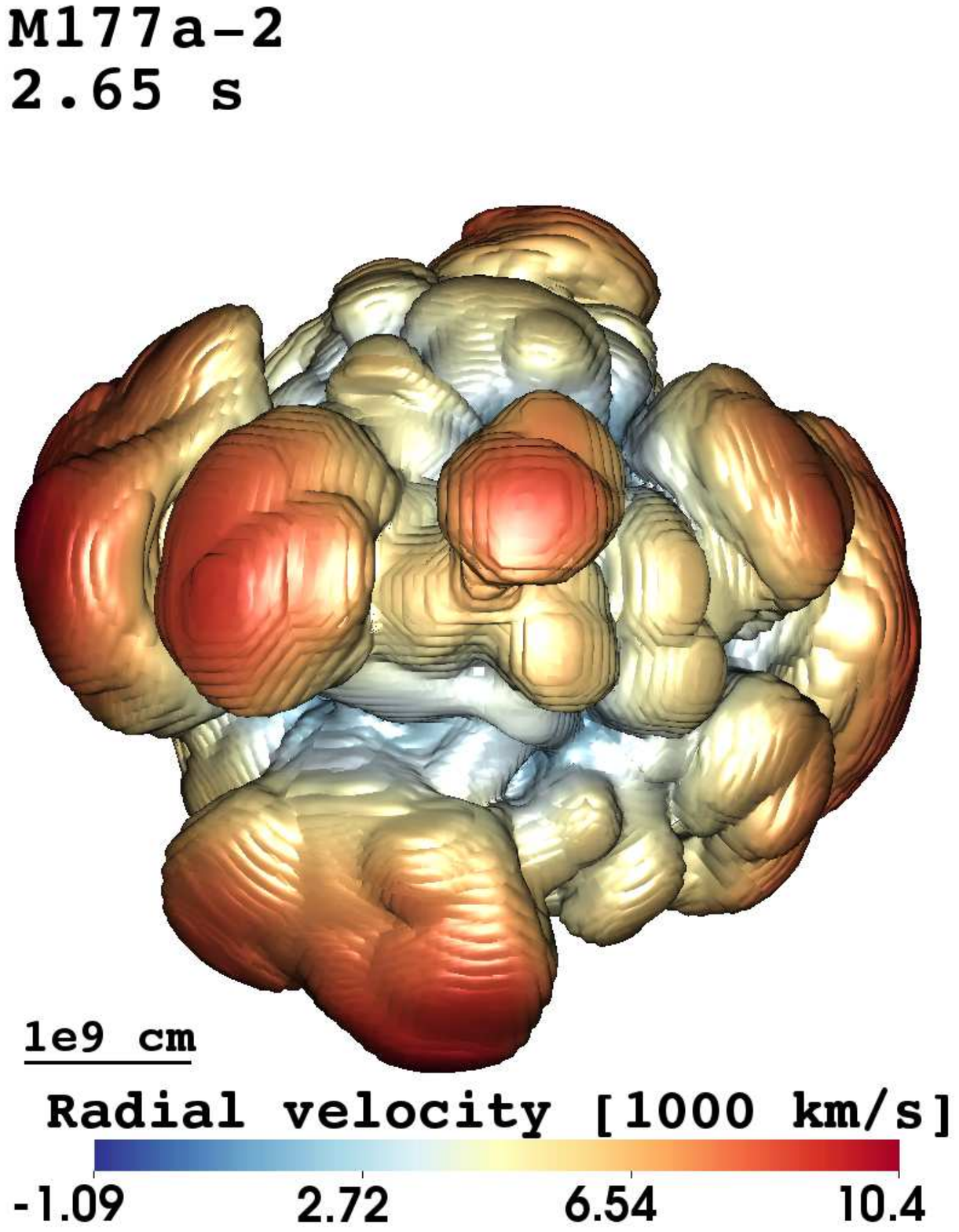}
   \includegraphics[width=0.23\hsize, clip, trim=60 60 60 60]{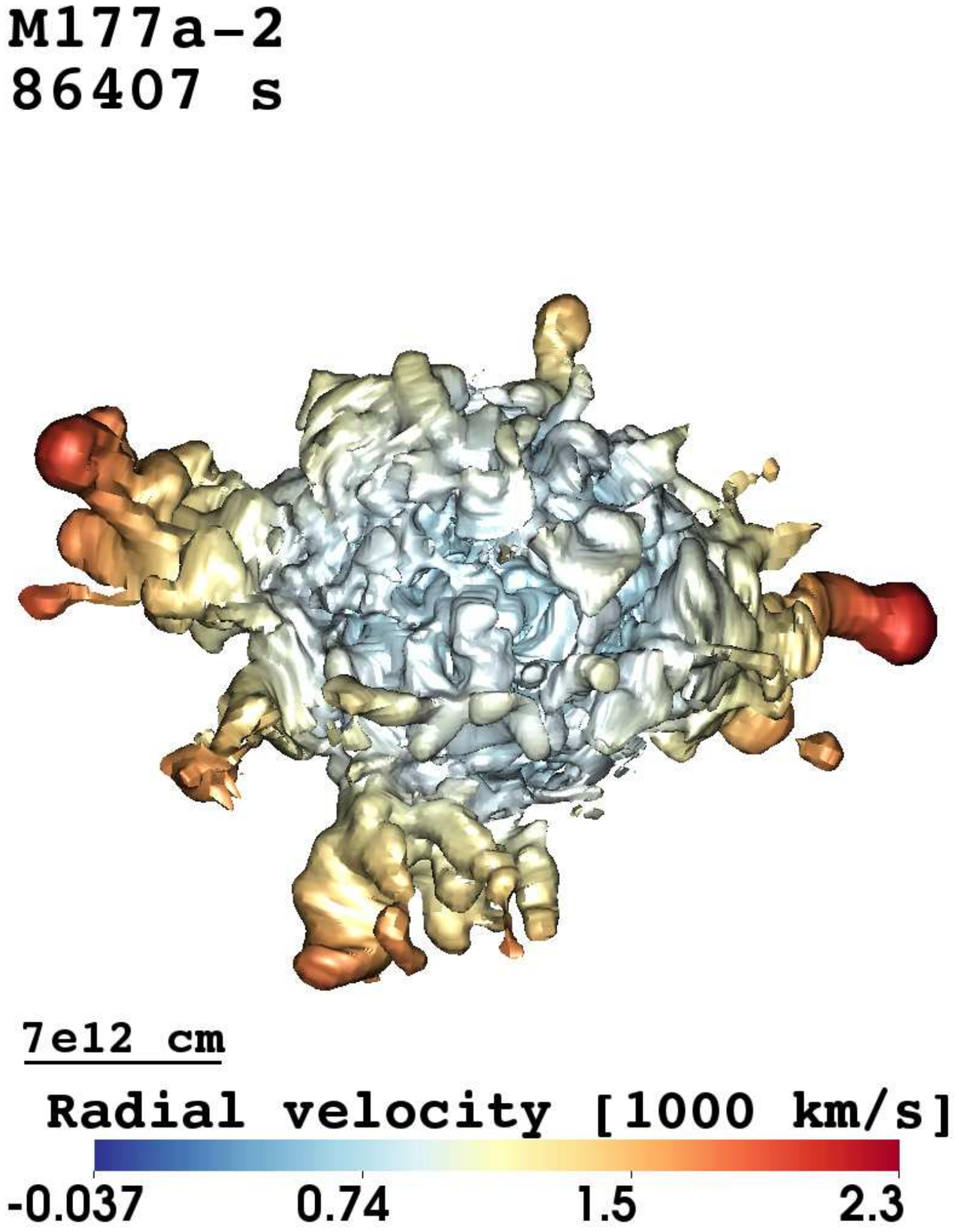}
   \hspace{0.5cm}
   \includegraphics[width=0.23\hsize, clip, trim=60 60 60 60]{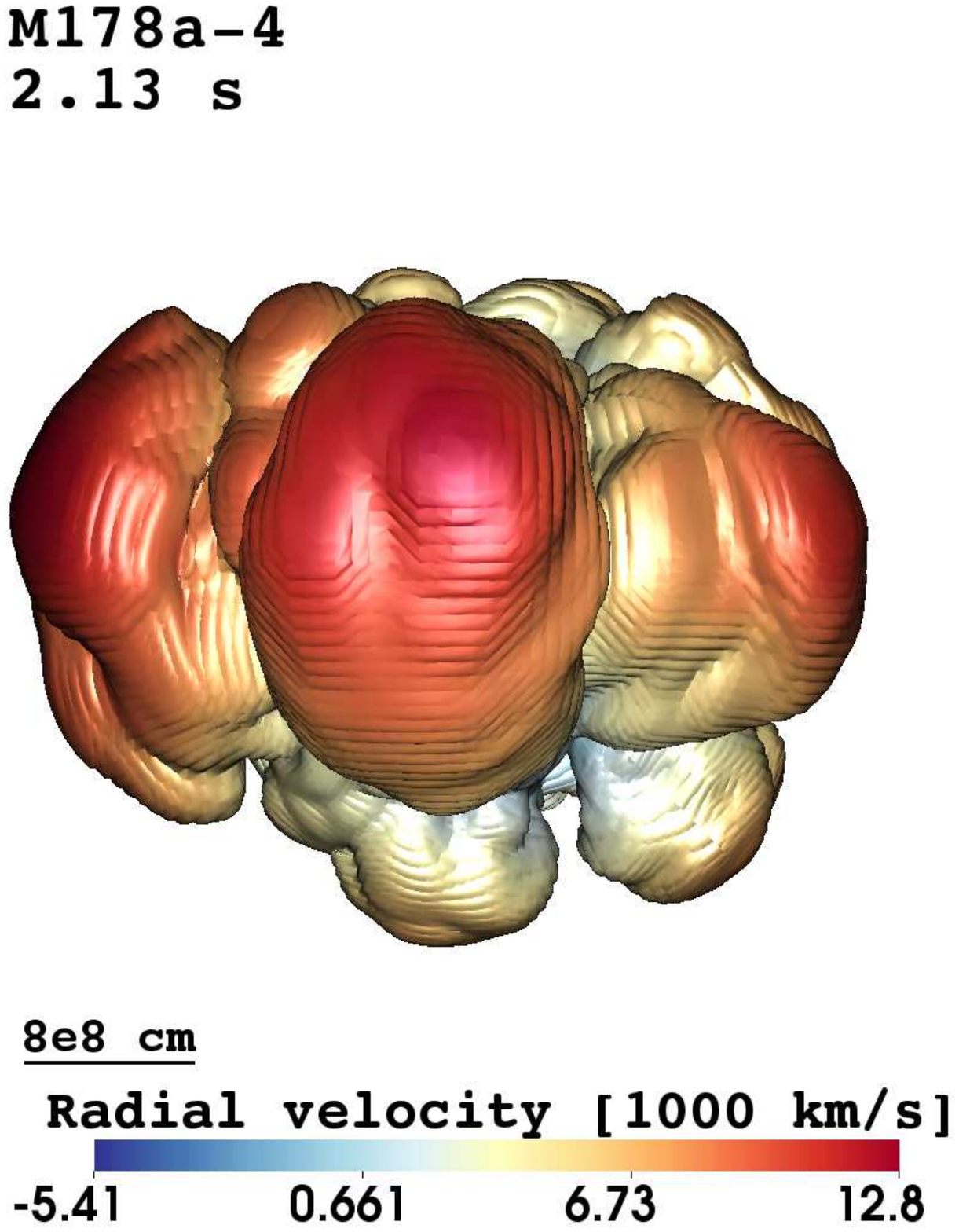}
   \includegraphics[width=0.23\hsize, clip, trim=60 60 60 60]{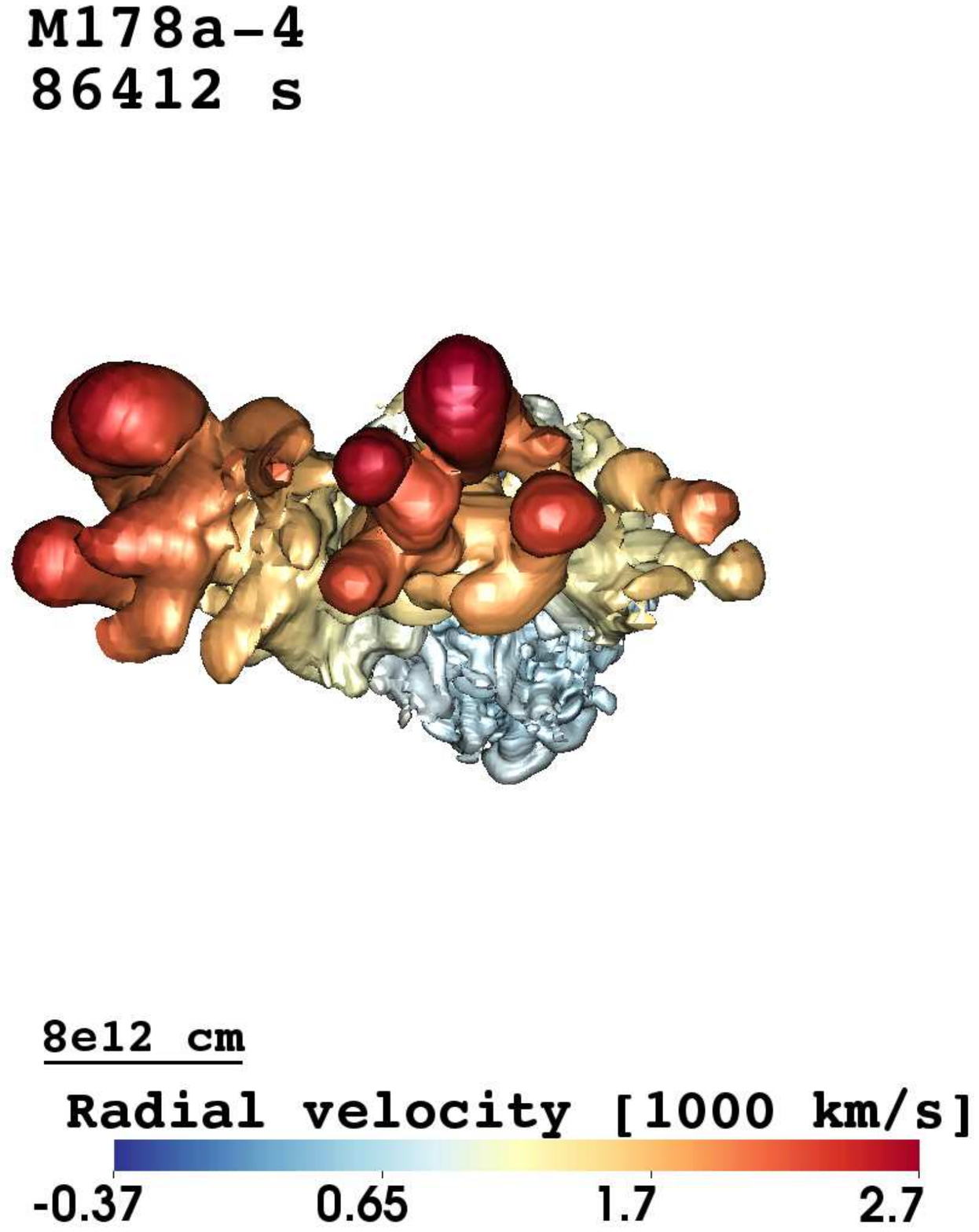}\\
   \caption{%
   Morphology of radioactive $^{56}$Ni-rich matter produced by iron-group
      nucleosynthesis in shock-heated and neutrino-heated ejecta.
   The snapshots display isosurfaces where the mass fraction of $^{56}$Ni plus
      the neutron-rich tracer nucleus equals $3\%$.
   The isosurfaces are shown for our six reference 3D explosion models
      at two different epochs: shortly before the SN shock crosses the
      (C+O)/He composition interface in the progenitor star (first and third
      columns), and long after shock breakout (second and fourth columns).
   In the top left corner of each panel, we give the name of the model and
      the post-bounce time of the snapshot.
   The colors represent the radial velocity on the isosurface, and the
      color-coding is defined at the bottom of each panel.
   The size of the displayed volume and of the plume-like structures can be
      estimated from the yardsticks given in the lower left corner of each
      panel.
   There are striking differences in the final morphology of the $^{56}$Ni-rich
      ejecta between these models, all of which have comparable and effectively
      SN~1987A-like explosion energies.
   These differences arise from progenitor-specific differences in
      their stellar density and composition structures, and from the influence
      of the latter on the unsteady SN shock propagation.
   From an inspection of the final morphology, it is possible to distinguish
      two groups of models: M15-8b-1, M16-7b-2, and M17-7a-2 with
      a small asphericity and M15-7b-3, M16-4a-1, and M17-8a-4 with a
      large asphericity.
   }
   \label{fig:3D_models}
\end{figure*}
\begin{figure*}
\centering
   \includegraphics[width=0.49\hsize, clip, trim=22 155 47 319]{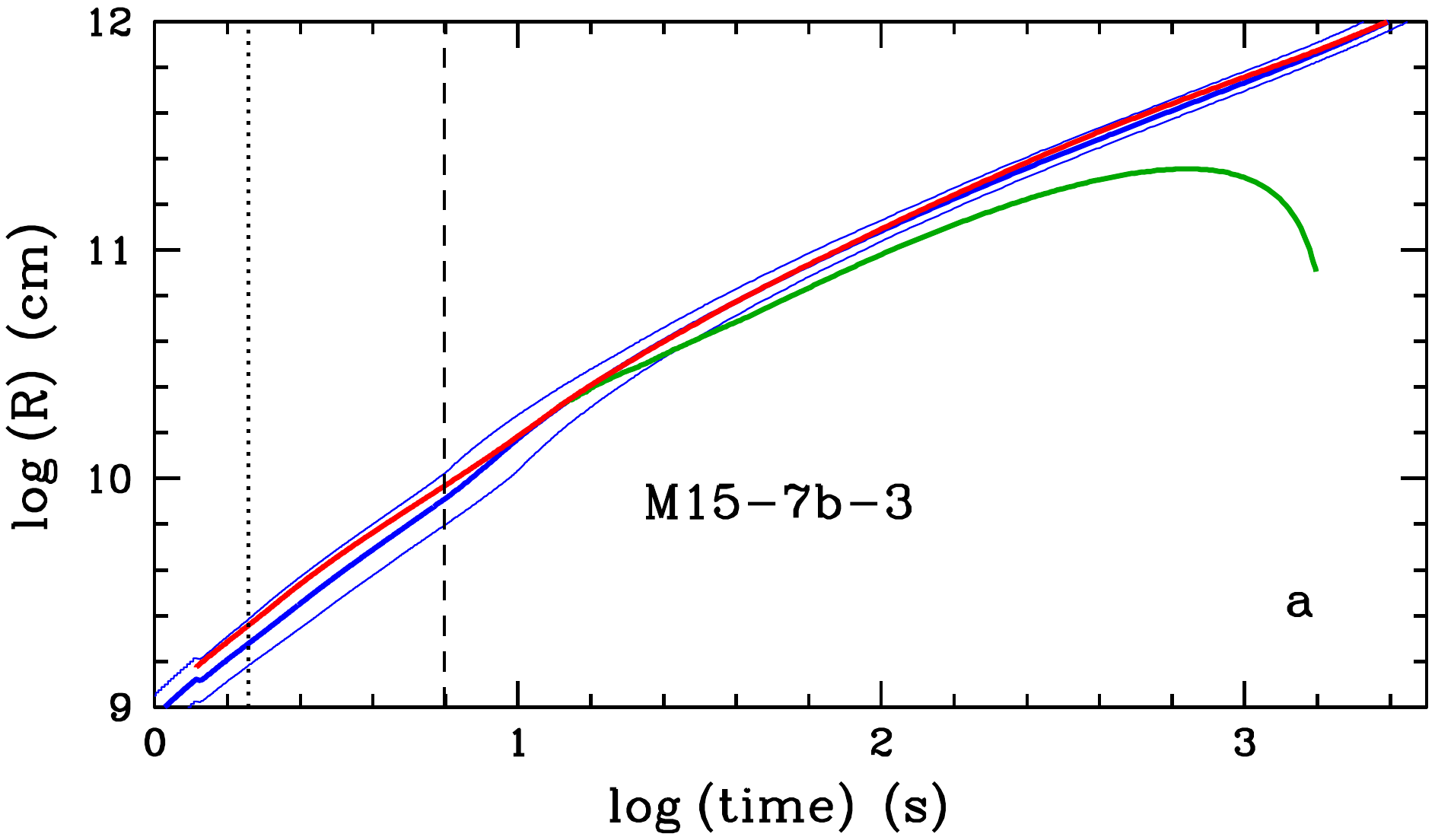}
   \hspace{0.05cm}
   \includegraphics[width=0.49\hsize, clip, trim=22 155 47 319]{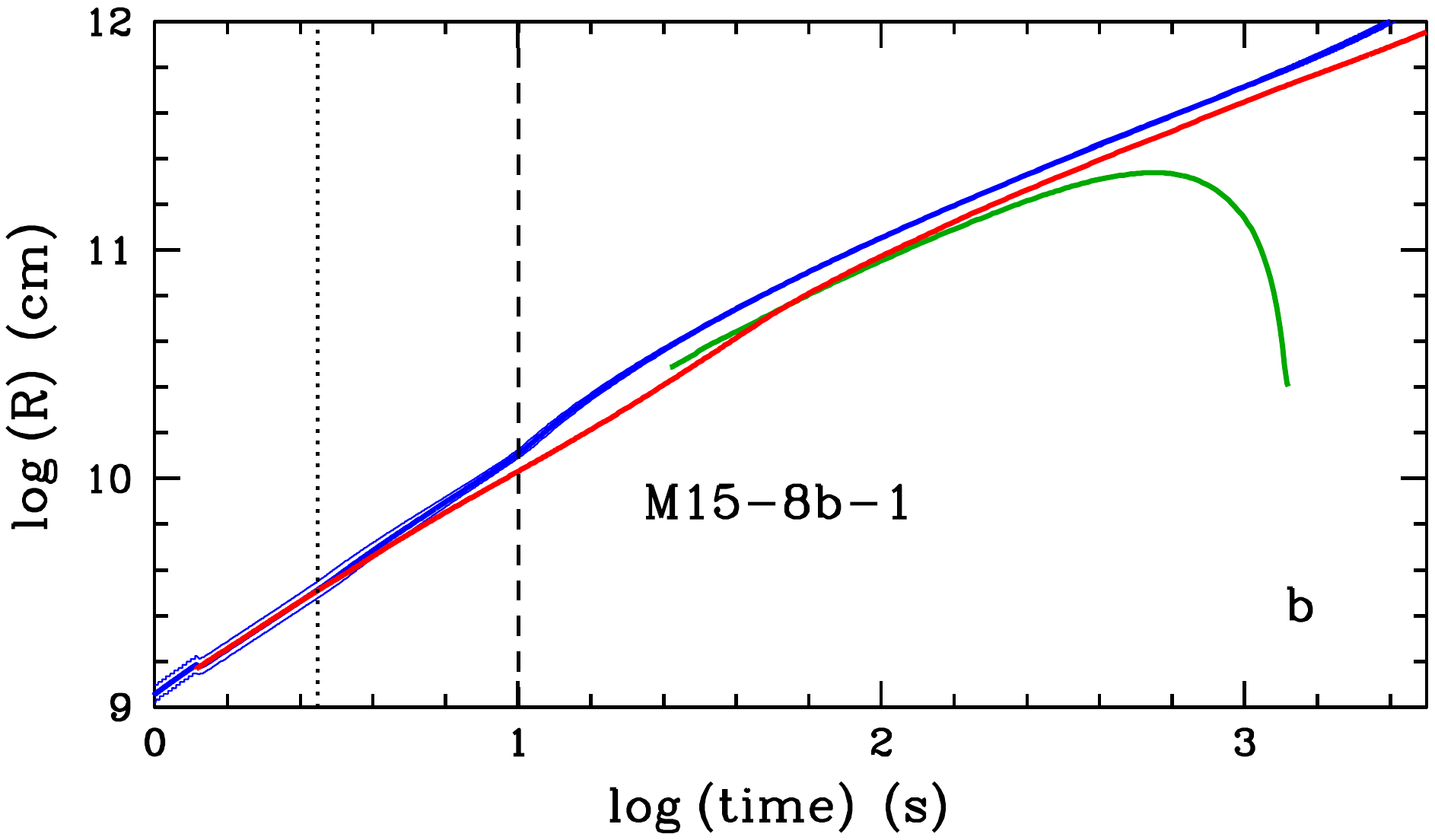}\\
   \vspace{0.5cm}
   \includegraphics[width=0.49\hsize, clip, trim=22 155 47 319]{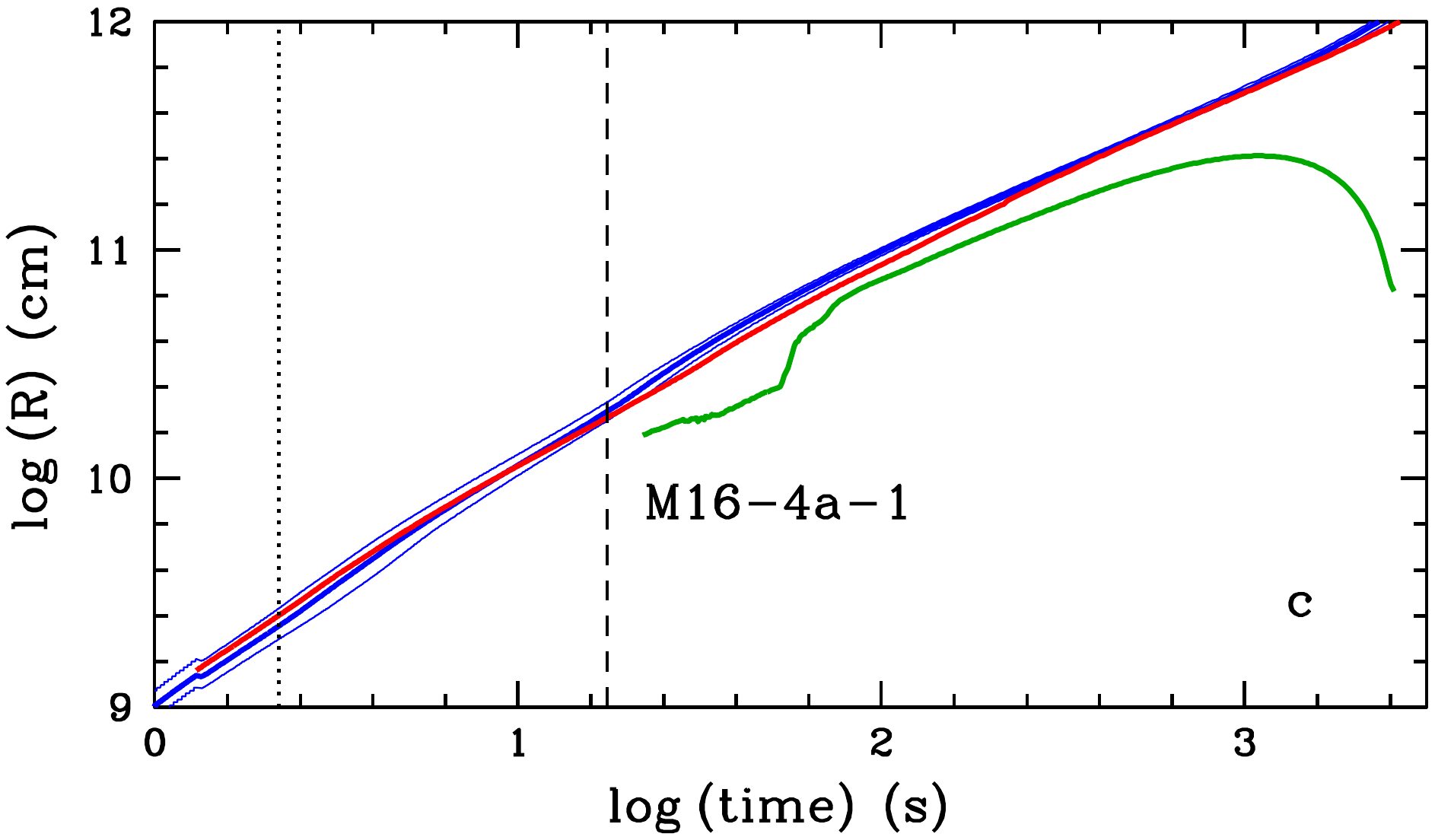}
   \hspace{0.05cm}
   \includegraphics[width=0.49\hsize, clip, trim=22 155 47 319]{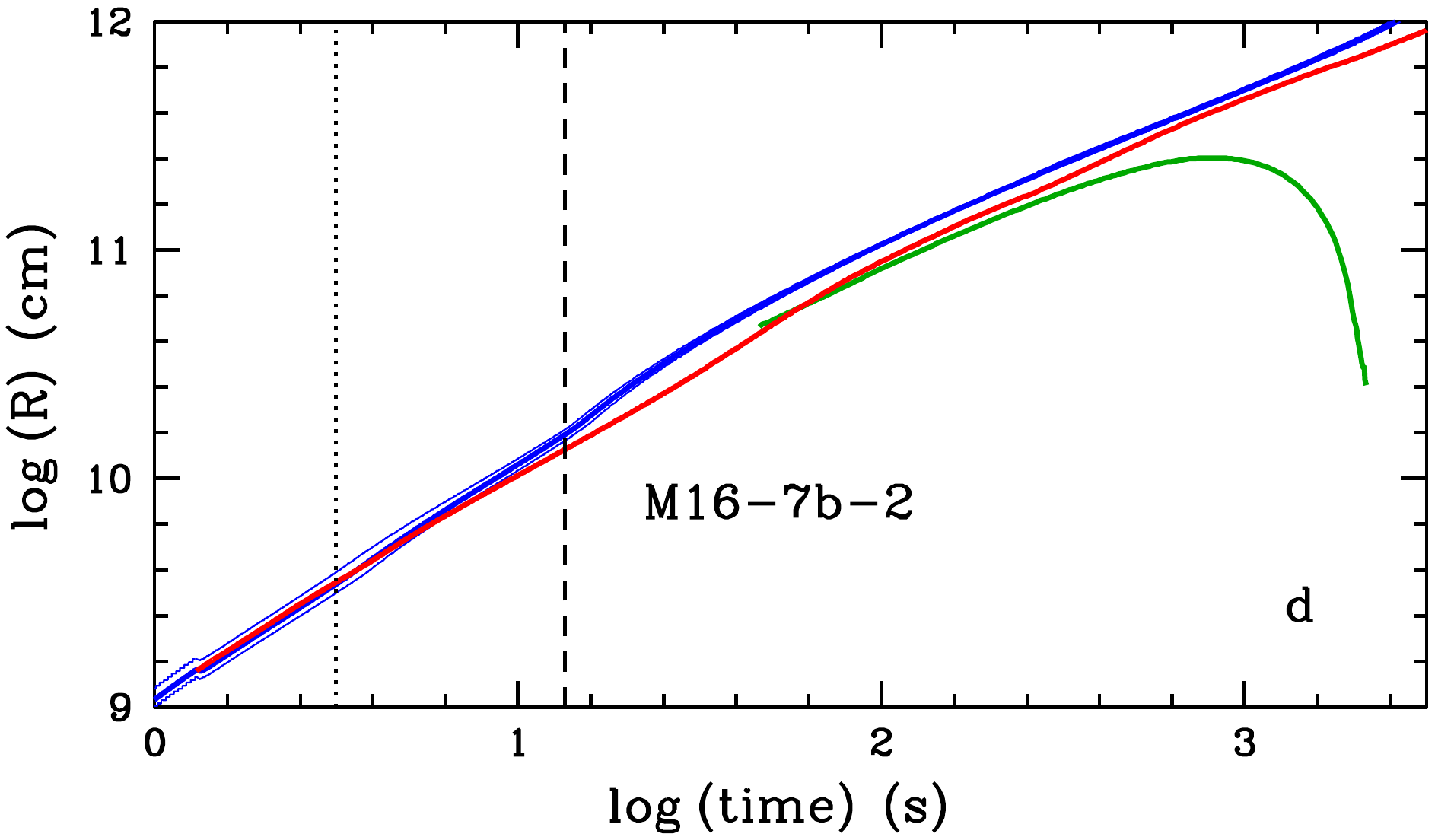}\\
   \vspace{0.5cm}
   \includegraphics[width=0.49\hsize, clip, trim=22 155 47 319]{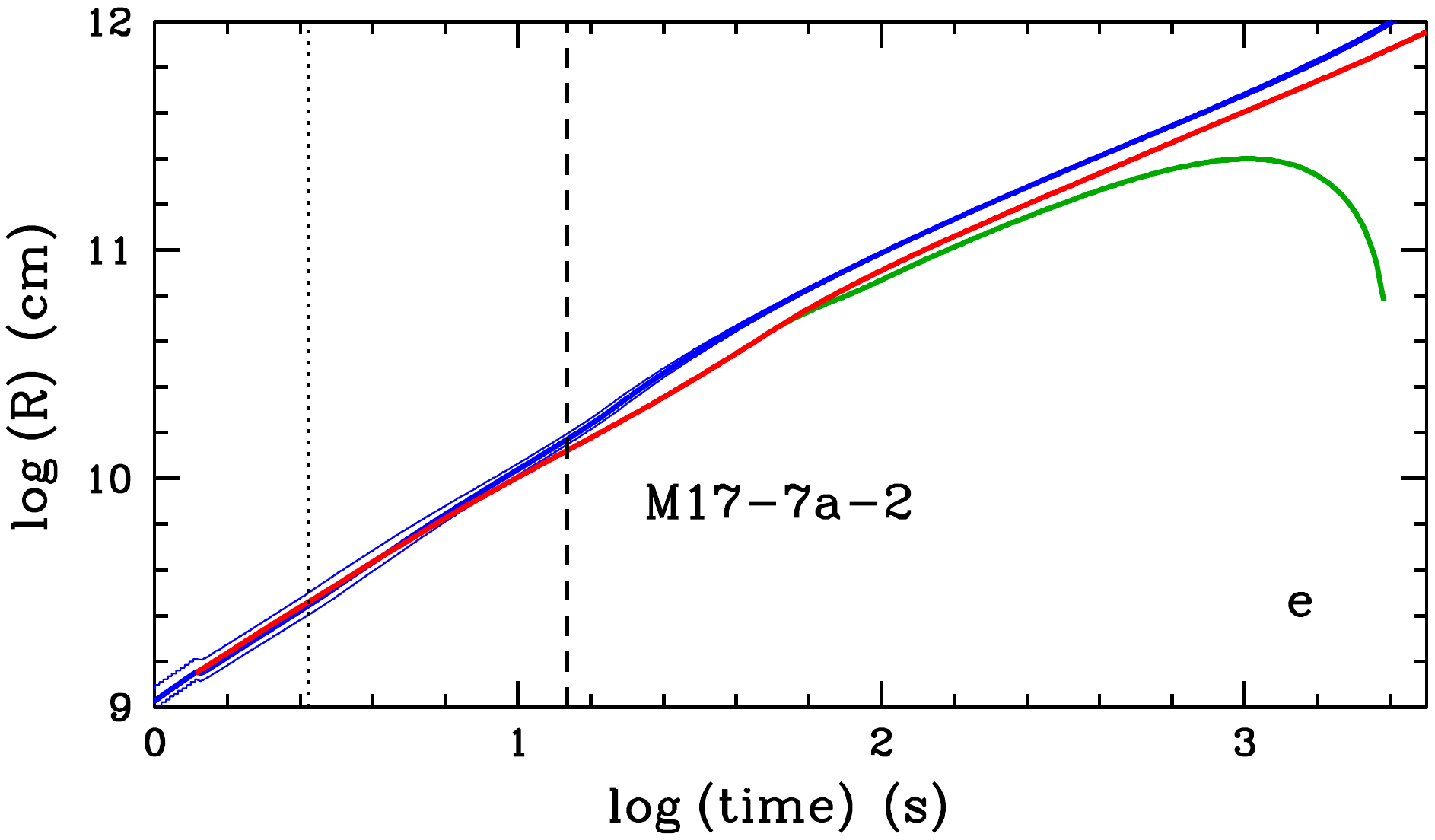}
   \hspace{0.05cm}
   \includegraphics[width=0.49\hsize, clip, trim=22 155 47 319]{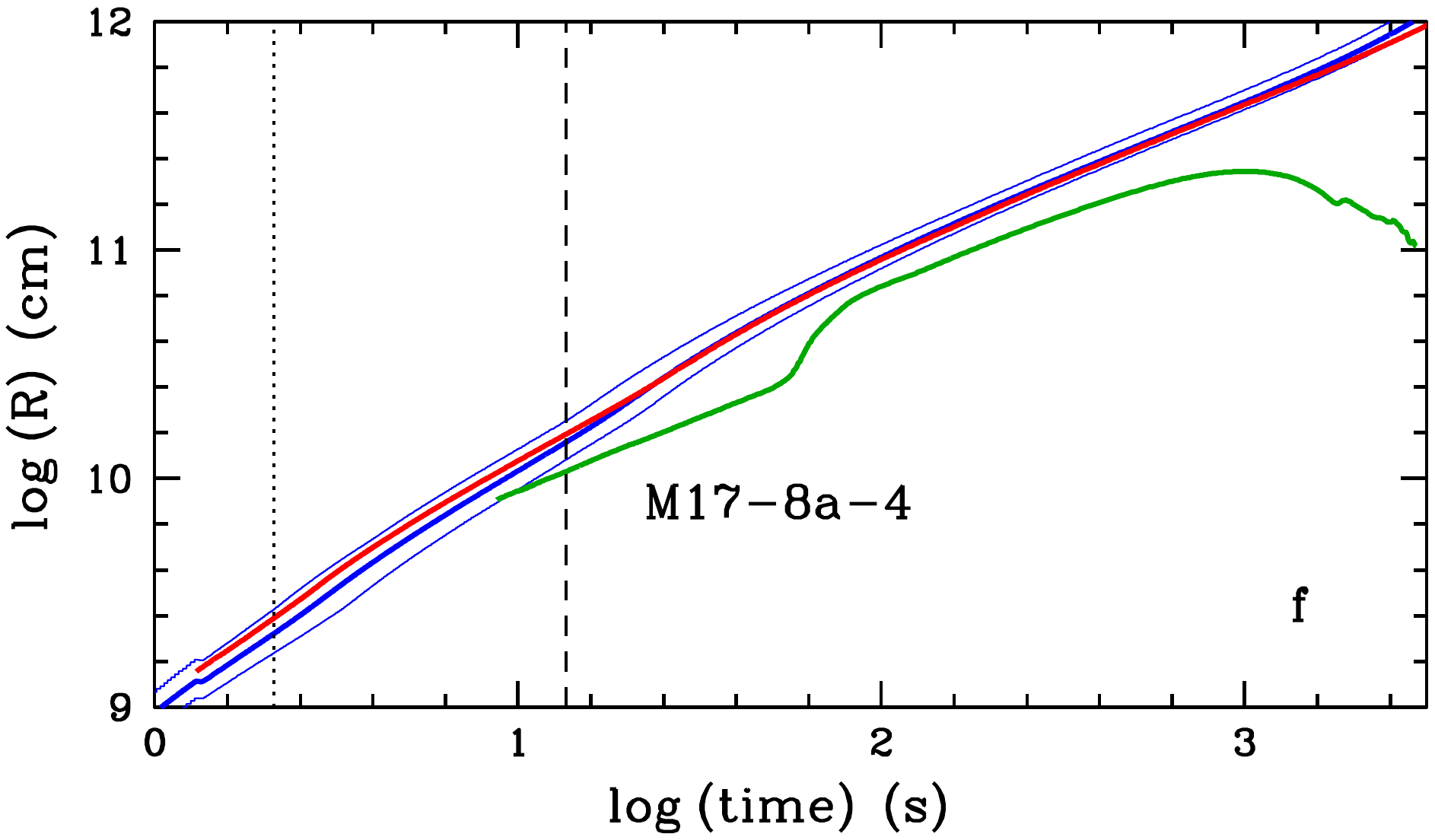}\\
   \caption{%
   Time evolution of the angle-averaged radius of the SN shock
      (blue thick lines), the minimum and maximum radii of the SN shock
      (blue thin lines), the angle-averaged radius of the reverse shock
      (green lines), and the maximum radius on the surface
      where the mass fraction of $^{56}$Ni plus neutron-rich tracer
      nucleus equals $3\%$ (red lines) for our six reference 3D explosion
      models M15-7b-3, M15-8b-1, M16-4a-1, M16-7b-2, M17-7a-2, and M17-8a-4.
   See caption of Figure~\ref{fig:vnivsh} for details.
   }
   \label{fig:rnirsh}
\end{figure*}
\begin{figure*}[t]
\centering
   \includegraphics[width=0.33\hsize, clip, trim=37 163 323 213]{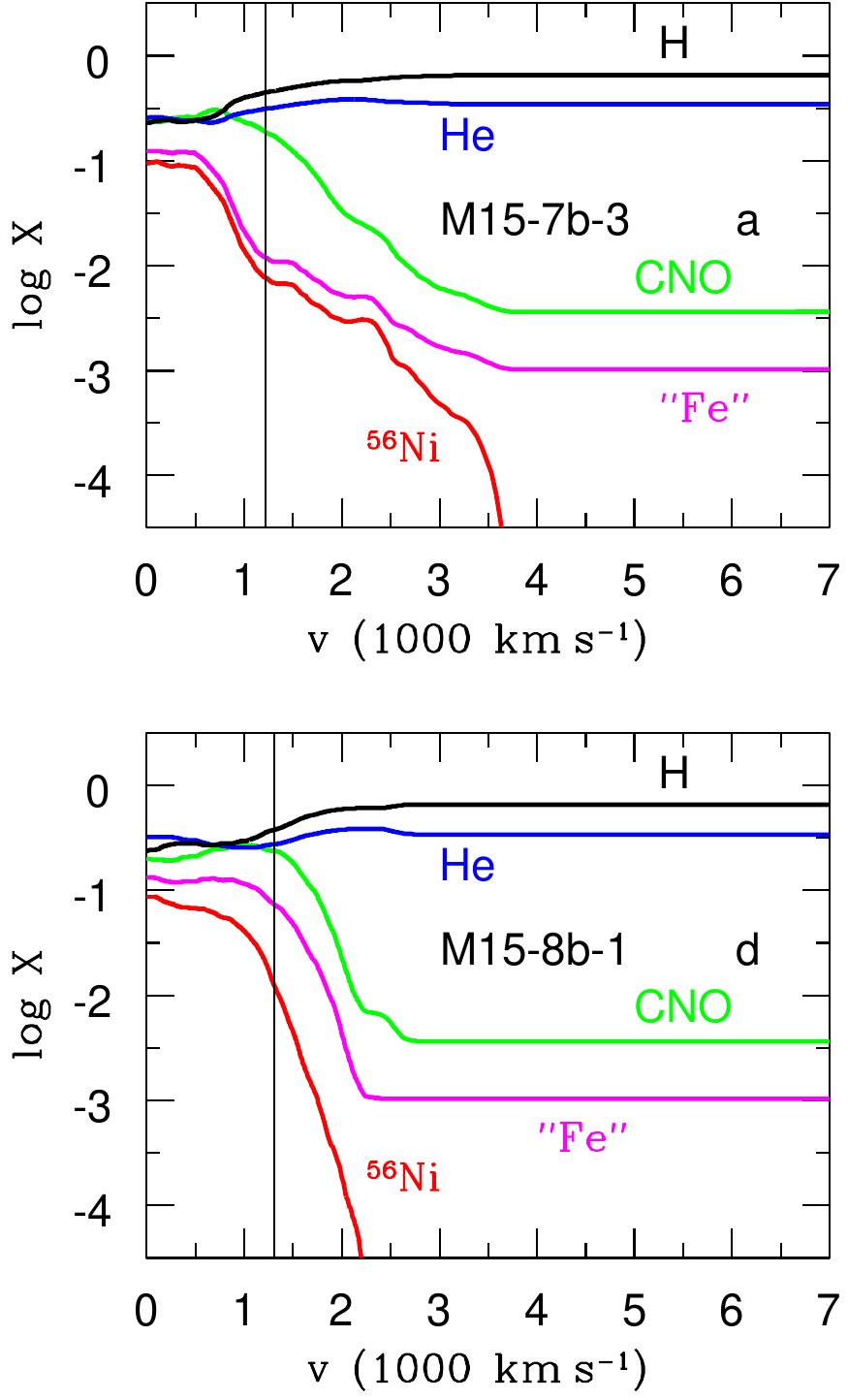}
   \includegraphics[width=0.33\hsize, clip, trim=37 163 323 213]{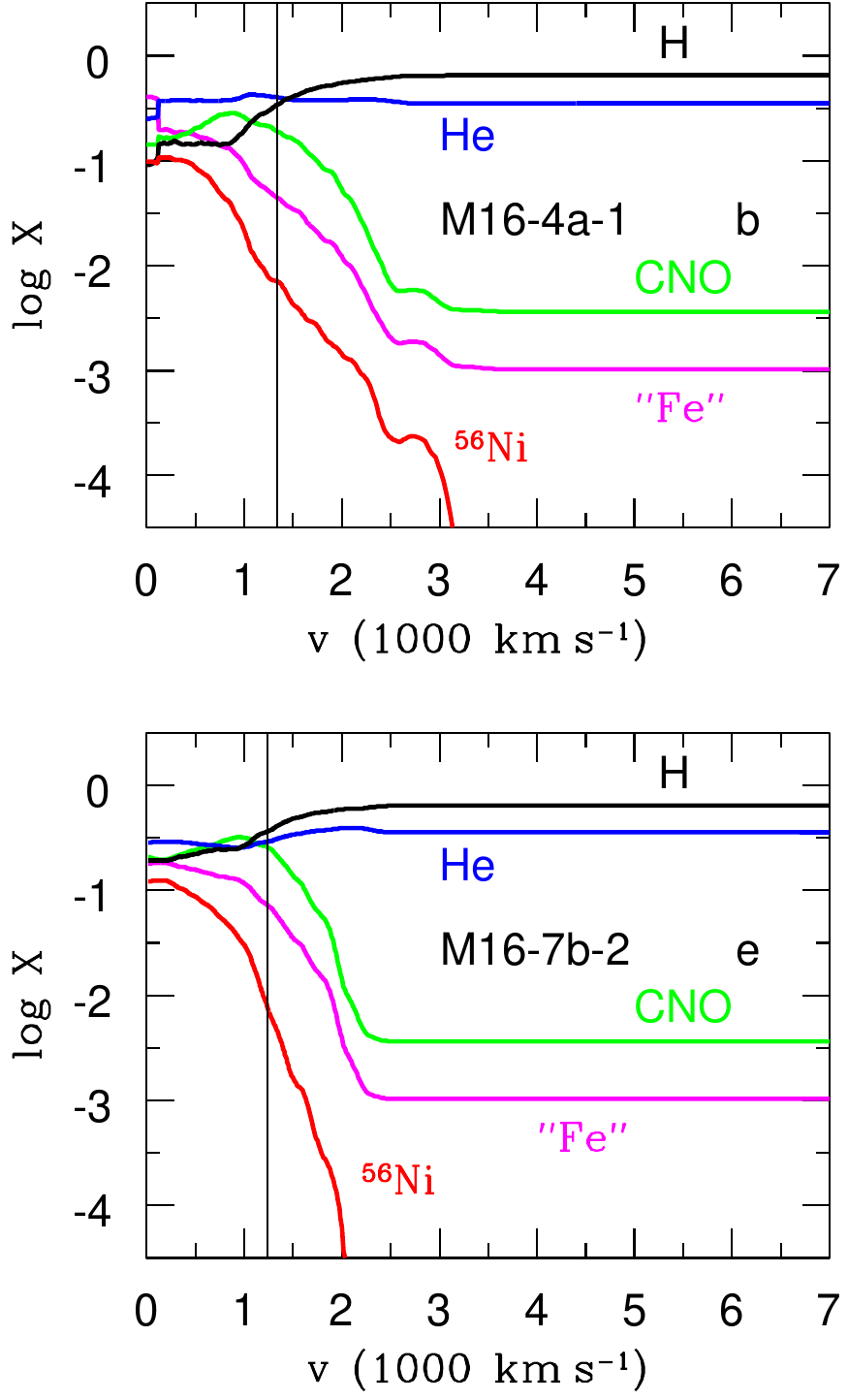}
   \includegraphics[width=0.33\hsize, clip, trim=37 163 323 213]{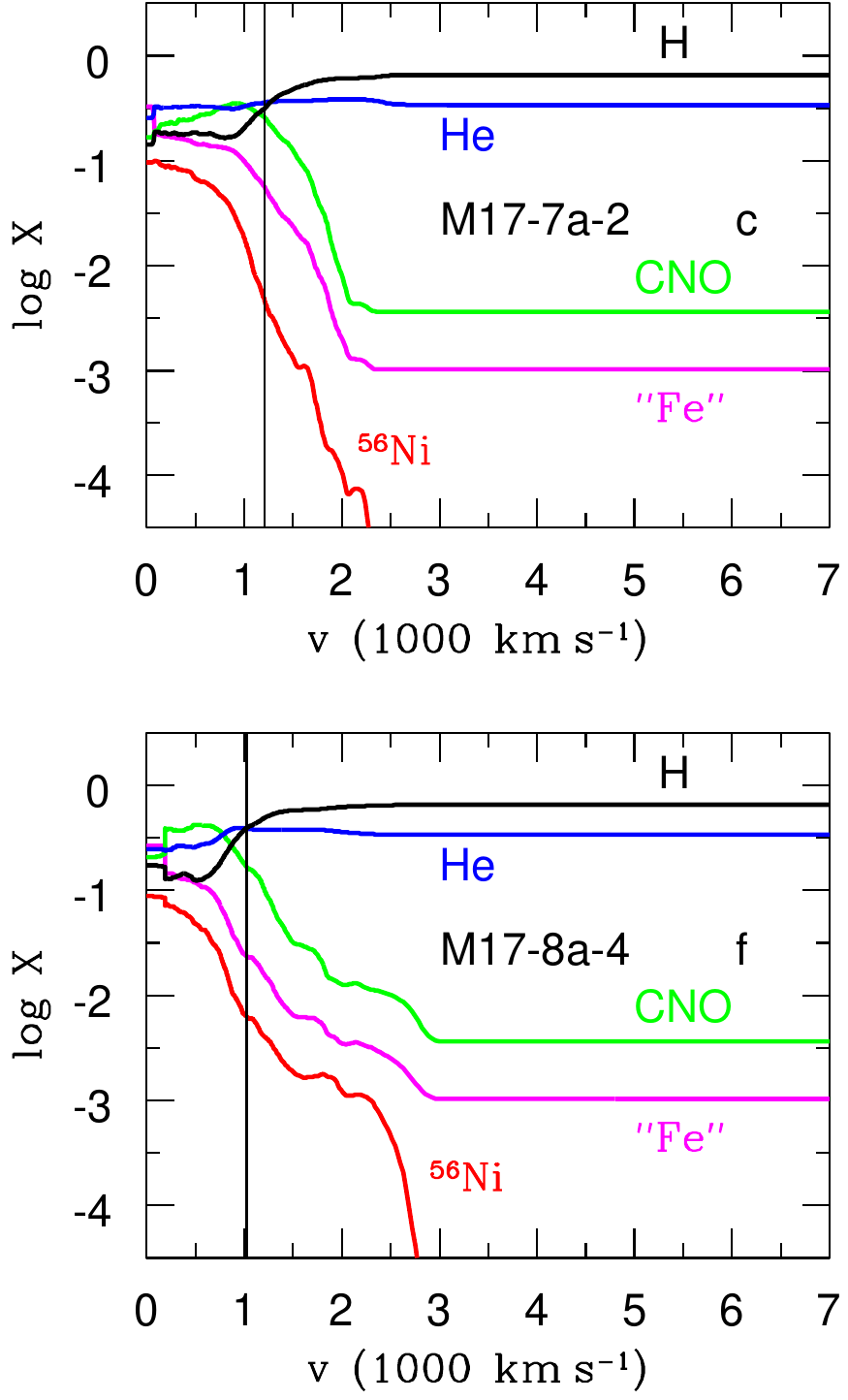}\\
   \caption{%
   Mass fractions of hydrogen (black line),
      helium (blue line), CNO group elements (green line),
      ``iron-group'' elements containing the $\alpha$-nuclei from $^{28}$Si
      to $^{52}$Fe except for $^{56}$Ni (magenta line),
      and radioactive $^{56}$Ni (red line) as functions of velocity
      at day 50 in the averaged 3D explosion models M15-7b-3 (a),
      M16-4a-1 (b), M17-7a-2 (c), M15-8b-1 (d), M16-7b-2 (e),
      and M17-8a-4 (f) (Table~\ref{tab:hydmod}).
   Thin vertical lines indicate ejecta velocities at the location of
      the outer edge of the pre-SN helium cores.
   }
   \label{fig:mfvel}
\end{figure*}

Let us first consider the reference 3D explosion models M15-7b-3,
   M15-8b-1, M16-4a-1, M16-7b-2, M17-7a-2, and M17-8a-4, which explode with
   comparable energies (Table~\ref{tab:hydmod}).
The SN shock wave first crosses the Si/O interface and then reaches the
   (C+O)/He interface, at which time the maximum speed of the bulk mass of
   the $^{56}$Ni-rich matter, $v_\mathrm{Ni}^{\mathrm{CO}}$, 
   spreads over a rather wide range from 9210\,km\,s$^{-1}$ to
   14\,380\,km\,s$^{-1}$ (Table~\ref{tab:nimixing}; first and third columns
   in Figure~\ref{fig:3D_models}), depending on the progenitor.
At this stage the nickel-rich ejecta consist of plume-like or bubble-like
   structures, some of which move outward with relatively high velocities
   compared to others.
Note that during this phase the maximum radial velocities of these structures
   are larger than the average velocity of the SN shock in models M15-7b-3,
   M16-4a-1, and M17-8a-4, and, in contrast, are lower in models M15-8b-1,
   M16-7b-2, and M17-7a-2 (Figure~\ref{fig:vnivsh}).
Figure~\ref{fig:vnivsh} also shows that in the first three models the average
   shock velocity and the maximum radial velocity on the surface where the
   mass fraction of $^{56}$Ni plus neutron-rich tracer nucleus equals $3\%$
   are most similar (even equal in models  M15-7b-3 and M17-8a-4) at the time
   when the shock crosses the He/H interface.
After the SN shock has passed the He/H interface, a reverse shock forms
   below the He/H interface.
In models M15-7b-3, M16-4a-1, and M17-8a-4 a few fastest RT plumes
   expanding with increasing speed begin to decelerate shortly after
   the forward shock has entered the helium shell.
They remain close to the forward
   shock, and, consequently, avoid strong deceleration by the reverse
   shock, because they propagate ahead of the location where the reverse
   shock forms (Figure~\ref{fig:rnirsh}).
As a result, the fastest $^{56}$Ni-rich clumps move with velocities of about
   2780\,km\,s$^{-1}$, 3140\,km\,s$^{-1}$, and 3630\,km\,s$^{-1}$ at late times
   in models M17-8a-4, M16-4a-1, and M15-7b-3, respectively
   (Figure~\ref{fig:3D_models}, second and fourth columns;
   Figure~\ref{fig:mfvel}).
In contrast, in models M16-7b-2, M15-8b-1, and M17-7a-2 the $^{56}$Ni-rich
   plumes collide with the reverse shock, move well behind the forward
   shock (Figure~\ref{fig:rnirsh}), and are compressed in the dense shell
   that piles up between the forward and reverse shocks.
Thus they exhibit a considerably more isotropic overal shape with a much lower
   velocity contrast between the fastest and slowest structures
   (Figure~\ref{fig:3D_models}, second and fourth columns).
In models M16-7b-2, M15-8b-1, and M17-7a-2 the peak velocities are only about
   2030\,km\,s$^{-1}$, 2200\,km\,s$^{-1}$, and 2280\,km\,s$^{-1}$, respectively
   (Figure~\ref{fig:mfvel}).

A striking difference in the final morphology of
   the $^{56}$Ni-rich ejecta results in our reference 3D explosion models
   (Figure~\ref{fig:3D_models}, second and fourth columns) from
   the different structures of the binary-merger progenitors
   (Figure~\ref{fig:denmr}), and their influence on the unsteady
   SN shock propagation and the development of RT instabilities at the
   (C+O)/He and He/H composition interfaces (Figure~\ref{fig:rtgrowth}).
The final morphology of models M15-8b-1, M16-7b-2, and M17-7a-2 (the first
   group) exhibits a small asphericity in contrast to models M15-7b-3,
   M16-4a-1, and M17-8a-4 (the second group) which are largely aspherical.
This difference may be interpreted as a consequence of the different strength
   of the interaction between the $^{56}$Ni-rich ejecta and the reverse shock
   at the He/H composition interface.
In reality, the extent of $^{56}$Ni mixing, measured by the dimensionless
   ratio of the maximum velocity of the bulk mass of $^{56}$Ni ejecta to
   the mean velocity of the bulk mass of $^{56}$Ni (averaged with the
   $^{56}$Ni mass fraction as weight function) increases from the first to
   the second group of models (Figure~\ref{fig:mfvel},
   Table~\ref{tab:nimixing}).
In other words, the degree of asymmetry in the final morphology of
   the $^{56}$Ni-rich ejecta correlates with the extent of $^{56}$Ni mixing
   in velocity space.

\begin{figure}[t]
\centering
   \includegraphics[width=\hsize, clip, trim=21 156 43 97]{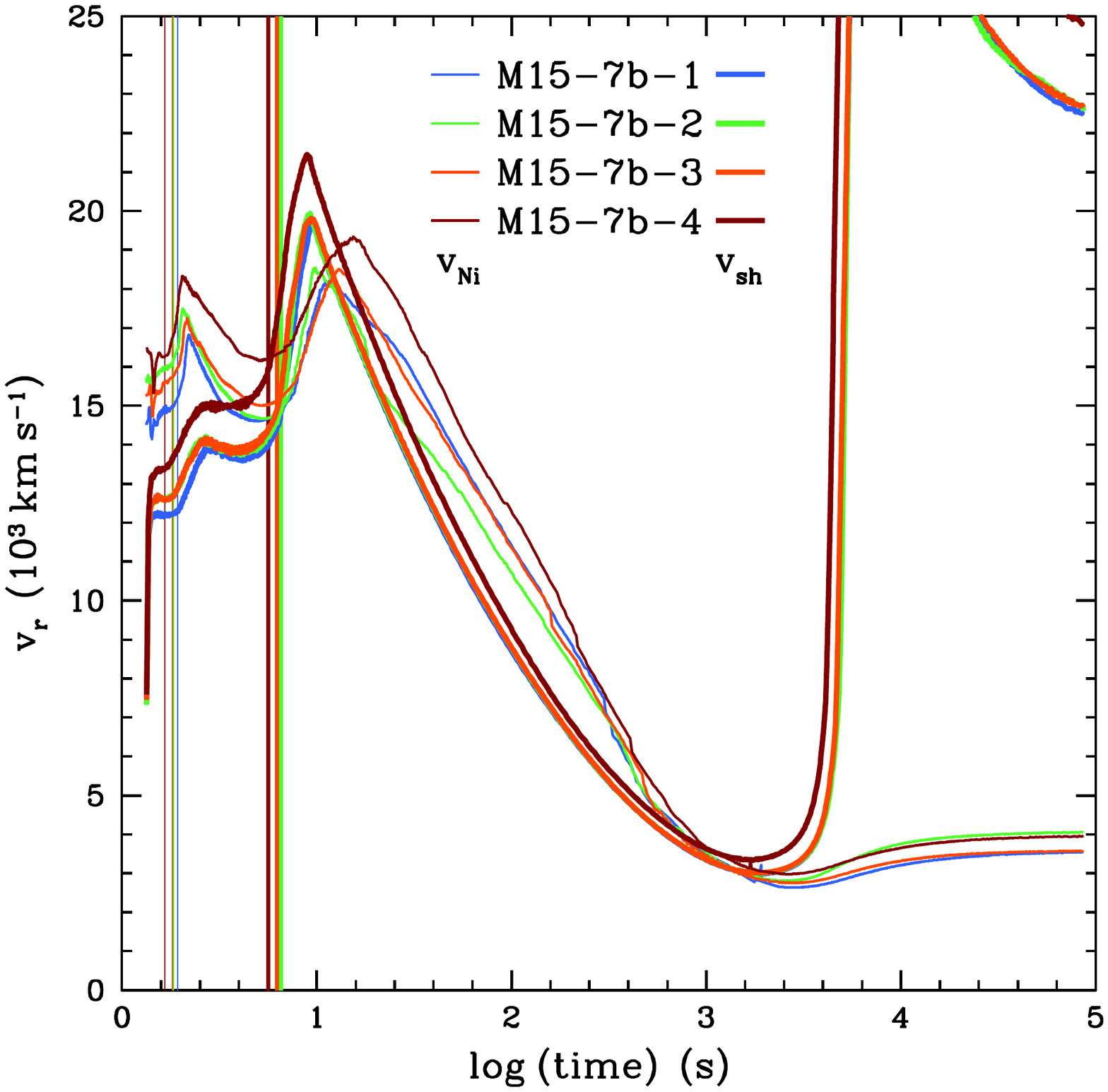}
   \caption{%
   Time evolution of the average velocity of the SN shock, $v_\mathrm{sh}$
      (thick lines), and of the maximum radial velocity on the surface
      where the mass fraction of $^{56}$Ni plus the neutron-rich tracer
      nucleus equals $3\%$, $v_{\mathrm{Ni}}$ (thin lines), for models
      M15-7b-1, M15-7b-2, M15-7b-3, and M15-7b-4.
   The vertical thin and thick lines mark the times when the shock crosses
      the (C+O)/He and He/H composition interfaces, respectively.
   }
   \label{fig:vnivsh_157b}
\end{figure}
\begin{figure*}
\centering
   \includegraphics[width=0.23\hsize, clip, trim=60 60 60 60]{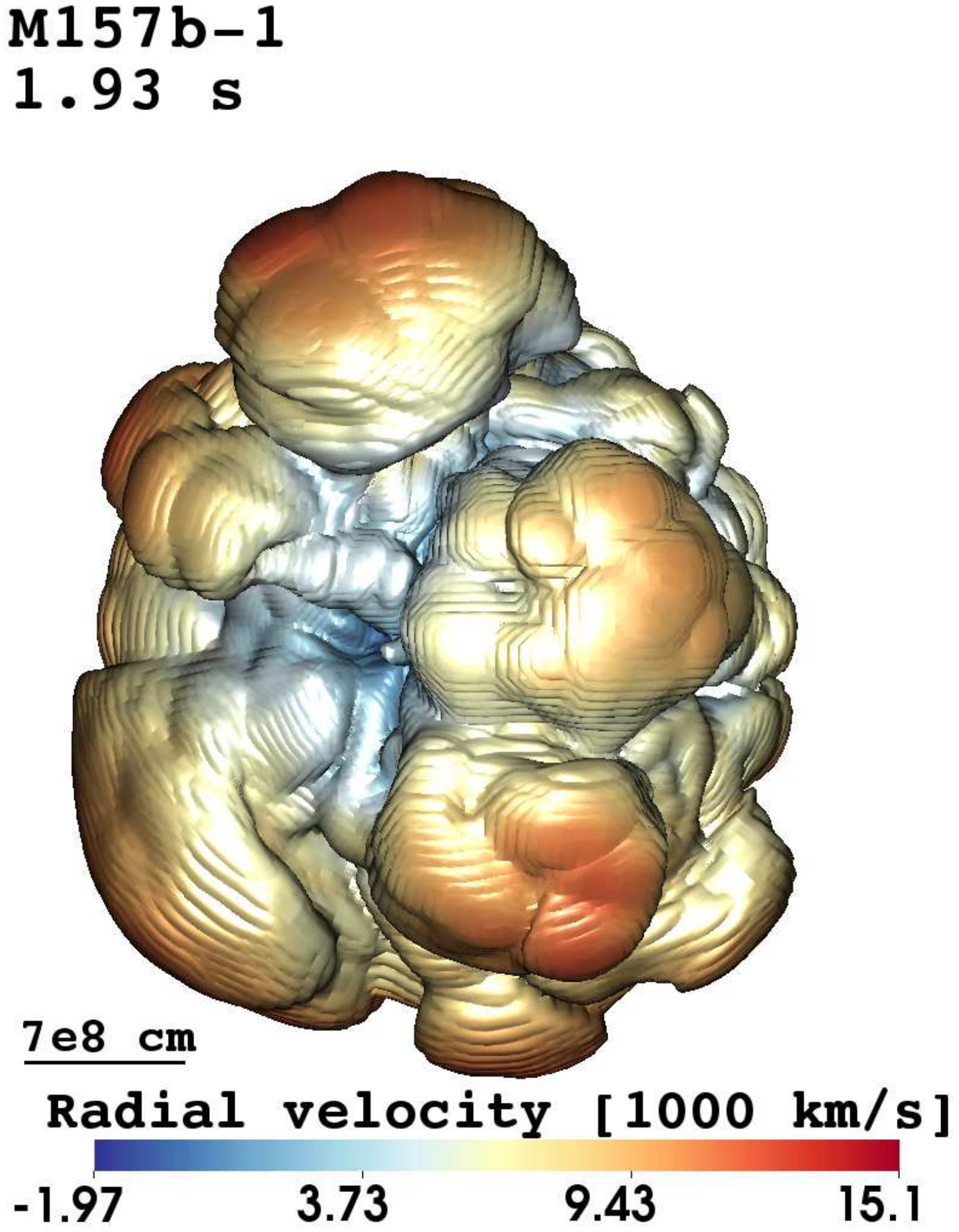}
   \includegraphics[width=0.23\hsize, clip, trim=60 60 60 60]{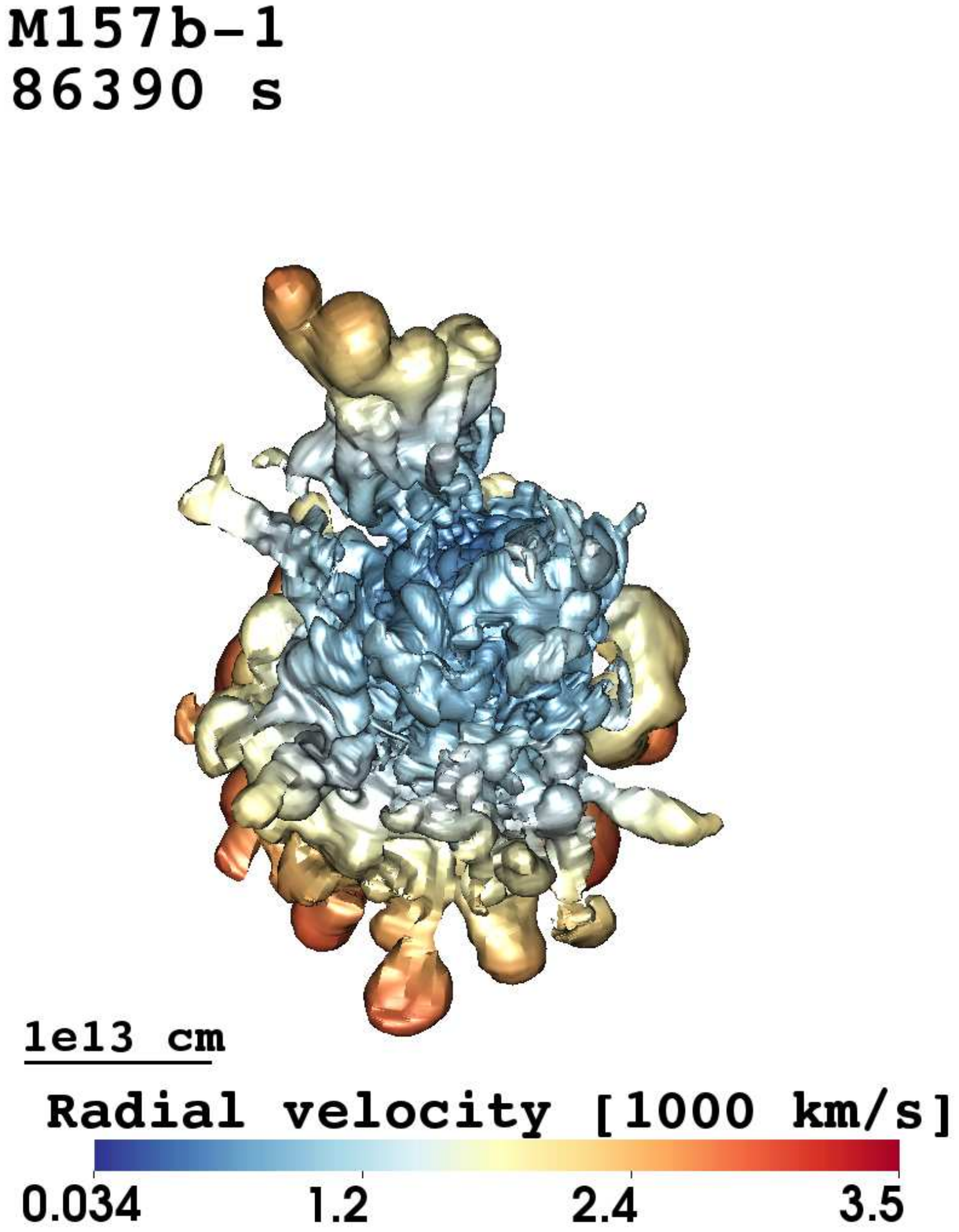}
   \hspace{0.5cm}
   \includegraphics[width=0.23\hsize, clip, trim=60 60 60 60]{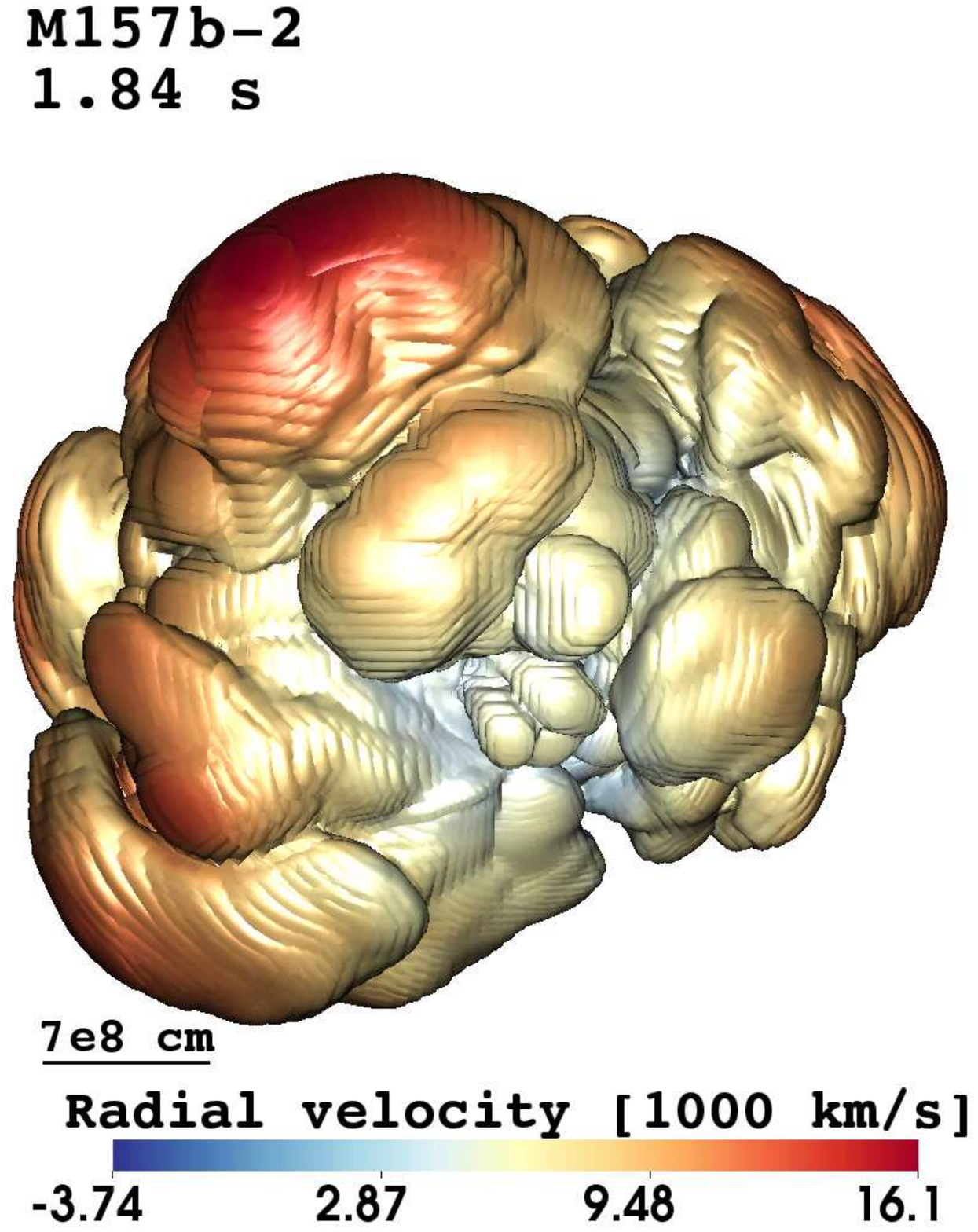}
   \includegraphics[width=0.23\hsize, clip, trim=60 60 60 60]{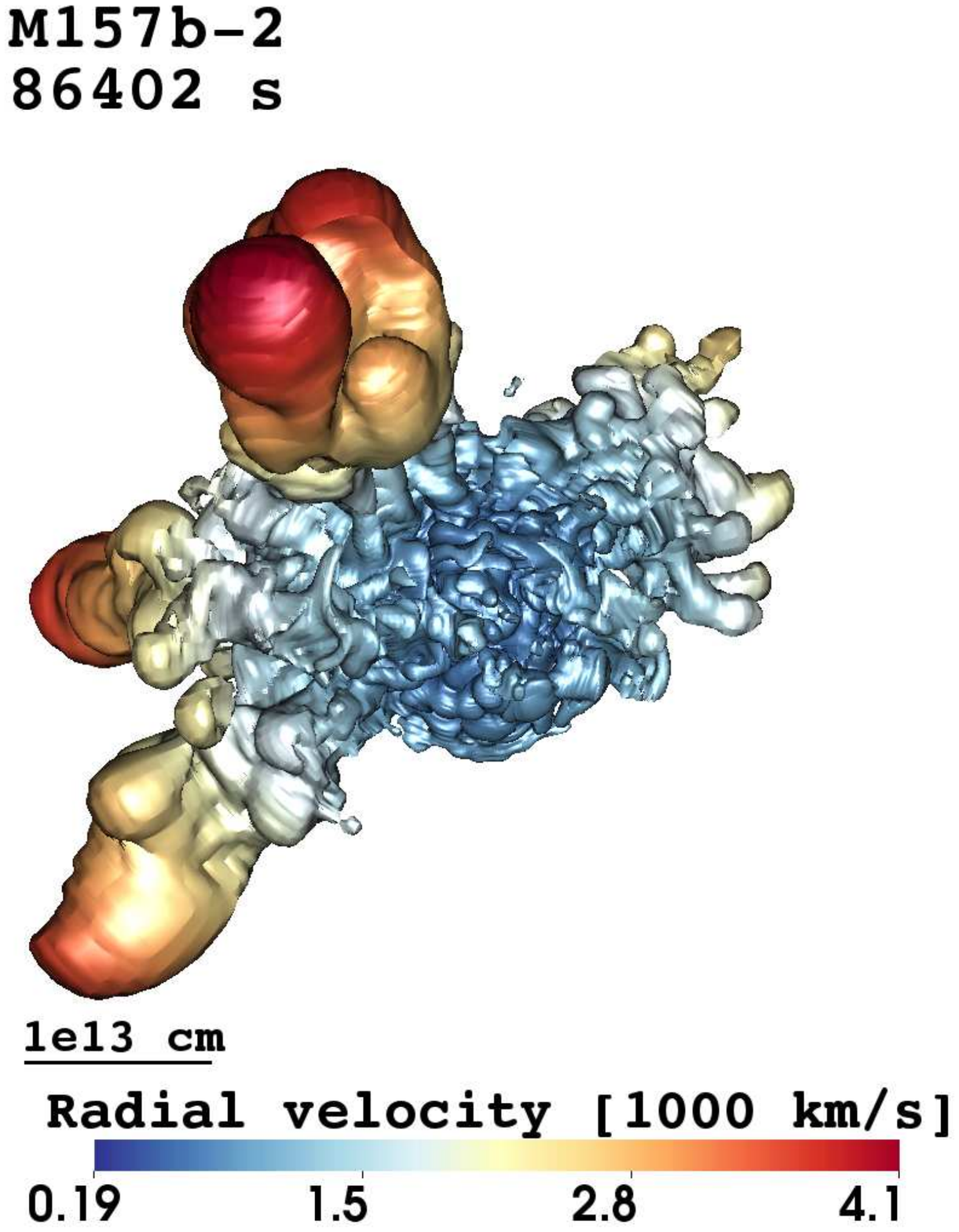}
   \vspace{0.5cm}
   \includegraphics[width=0.23\hsize, clip, trim=60 60 60 60]{M157b-3-early.pdf}
   \includegraphics[width=0.23\hsize, clip, trim=60 60 60 60]{M157b-3-late.pdf}
   \hspace{0.5cm}
   \includegraphics[width=0.23\hsize, clip, trim=60 60 60 60]{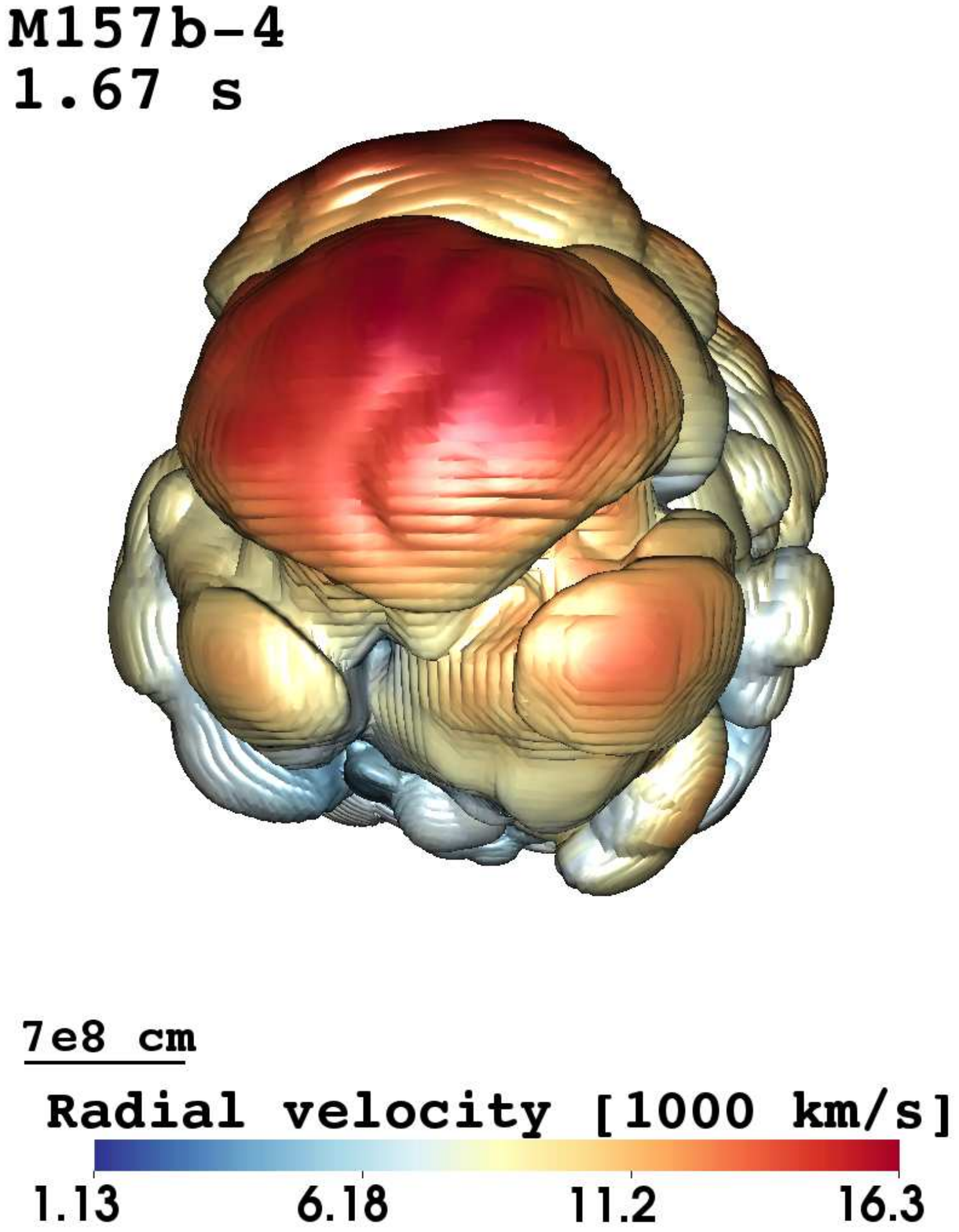}
   \includegraphics[width=0.23\hsize, clip, trim=60 60 60 60]{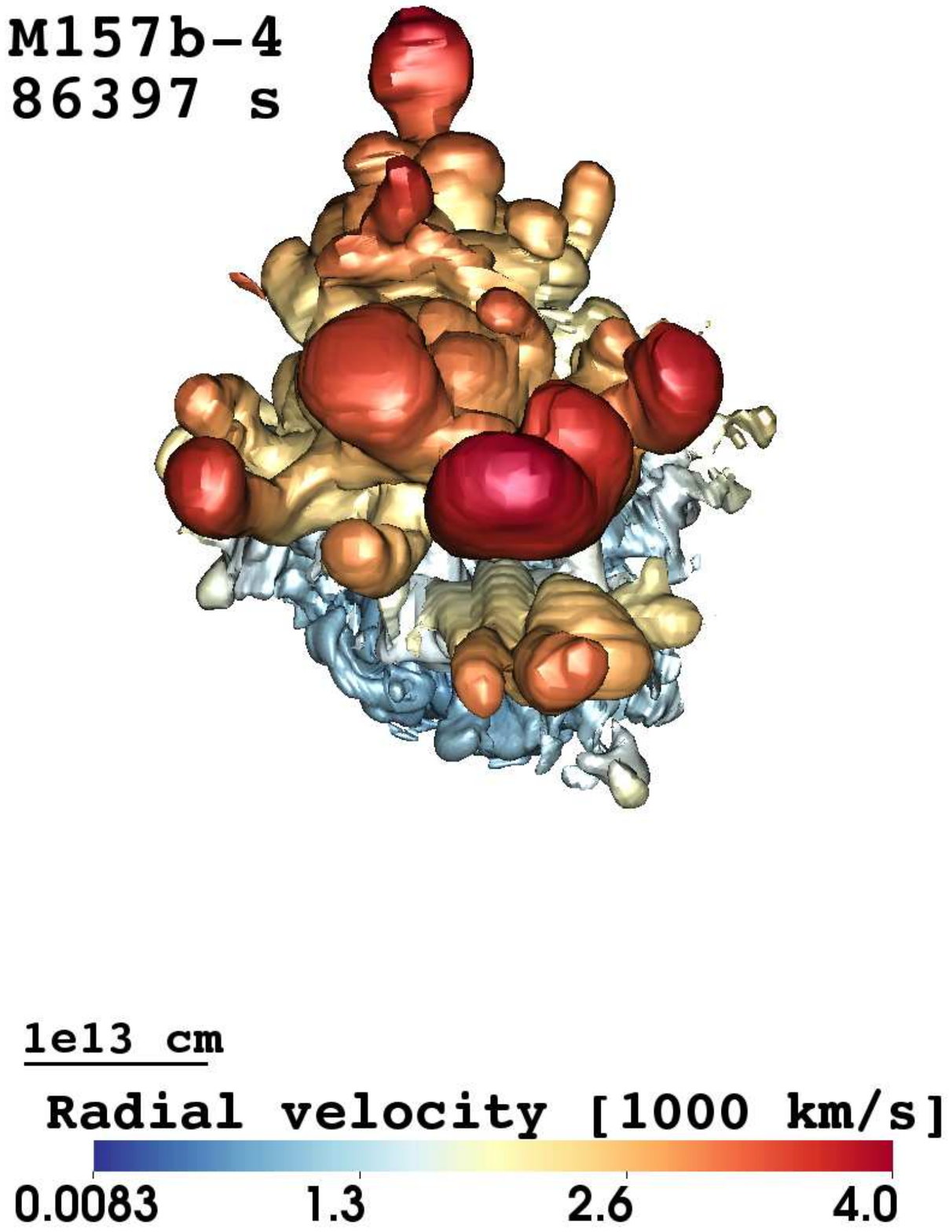}
   \caption{%
   Morphology of radioactive $^{56}$Ni-rich ejecta in models M15-7b-1,
      M15-7b-2, M15-7b-3, and M15-7b-4.
   See caption of Figure~\ref{fig:3D_models} for details.
   }
   \label{fig:3D_157b}
\end{figure*}
%


\subsection{Variations of explosion energy and stochasticity}
\label{sec:results-stoch}
%
3D neutrino-driven explosion simulations of SN~1987A based on the single-star
   pre-SNe showed that the extent of $^{56}$Ni mixing, measured by the maximum
   radial velocity of the bulk mass of the $^{56}$Ni-rich ejecta,
   is proportional to the explosion energy \citep{UWJM_15, UWJ_19}.
Such a correlation is also observed for 3D neutrino-driven explosion
   simulations based on the binary-merger pre-SNe with some exceptions
   (Table~\ref{tab:hydmod}).
The most prominent exception is demonstrated by models M15-7b-1, M15-7b-2,
   M15-7b-3, and M15-7b-4 which are based on the same progenitor model M15-7b
   and explode with a monotonically growing energy (Table~\ref{tab:hydmod}).
This sequence of models provides an opportunity to explore the dependence of
   their mixing properties not only on the explosion energy itself, but also
   in models exploding with sufficiently similar energies, on the initial
   asymmetry of the $^{56}$Ni-rich ejecta, created by the stochastic
   growth of convective and SASI mass motions.
In order to realize this opportunity, we analyze the time evolution of the
   radial velocity of the SN shock and the maximum radial velocity of the
   radioactive $^{56}$Ni-rich ejecta (Figure~\ref{fig:vnivsh_157b}) and
   the morphology of this matter (Figure~\ref{fig:3D_157b}).

Models M15-7b-1 and M15-7b-4 have considerably different explosion energies and
   obey the mentioned correlation between the explosion energy and the extent of
   $^{56}$Ni mixing (Table~\ref{tab:hydmod}).
Accordingly, the average velocity of the SN shock and the maximum radial
   velocity of the $^{56}$Ni-rich ejecta are both higher at all times in
   the more energetic explosion of model M15-7b-4 compared to model M15-7b-1
   (Figure~\ref{fig:vnivsh_157b}).
We then focus on the interesting results obtained for models M15-7b-2 and
   M15-7b-3 exploding with nearly the same energies of 1.428\,B and 1.432\,B,
   but whose $^{56}$Ni-rich ejecta are mixed to quite different maximum
   velocities of the bulk mass of $^{56}$Ni of 3406\,km\,s$^{-1}$ and
   2980\,km\,s$^{-1}$, respectively (Table~\ref{tab:hydmod}).
Such a significant difference in the maximum velocity of
   the bulk mass of $^{56}$Ni cannot be explained by a negligible difference
   in the explosion energy.

Along with the explosion energy, the initial asymmetry of the $^{56}$Ni-rich
   ejecta is another important property of SN explosions that determines the
   maximum radial velocity of $^{56}$Ni-rich ejecta.
\citet{WMJ_15} found that there is a clear correlation between the asymmetries
   of the $^{56}$Ni-rich ejecta at late times and the early-time asymmetries
   originating from hydrodynamic instabilities generated during the onset of
   the SN explosion.
In other words, the fundamental features of the morphology of the
   $^{56}$Ni-rich ejecta are imprinted by the ``neutrino
   engine'' during the first second of the explosion.
When the SN shock approaches the (C+O)/He composition interface,
   the maximum radial velocity of the $^{56}$Ni-rich ejecta in model M15-7b-2
   exceeds that of model M15-7b-3 (Figure~\ref{fig:vnivsh_157b}) despite
   their practically equal explosion energies (Table~\ref{tab:hydmod}).
This fact may be interpreted in terms of a specific morphology of the
   $^{56}$Ni-rich ejecta in these models, which is seeded by different
   random perturbations and amplified by the stochastic growth of the
   hydrodynamic instabilities in the neutrino-heating layer.
Inspecting the early-time morphology of the $^{56}$Ni-rich ejecta of models
   M15-7b-2 and M15-7b-3 in Figure~\ref{fig:3D_157b}, we see large
   mushroom-head-like structures in directions of the biggest initial
   convective plumes observed at the onset of the explosion.
These mushroom heads are more prominent and faster in model M15-7b-2 than
   in M15-7b-3.
At late times these large-scale, coherent structures have grown to extended
   RT fingers with the maximum speed of the bulk mass of the $^{56}$Ni-rich
   matter in model M15-7b-2 being larger than that in model M15-7b-3, similar
   to the situation at early times.
This essential property and difference between
   models M15-7b-2 and M15-7b-3 is facilitated by the fact that the fast
   RT plumes in model M15-7b-2 enter the helium shell earlier than in M15-7b-3,
   which is indicated by the earlier deceleration of the $^{56}$Ni-rich ejecta
   in model M15-7b-2.
Consequently, the RT plumes are closer to the SN shock in model M15-7b-2 and
   undergo a weaker interaction with the reverse shock that forms below
   the He/H composition interface (Figure~\ref{fig:vnivsh_157b}).
Thus, a comparison of models M15-7b-1, M15-7b-2, M15-7b-3, and M15-7b-4
   leads to the generalized conclusion that the maximum velocity of the bulk
   mass of $^{56}$Ni is proportional to the explosion energy with some
   stochastic uncertainty for a given progenitor model.

The morphology of the $^{56}$Ni-rich ejecta in models M15-7b-1, M15-7b-2,
   M15-7b-3, and M15-7b-4 (Figure~\ref{fig:3D_157b}) exploding with
   a monotonically growing energy (Table~\ref{tab:hydmod}) confirms two
   previous findings.
First, these models demonstrate the correlation between the early-time
   asymmetries of the $^{56}$Ni-rich ejecta and the asymmetries at late times
   discussed already by \citet{WMJ_15}.
Second, the size of the asymmetry at early times correlates with the degree of
   asymmetry in the final morphology of the $^{56}$Ni-rich ejecta, which
   in turn correlates with the extent of $^{56}$Ni mixing
   in velocity space: the global deformation of M15-7b-1 is much less extreme,
   models M15-7b-2 and M15-7b-3 have a very pronounced asphericity with
   a bipolar elongation, and the morphology of model M15-7b-4 is distinguished
   by a strongly one-sided distribution of the $^{56}$Ni-rich ejecta,
   suggesting a large dipole asymmetry of the SN shock at the onset of
   the explosion.\\


%
\begin{deluxetable*}{ l c c c c c c c c c c }
\tabletypesize{\small}
\tablewidth{0pt}
\tablecaption{$^{56}$Ni mixing and hydrodynamic properties of binary-merger
              progenitors%
\label{tab:nimixing}}
\tablehead{
\colhead{Model} & \colhead{$M_\mathrm{He}^{\,\mathrm{core}}$}
       & \colhead{$t^{\mathrm{CO}}$}
       & \colhead{$t^{\mathrm{He}}$}
       & \colhead{$\langle v \rangle_\mathrm{Ni}^{\mathrm{150}}$}
       & \colhead{\phantom{m}$v_\mathrm{Ni}^{\mathrm{150}}$}
       & \colhead{\phantom{m}$v_\mathrm{Ni}^{\mathrm{CO}}$}
       & \colhead{$\log\,(\xi/\xi_{0})^\mathrm{CO}$}
       & \colhead{$\log\,(\xi/\xi_{0})^\mathrm{He}$}
       & \colhead{$t_\mathrm{RS}$}
       & \colhead{$R_\mathrm{RS}$} \\
\colhead{} & \colhead{$(M_{\sun})$}
           & \multicolumn{2}{c}{(s)}
           & \multicolumn{3}{c}{($10^3$\,km\,s$^{-1}$)}
           & \colhead{} & \colhead{} & \colhead{(s)} & \colhead{$(R_{\sun})$}
}
\startdata
 M15-7b-1 & 2.90 & 1.933 & \phantom{e}6.46  & 1.344 & 2.629 & 14.51           & 1.16 & 14.4           & 19.0 & 0.607 \\
 M15-7b-2 & 2.90 & 1.844 & \phantom{e}6.58  & 1.351 & 2.954 & 14.38           & 1.16 & 14.4           & 18.9 & 0.605 \\
 M15-7b-3 & 2.90 & 1.814 & \phantom{e}6.26  & 1.364 & 2.579 & 14.83           & 1.16 & 14.4           & 18.9 & 0.606 \\
 M15-7b-4 & 2.90 & 1.668 & \phantom{e}5.64  & 1.500 & 3.344 & 16.36           & 1.16 & 14.4           & 17.5 & 0.600 \\
 M15-8b-1 & 2.95 & 2.811 & 10.03            & 0.992 & 1.609 & \phantom{e}9.78 & 1.48 & 11.6           & 10.6 & 0.559 \\
 M15-8b-2 & 2.95 & 3.041 & 11.39            & 0.816 & 1.282 & \phantom{e}8.32 & 1.48 & 11.6           & 11.5 & 0.556 \\
 M16-4a-1 & 4.10 & 2.200 & 17.63            & 0.951 & 1.997 & 10.79           & 4.49 & \phantom{e}8.1 & 35.9 & 1.049 \\
 M16-4a-2 & 4.10 & 2.276 & 19.14            & 0.790 & 1.721 & 10.07           & 4.49 & \phantom{e}8.1 & 41.3 & 1.059 \\
 M16-7b-1 & 3.41 & 3.301 & 14.24            & 0.811 & 1.356 & \phantom{e}8.74 & 2.15 & 10.0           & 20.4 & 0.791 \\
 M16-7b-2 & 3.41 & 3.154 & 13.45            & 0.881 & 1.485 & \phantom{e}9.42 & 2.15 & 10.0           & 18.2 & 0.760 \\
 M17-7a-1 & 4.25 & 2.659 & 13.82            & 0.831 & 1.526 & \phantom{e}9.21 & 2.32 & \phantom{e}7.7 & 28.6 & 0.890 \\
 M17-7a-2 & 4.25 & 2.653 & 13.64            & 0.819 & 1.446 & \phantom{e}9.21 & 2.32 & \phantom{e}7.7 & 28.3 & 0.889 \\
 M17-8a-3 & 4.23 & 2.064 & 13.87            & 0.941 & 2.068 & 11.12           & 4.80 & \phantom{e}9.0 & 45.1 & 1.003 \\
 M17-8a-4 & 4.23 & 2.130 & 13.58            & 0.943 & 2.147 & 11.09           & 4.80 & \phantom{e}9.0 & 43.2 & 1.001 \\
\enddata
\tablecomments{%
Columns 1 and 2 give the name of the 3D hydrodynamic explosion model
   and the helium-core mass of the corresponding pre-SN model.
Columns 3 and 4 list the times when the SN shock in the 3D explosion models
   crosses the C+O/He and He/H composition interfaces, respectively.
Columns 5, 6, and 7 provide the averaged characteristic velocities for
   the bulk of $^{56}$Ni containing $96\%$ of the total $^{56}$Ni mass in
   the 3D simulations:
$\langle v \rangle_\mathrm{Ni}^{\mathrm{150}}$ is the mean velocity
   of the bulk mass of $^{56}$Ni, averaged with the $^{56}$Ni mass fraction
   as weight function at day 150;
$v_\mathrm{Ni}^{\mathrm{150}}$ is the maximum velocity of the bulk mass of
   $^{56}$Ni at the same epoch; and
$v_\mathrm{Ni}^{\mathrm{CO}}$ is the maximum velocity of the bulk mass of
   $^{56}$Ni at the moment just after the SN shock has passed the (C+O)/He
   composition interface.
Columns $8-11$ give hydrodynamic properties of 1D explosion models:
$(\xi/\xi_{0})^\mathrm{CO}$ and $(\xi/\xi_{0})^\mathrm{He}$ are the maximum
   time-integrated RT growth factors in the close vicinity of the (C+O)/He
   and He/H composition interfaces, respectively (Figure~\ref{fig:rtgrowth});
$t_\mathrm{RS}$ is the formation time of the reverse shock below
   the He/H composition interface, which is measured from the moment when the
   SN shock crosses this composition interface; and
$R_\mathrm{RS}$ is the radius of the reverse shock at the formation epoch.
}
\end{deluxetable*}
%
\subsection{Extent of $^{56}$Ni mixing and properties of progenitors}
\label{sec:results-1Dpran}
%
\citet{UWJ_19} interpreted the complex phenomenon of RT mixing (see
   Section~\ref{sec:results-3Dmix}) in a simple phenomenological approach and
   found that the efficiency of outward radioactive $^{56}$Ni mixing in the
   framework of the 3D neutrino-driven simulations depends mainly on the
   following two hydrodynamic properties of the single-star BSG progenitor
   models: a high growth factor of RT instabilities at the (C+O)/He composition
   interface, and a weak interaction of fast RT plumes with the reverse shock
   occurring below the He/H composition interface.
It is interesting now to carry out a similar analysis of the binary-merger
   progenitors which have a RSG-like structure inside the He/H composition
   interface of their primaries.

This phenomenological approach attempts to capture multidimensional
   effects of RT mixing at the (C+O)/He composition interface by
   describing the evolution of the $^{56}$Ni velocity by means of a
   velocity growth factor which is proportional to the logarithm of
   the time-integrated RT growth factor (see Equation~\ref{eq:rtsol}).
To this end, we calculated the maximum (i.e., final) value of this
   growth factor, $(\xi/\xi_{0})^\mathrm{CO}$, in the close vicinity
   of the (C+O)/He composition interface (Figure~\ref{fig:rtgrowth},
   Table~\ref{tab:nimixing}).

The fast RT plumes that grow from the C+O core outward through the
   helium shell can interact with the reverse shock developing below
   the He/H composition interface.
Whether such an interaction happens depends on the ratio of two
   timescales, namely (1) the time it takes to form the reverse shock,
   after the SN shock has crossed the He/H composition interface,
   $t_\mathrm{RS}$, and (2) the time needed by RT plumes originating
   from the (C+O)/He interface to reach the radius, $R_\mathrm{RS}$,
   where the reverse shock forms.
The latter timescale is given by the ratio of $R_\mathrm{RS}$ and the
   maximum velocity of the bulk mass of $^{56}$Ni when the SN shock
   passes the (C+O)/He composition interface,
   $v_\mathrm{Ni}^{\mathrm{CO}}$, multiplied by the growth factor
   $(\xi/\xi_{0})^\mathrm{CO}$, which takes into account the
   subsequent evolution of this velocity.

The greater the ratio of the two timescales, the weaker the
   interaction and the higher the terminal velocity of fast RT plumes
   in the hydrogen envelope.
In Table~\ref{tab:nimixing} we give the corresponding quantities that
   determine the amount of mixing of radioactive $^{56}$Ni for all
   binary-merger models.

\begin{figure*}
\centering
   \includegraphics[width=0.49\hsize, clip, trim=17 153 35 318]{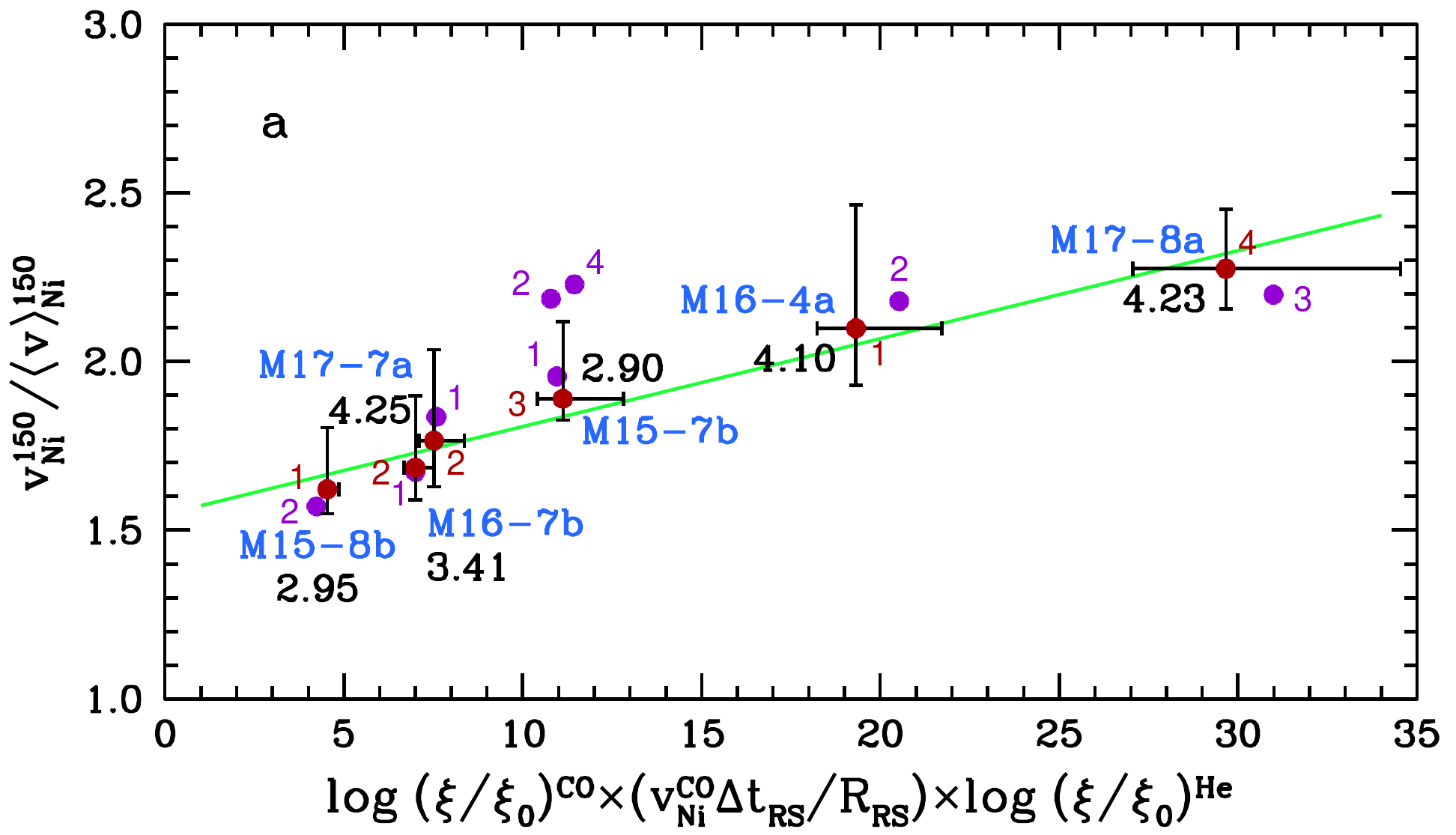}
   \hspace{0.05cm}
   \includegraphics[width=0.49\hsize, clip, trim=17 153 35 318]{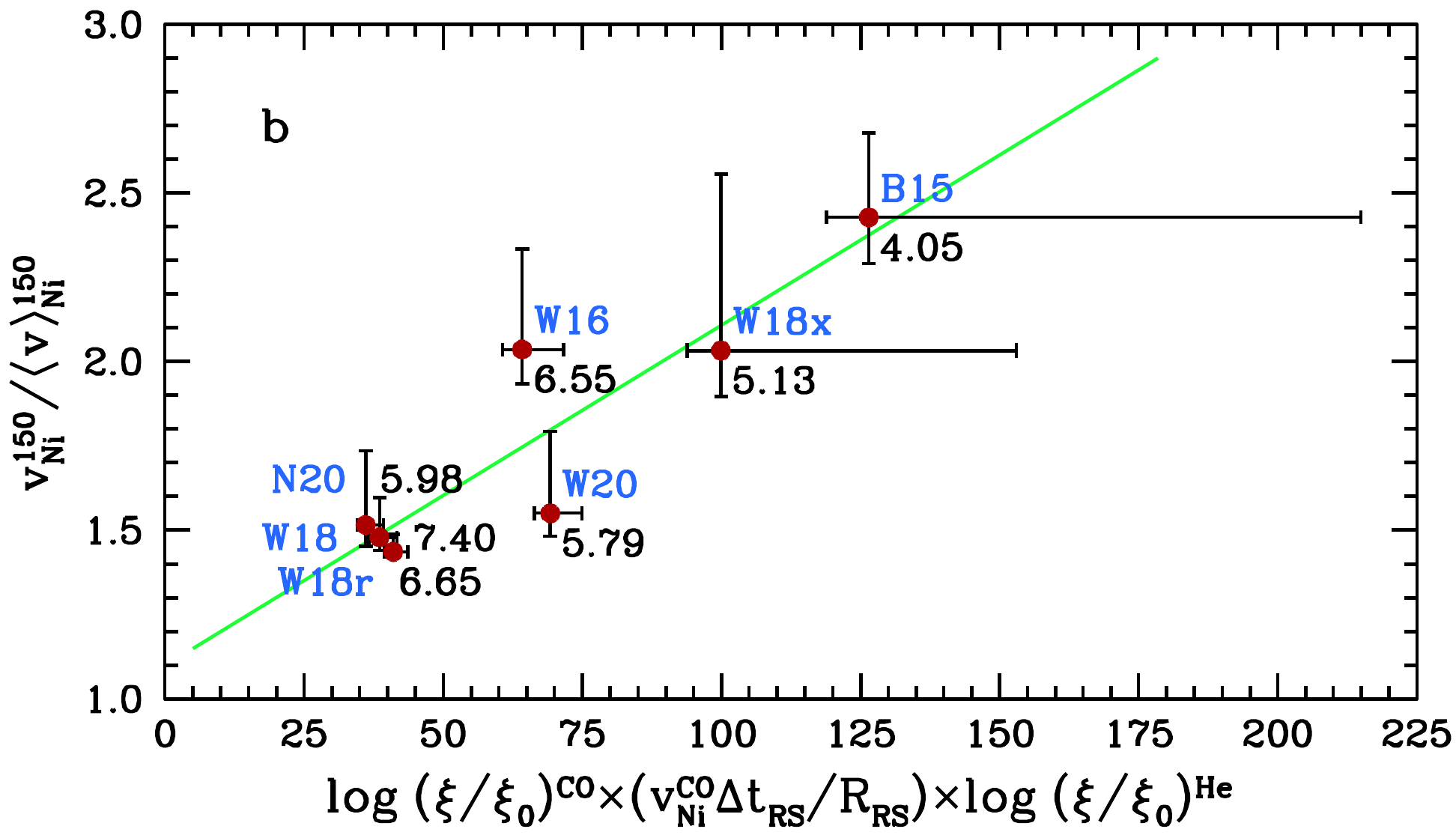}\\
   \vspace{0.25cm}
   \includegraphics[width=0.49\hsize, clip, trim=17 153 35 318]{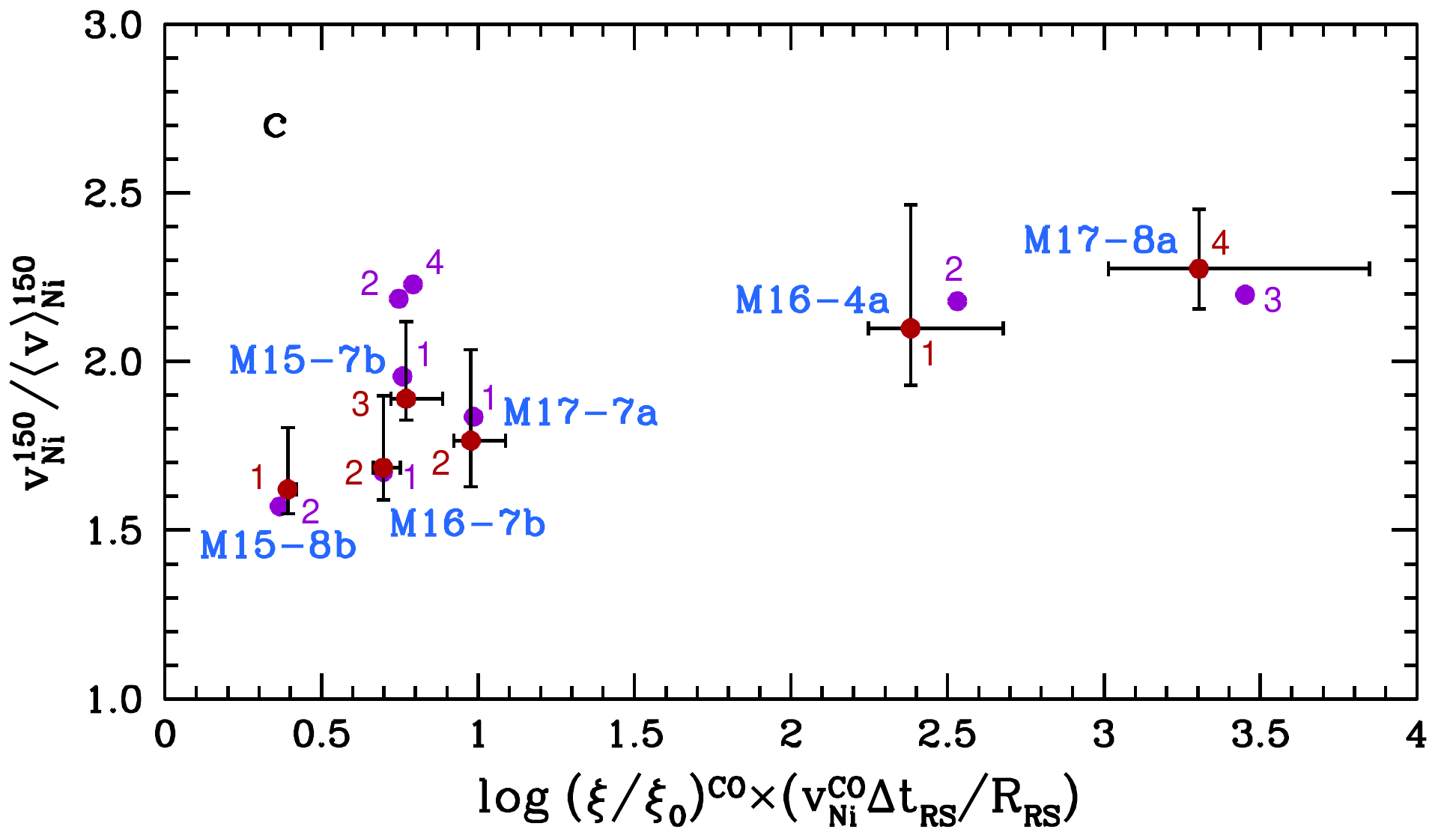}
   \hspace{0.05cm}
   \includegraphics[width=0.49\hsize, clip, trim=17 153 35 318]{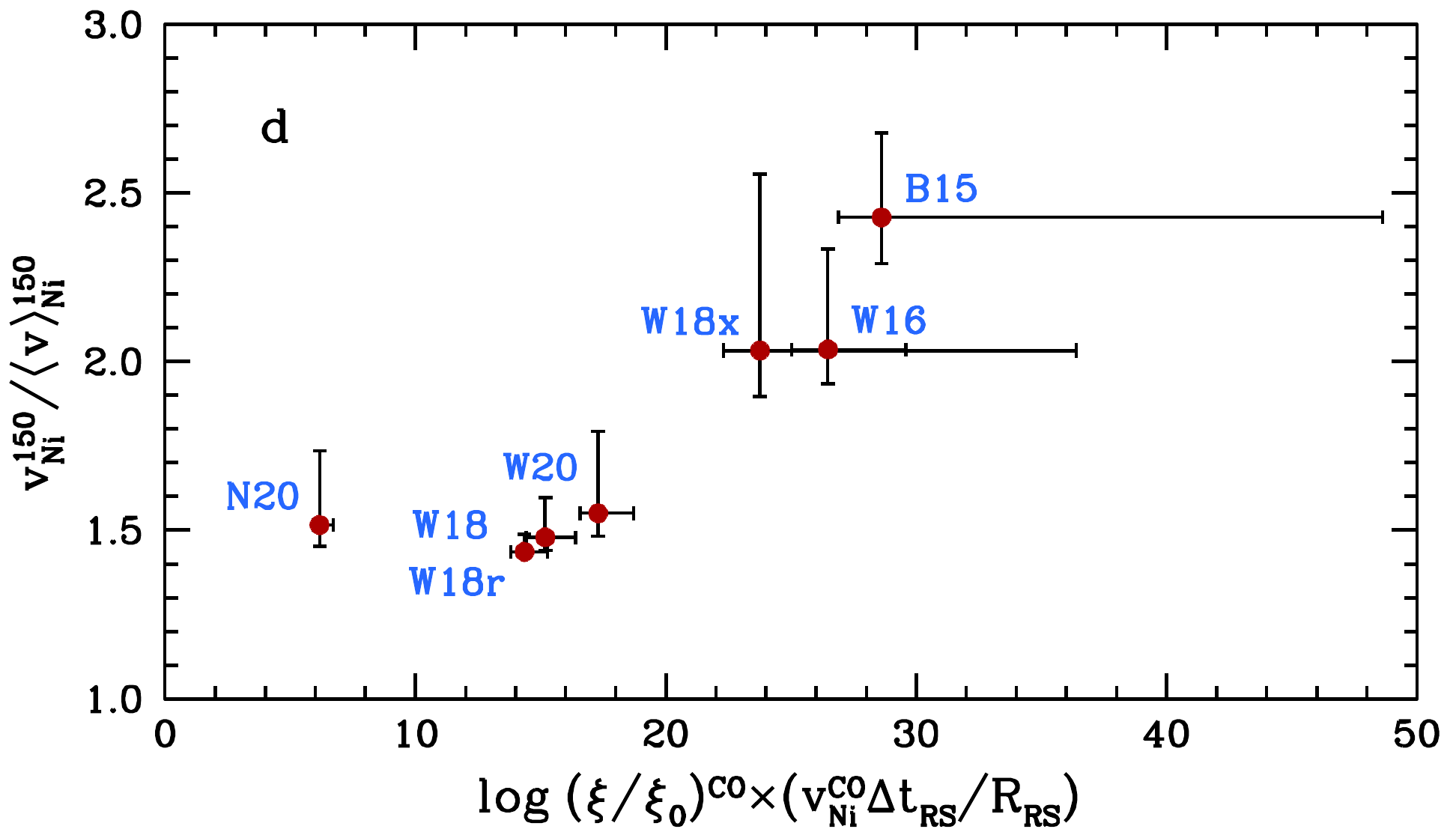}\\
   \vspace{0.25cm}
   \includegraphics[width=0.49\hsize, clip, trim=17 153 35 318]{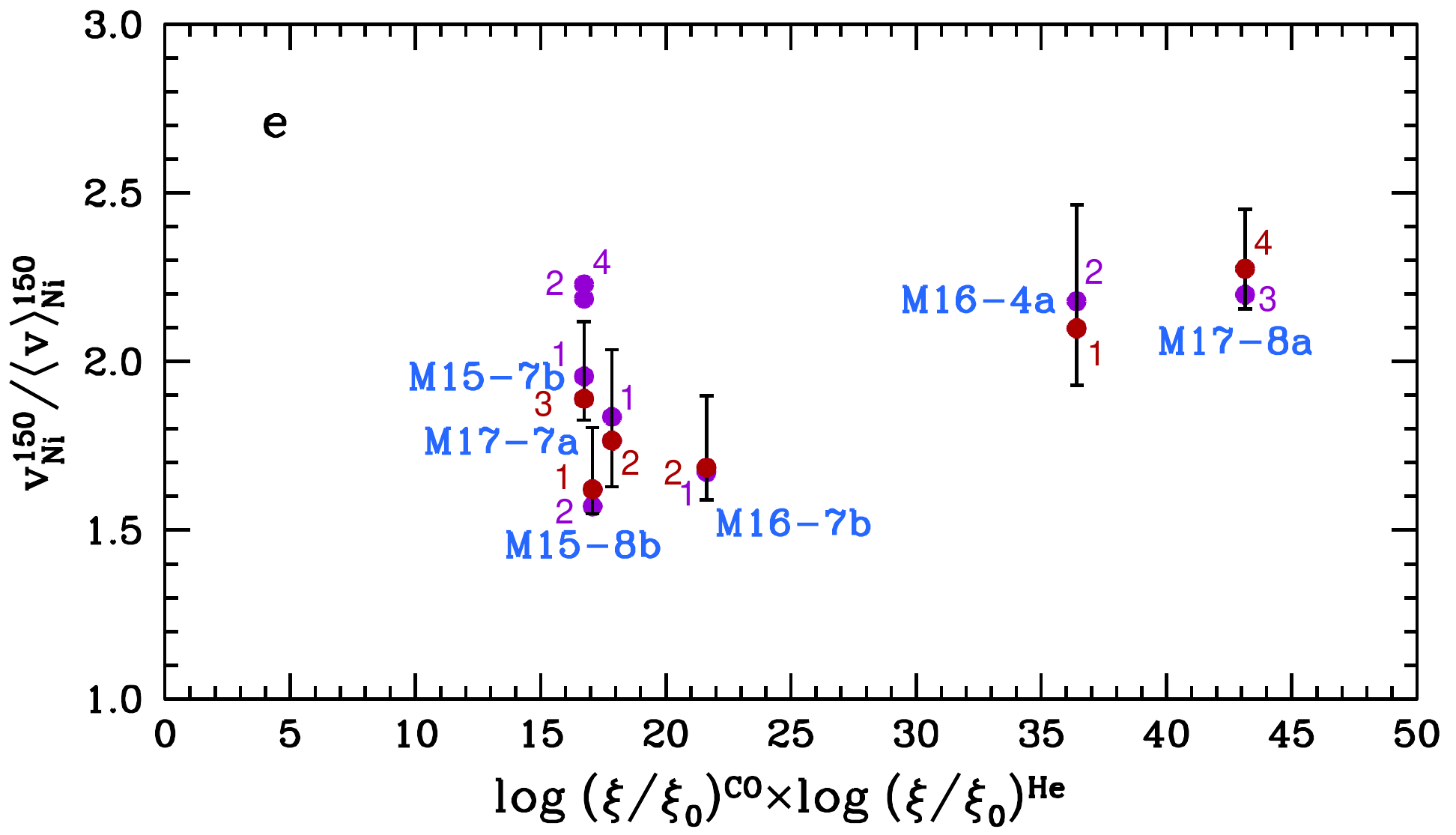}
   \hspace{0.05cm}
   \includegraphics[width=0.49\hsize, clip, trim=17 153 35 318]{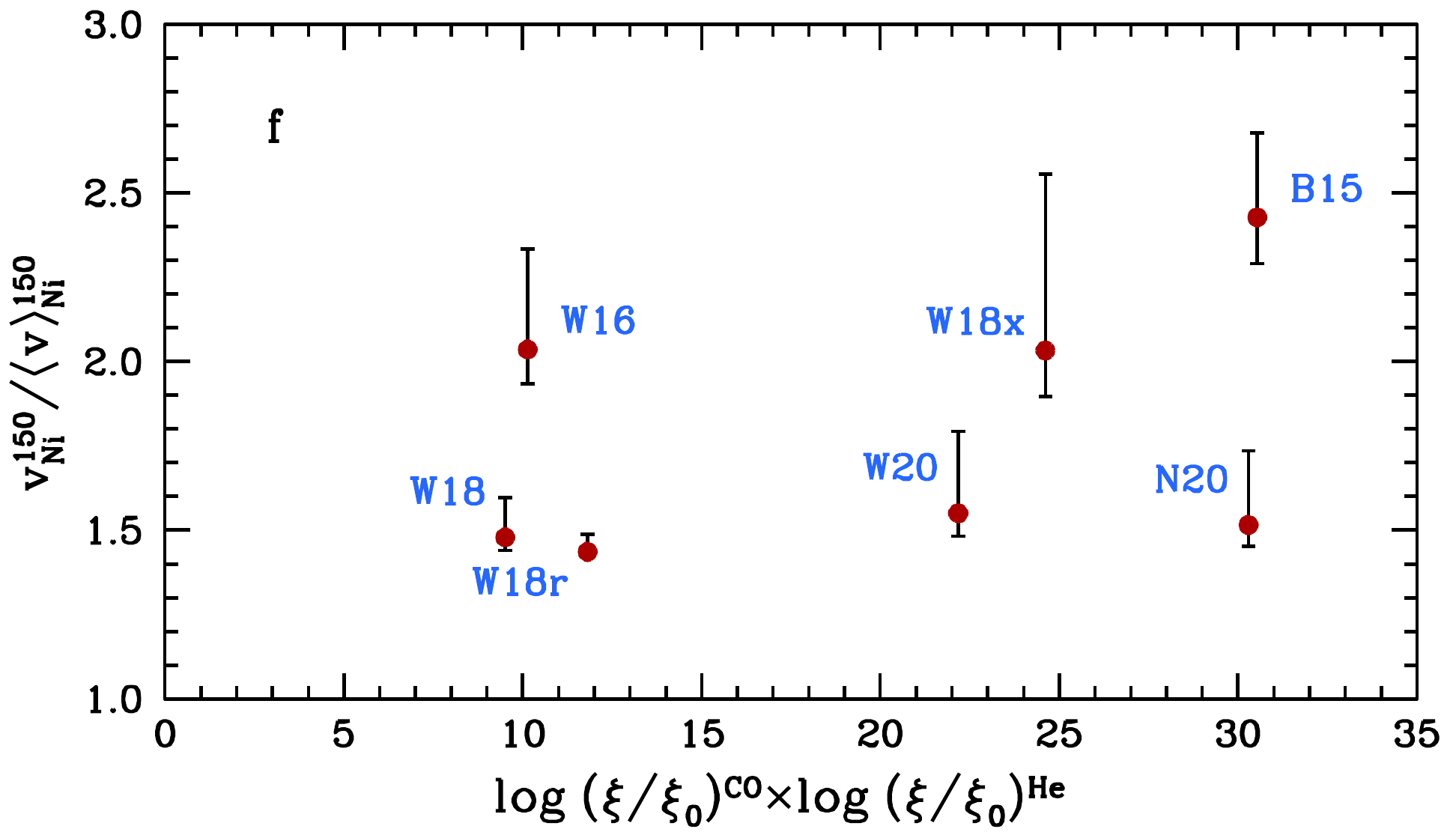}\\
   \caption{%
   Dependence of the normalized extent of $^{56}$Ni mixing in velocity space
      on the hydrodynamic explosion properties linked to the structure of the
      binary-merger (left panels) and single-star (right panels) progenitors,
      listed in Table~\ref{tab:nimixing} and Table~3 in \citet{UWJ_19},
      respectively.
   Names of the progenitor models are printed in blue.
   Numbers in black give the helium-core masses of the corresponding
      progenitor models.
   Red circles and figures show the reference models which explode with
      similar energies, and dark-violet circles and numbers for
      the binary-merger models represent the remaining models with different
      explosion energies.
   The uncertainties in the $^{56}$Ni mixing are estimated from estimates
      of the velocities of the $^{56}$Ni ejecta containing $93\%$ and $99\%$
      of the total $^{56}$Ni mass.
   Green lines are least square fits which are based on the reference
      models.
   They prove a correlation between the extent of $^{56}$Ni
      mixing in velocity space and the hydrodynamic properties of the
      explosion models (for definitions of the quantities on the horizontal
      and vertical axis see the comments of Table~\ref{tab:nimixing}).
   }
   \label{fig:vnipsn}
\end{figure*}
Following \citet{UWJ_19}, we measure the extent of $^{56}$Ni mixing by
   the dimensionless ratio of the maximum velocity of the bulk mass of
   $^{56}$Ni ejecta containing $96\%$ of the total $^{56}$Ni mass at day 150,
   $v_\mathrm{Ni}^{\mathrm{150}}$, to the weighted mean velocity of the bulk
   mass of $^{56}$Ni at the same epoch,
   $\langle v \rangle_\mathrm{Ni}^{\mathrm{150}}$, whereby we eliminate
   the influence of slightly different explosion energies of the models.
Because the outward mixing of $^{56}$Ni depends on the growth factor of
   RT plumes at the (C+O)/He composition interface as well as on the
   interaction of those plumes with the reverse shock from the He/H composition
   interface, we use the product of the corresponding growth factors
   (as explained above) on the abscissa in Figure~\ref{fig:vnipsn}(c).
For comparison, we provide the corresponding plot for the single-star models
   in Figure~\ref{fig:vnipsn}(d).
From an inspection of these figures, it is evident that both the reference
   binary-merger models and the single-star models exhibit in a quite similar
   way a correlation between the extent of outward radioactive $^{56}$Ni mixing
   and the hydrodynamic explosion properties linked to the structure of
   the progenitor models. 
But there are two relevant exceptions.
First, the binary-merger model M15-7b-3 lies well above the correlation
   relation suggested by Figure~\ref{fig:vnipsn}(c).
Second, the single-star model N20 lies far off the correlation
   suggested by Figure~\ref{fig:vnipsn}(d).

At this point let us recall that the progenitor model M15-7b has the largest
   time-integrated RT growth factor near the He/H composition interface
   among all binary-merger progenitor models under consideration
   (Figure~\ref{fig:rtgrowth} and Table~\ref{tab:nimixing}), and
   the single-star progenitor model N20 is a RSG progenitor
   rather than a BSG one with respect to the structure of the helium core
   as we have pointed out before.
Hence, the locations of these models off the expected correlation lines
   in Figures~\ref{fig:vnipsn}(c) and (d) might suggest a noticeable influence
   of the high growth factor of RT instabilities at the He/H composition
   interface on the outward mixing of radioactive $^{56}$Ni.
In other words, we have to perform a more careful analysis to account for
   this influence and to further improve the correlations between nickel
   mixing and explosion properties of binary-merger and single-star
   progenitors, respectively. 

In our phenomenological approach, which captures multidimensional effects of
   RT mixing at a composition interface and describes the evolution of
   the nickel velocity (Equation~\ref{eq:rtsol}), the terminal maximum velocity
   of the bulk mass of radioactive $^{56}$Ni is proportional to the natural
   logarithm of the time-integrated RT growth factor at this interface and
   to the initial value of the radial velocity of $^{56}$Ni.
Considering the mixing processes around the He/H composition interface,
   it is therefore natural to use the velocity of the bulk mass of
   $^{56}$Ni -- which results from the acceleration of nickel-rich RT
   plumes at the (C+O)/He interface and their subsequent interaction with
   the reverse shock from the He/H interface (both are accounted for by
   the quantity on the horizontal axis of Figures~\ref{fig:vnipsn}(c) and
   (d)) -- as the initial value of the radial velocity of $^{56}$Ni at
   the He/H interface in Equation~\ref{eq:rtsol}.
Thus, multiplying the quantity at the abscissa of
   Figures~\ref{fig:vnipsn}(c) and (d) by the logarithm of the time-integrated RT growth factor at 
   the He/H composition interface, we reasonably estimate the effect of
   a high growth factor at this interface on the extent of outward mixing
   of radioactive $^{56}$Ni.

The results of this procedure are shown in Figures~\ref{fig:vnipsn}(a)
   and (b) for the binary-merger and single-star explosion models,
   respectively.
We find a much better correlation between the normalized extent of
   $^{56}$Ni mixing and the product of the three hydrodynamic properties
   that favor the mixing of radioactive $^{56}$Ni during the explosions of
   the different progenitors.
Our modified measure for the efficiency of $^{56}$Ni mixing moves the
   locations of both the binary-merger model M15-7b-3 and the single-star
   model N20 to the ideal, linear correlation lines plotted in green.
The existence of these correlations confirms the decisive role of the three
   considered factors for the outward mixing of radioactive $^{56}$Ni
   during 3D neutrino-driven SN simulations of binary-merger and single-star
   progenitors.

Note that the growth factors at the (C+O)/He and He/H composition interfaces
   alone cannot explain the final $^{56}$Ni-mixing velocities because they
   result in a weak correlation with significant scatter for both
   the binary-merger and single-star models as it is shown in
   Figures~\ref{fig:vnipsn}(e) and (f), respectively.
At the same time, this kind of plot can give a crude and a priori
   (i.e., without any information from 3D simulations) estimate
   of the maximum velocity of the bulk mass of radioactive $^{56}$Ni
   in explosions of different progenitors on the basis of -- and this is
   the important point here -- only the growth factors (derived from
   linear theory) of RT instabilities at the composition interfaces
   after shock passage.

From an inspection of Figures~\ref{fig:vnipsn}(a) and (b), it is evident that
   both the binary-merger and single-star progenitors obey a correlation
   between the extent of outward radioactive $^{56}$Ni mixing and the product
   of the hydrodynamical quantities on the abscissa equally well.
The dimensionless extent of $^{56}$Ni mixing varies within 1.6--2.3 and
   1.4--2.4, respectively, and the values on the abscissa fall into
   two non-overlapping intervals of 4.5--30 and 36--127.
The slopes of the correlation lines are 0.026 and 0.010 for the reference
   binary-merger and single-star models, respectively.
The significant difference between both the abscissa values and the slopes of
   the correlation functions is presumably related to the fact that we only
   performed a scaling analysis of the outward mixing of radioactive $^{56}$Ni
   during 3D neutrino-driven simulations.
The difference might be eliminated by a more detailed theory that is able
   to predict the dependence of the mixing efficiency on the relevant
   hydrodynamical properties in a quantitative manner. 

\citet{UWJ_19} showed that their set of BSG explosion models based on
   single-star progenitors exhibits $^{56}$Ni mixing with an efficiency
   that scales roughly inversely with the helium-core mass, which
   varied from $4.05\,M_{\sun}$ to $7.40\,M_{\sun}$, i.e., having
   a spread of $3.35\,M_{\sun}$
   (see the numbers next to the model names in Figure~\ref{fig:vnipsn}(b)).
In contrast, the binary-merger models in Figure~\ref{fig:vnipsn}(a) fail to
   reveal any clear correlation between the amount of radial mixing of
   radioactive $^{56}$Ni and the helium-core mass of the progenitor model,
   which for the considered binary progenitors varies only within
   $2.90-4.25\,M_{\sun}$, i.e., having a spread of only $1.35\,M_{\sun}$.
The latter spread might be too narrow for detecting such a correlation.
However, it is possible to assert that a decrease in the fraction of the He
   core of the primary that was dredged up, $f_\mathrm{c}$, from $17.5\%$
   to $3.3\%$ (in models M15-8b-1, M16-7b-2, M17-8a-4, and M16-4a-1) 
   results in a growth of both the He-shell mass,
   $M_\mathrm{He}^{\,\mathrm{shell}}$, and the width of this shell,
   $\Delta R_\mathrm{He}^{\,\mathrm{shell}}$, (Table~\ref{tab:presnm}),
   and it also correlates with the extent of $^{56}$Ni mixing in velocity
   space (Figure~\ref{fig:vnipsn}(a)).

\subsection{Light-curve modeling}
\label{sec:results-lcmod}
%
\begin{figure*}[t]
\centering
   \includegraphics[width=0.33\hsize, clip, trim=20 163 323 213]{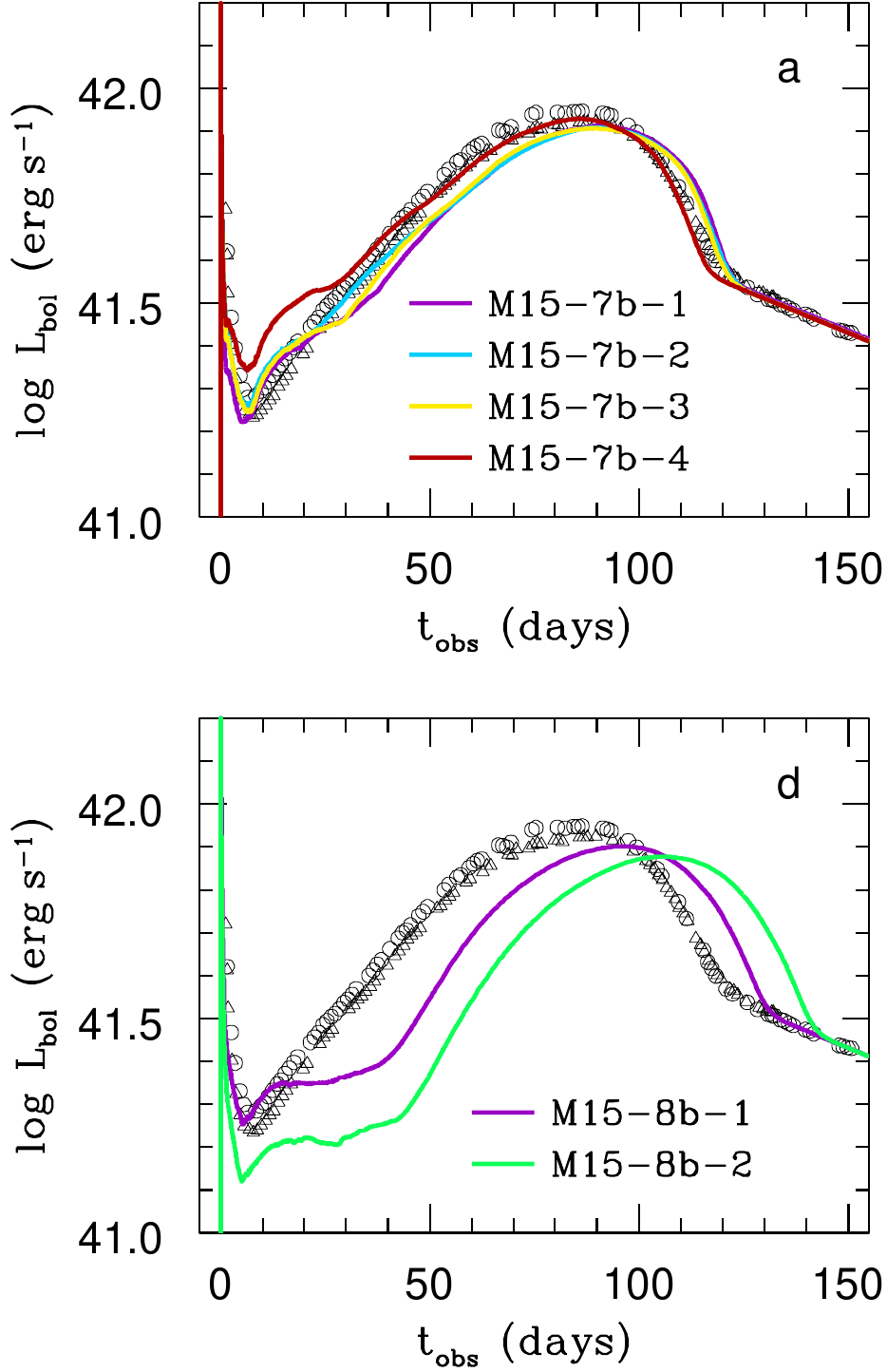}
   \includegraphics[width=0.33\hsize, clip, trim=20 163 323 213]{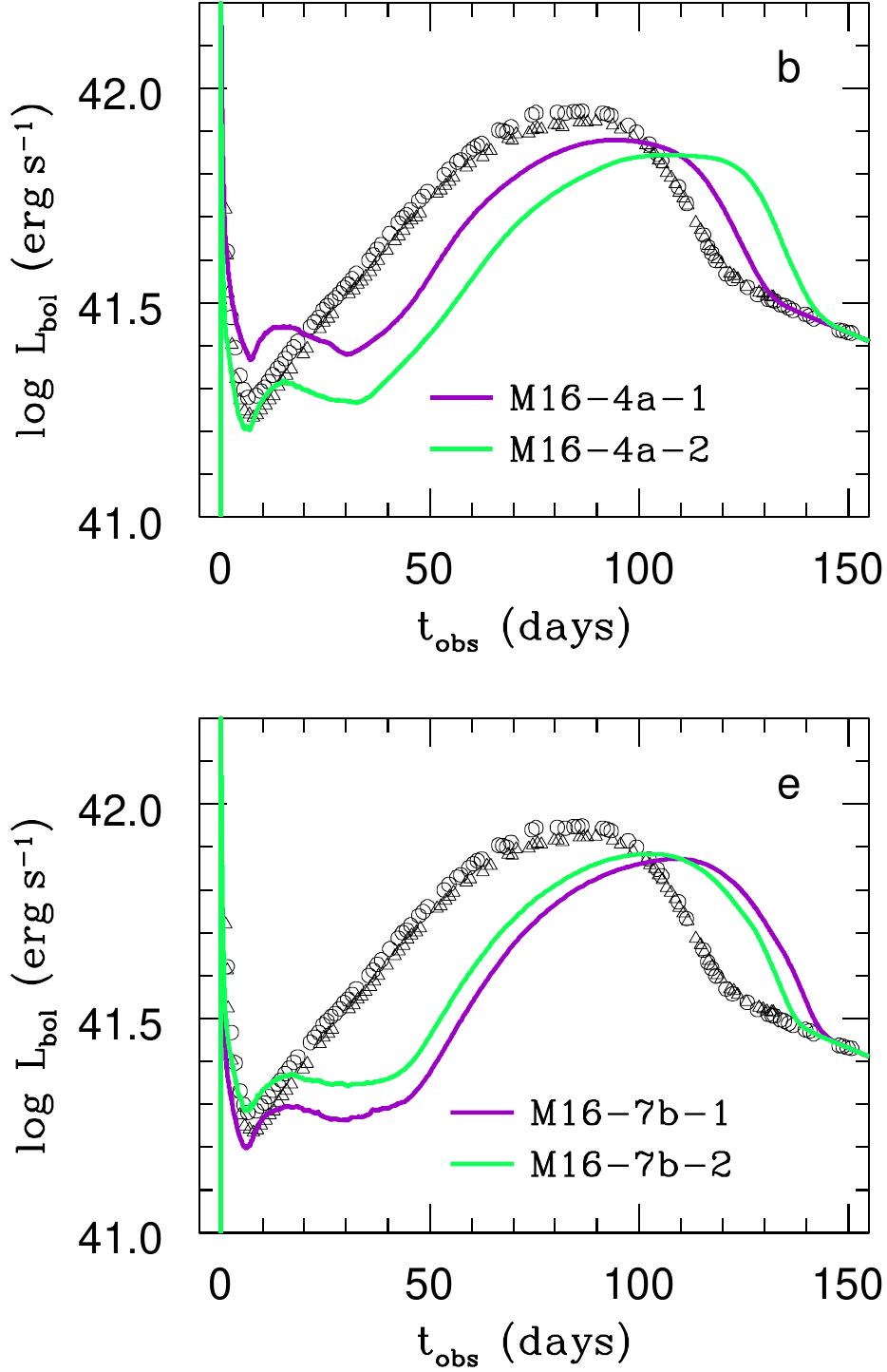}
   \includegraphics[width=0.33\hsize, clip, trim=20 163 323 213]{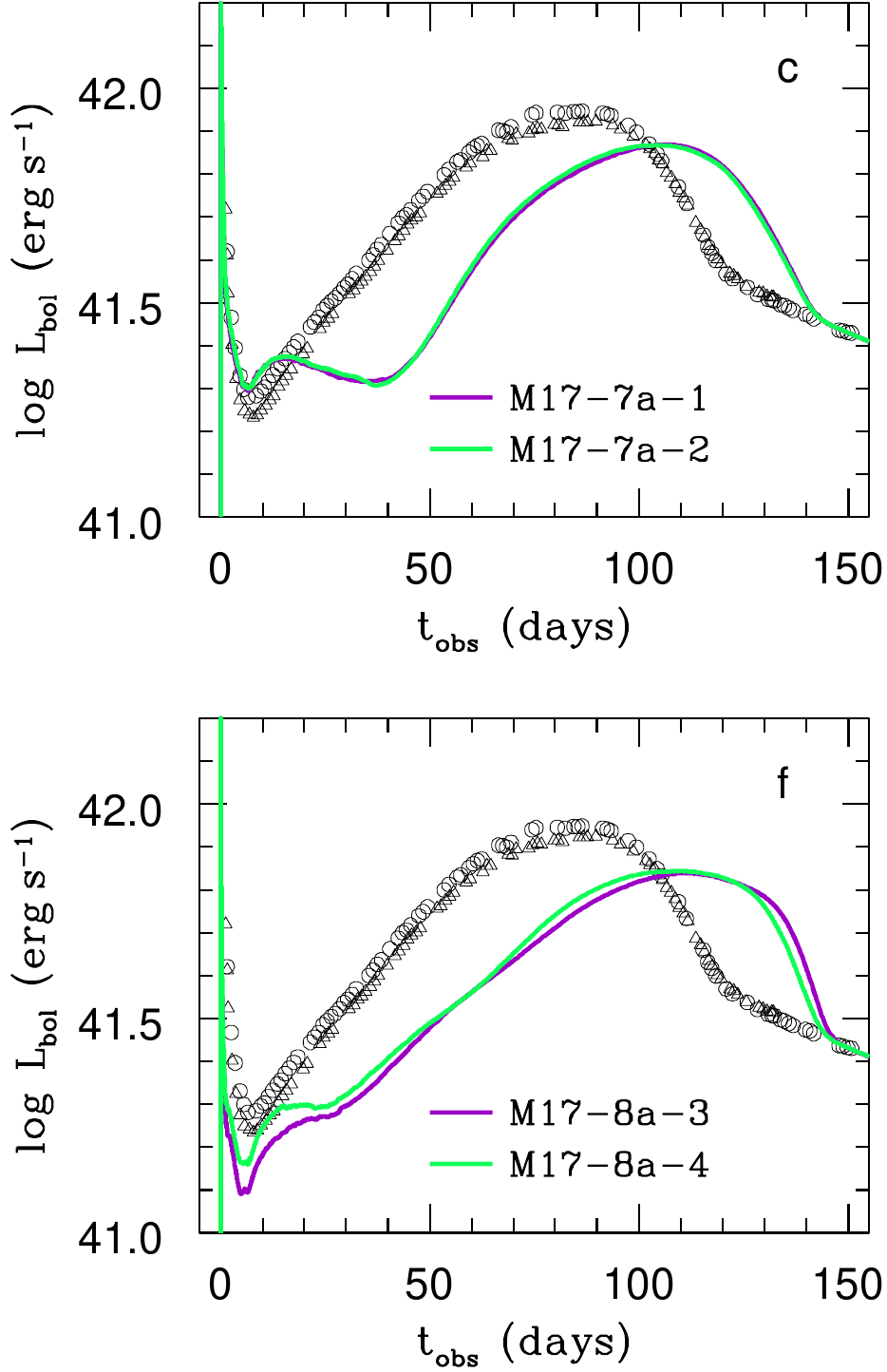}\\
   \caption{%
   Bolometric light curves of hydrodynamic explosion models
      (Table~\ref{tab:hydmod}) based on the pre-SN models M15-7b (a),
      M16-4a (b), M17-7a (c), M15-8b (d), M16-7b (e), and M17-8a (f)
      (Table~\ref{tab:presnm}).
   The computed light curves are compared with the observed bolometric
      luminosity of SN~1987A obtained by \citet{CMM_87, CWF_88} (open circles)
      and \citet{HSGM_88} (open triangles).
   }
   \label{fig:lcurves}
\end{figure*}
The results of our light-curve modeling based on the averaged 3D explosion
   simulations are illustrated by the bolometric light curves
   (Figure~\ref{fig:lcurves}).
They depend on the relevant basic properties
   of the progenitors (Table~\ref{tab:presnm}) and 3D explosion models 
   (Table~\ref{tab:hydmod}), and on the distributions of chemical composition
   in velocity space (Figure~\ref{fig:mfvel}).
In the context of the light-curve modeling, the structure of the outermost
   layers is responsible for the amplitude and width of the initial luminosity
   peak which forms during shock breakout from the stellar surface and the
   subsequent adiabatic cooling phase.
Analyzing both photometric and spectroscopic observations of SN~1987A,
   especially the sensitivity of the width of the initial peak to the radius
   of the pre-SN, \citet{Utr_05} found that the radius is $35\pm5\,R_{\sun}$.
The pre-SN radii of all binary-merger models considered in our study
   (Table~\ref{tab:presnm}) fall in this range and guarantee a good description
   of the initial luminosity peak during the first $\sim$7\,days
   (Figure~\ref{fig:lcurves}).

\citet{UWJM_15} studied the influence of the explosion energy
   and the ejected $^{56}$Ni mass on the calculated light curves for the
   3D explosion models based on the single-star progenitors.
All models in Figure~\ref{fig:lcurves} show the influence of the explosion
   energy on the calculated light curve for the explosions of the different
   binary-merger progenitors, except for two model pairs (M15-7b-2, M15-7b-3)
   and (M17-7a-1, M17-7a-2) with almost equal explosion energies.
It is well known that from about day 7 to day 30 the bolometric light curve
   is mainly determined by the properties of a cooling and recombination wave
   (CRW): the higher the ratio of explosion energy and ejecta mass,
   the higher the luminosity in the CRW phase because the ejecta expand and
   cool more quickly \citep[cf.][]{UWJM_15}.

After the CRW stage, when radiative diffusion takes place, the radioactive
   decay of $^{56}$Ni and $^{56}$Co nuclides becomes dominant in powering
   the luminosity.
Now the bolometric light curve depends on the amount of
   radioactive material and its distribution over the ejecta for a given
   progenitor.
In other words, there is an indirect influence of the explosive nucleosynthesis
   and mixing processes during the explosion on the light curve.
The total amount of radioactive $^{56}$Ni strongly affects the luminosity
   from about day 30 to the radioactive tail, and gives rise to the second
   dome-like maximum of the light curve of SN~1987A.
While the total amount of $^{56}$Ni mainly determines the energy radiated
   during this period and the highest luminosity of the broad dome-like
   maximum, the amount of outward $^{56}$Ni mixing affects the smoothness of
   the rising part of the bolometric light curve just after the CRW stage.
The more extreme the large-scale mixing of $^{56}$Ni in velocity space is,
   the earlier the luminosity starts to grow to the dome-shaped maximum.
The hydrodynamic models based on the pre-SN model M15-7b with maximum
   velocities of the bulk of ejected $^{56}$Ni equal to or exceeding about
   3000\,km\,s$^{-1}$ (Table~\ref{tab:hydmod}) most clearly demonstrate
   this dependence in comparison to the other hydrodynamic models
   (Figure~\ref{fig:lcurves}).
Along with the outward $^{56}$Ni mixing, hydrogen-rich matter is mixed
   down to the center.
It increases the optical depth of the inner ejecta,
   because Thomson scattering off free electrons dominates the opacity of
   highly ionized matter.
This, in turn, gives rise to a wide and luminous dome-shaped light-curve
   maximum (Figure~\ref{fig:lcurves}).

For SN~1987A, the total amount of $^{56}$Ni in the ejecta is well fixed by
   the bolometric luminosity in the radioactive tail.
Equating the observed bolometric luminosity in the radioactive tail to
   the gamma-ray luminosity gives a ``directly observed'' mass of radioactive
   $^{56}$Ni of $0.0722\,M_{\sun}$ with a distance modulus for the LMC of
   $m-M=18.5$\,mag and a reddening of $E(B-V)=0.17$\,mag \citep{UWJM_15}.
In all of our hydrodynamic models the initial $^{56}$Ni mass at the onset of
   light-curve modeling, $M^{i}_{\mathrm{Ni}}$, is chosen such that the
   $^{56}$Ni mass in the ejecta evaluated at day 150, $M^{f}_{\mathrm{Ni}}$,
   fits the observed luminosity in the radioactive tail.
Taking fallback of $^{56}$Ni into account requires $M^{i}_{\mathrm{Ni}}$
   to be greater than the value of $M^{f}_{\mathrm{Ni}}$, which, in turn,
   exceeds the directly observed mass (Table~\ref{tab:hydmod}) because of
   expansion-work effects \citep{Utr_07}.
The initial $^{56}$Ni masses of all 3D explosion models fall in
   between the minimum, $M_\mathrm{Ni}^{\,\mathrm{min}}$, and maximum,
   $M_\mathrm{Ni}^{\,\mathrm{max}}$, values listed in Table~\ref{tab:hydmod},
   and, consequently, these models are able to produce the amount of $^{56}$Ni
   that is needed to explain the SN~1987A observations.

\subsection{Comparison with observations}
\label{sec:results-compobs}
%
For a comparison of our results of 3D hydrodynamic simulations of
   neutrino-driven explosions with observations of SN~1987A, we consider
   the set of models M15-7b-3, M15-8b-1, M16-4a-1, M16-7b-2, M17-7a-2, and
   M17-8a-4 with comparable and SN~1987A-like explosion energies
   (Table~\ref{tab:hydmod}).
These hydrodynamic models are based on the corresponding compact blue
    binary-merger pre-SN 
    models with radii of 31.8\,$R_{\sun}$ to 37.3\,$R_{\sun}$
   (Table~\ref{tab:presnm}), which agree well with the photometric radius of
   the BSG Sanduleak\,$-69^{\circ}202$ star of 28.7\,$R_{\sun}$ to
   57.5\,$R_{\sun}$ \citep{ABKW_89}.
Moreover, the pre-SN radii fall into the range of 30\,$R_{\sun}$ to
   40\,$R_{\sun}$, which is allowed by the hydrodynamic modeling and
   the time-dependent atmosphere model that explain both the bolometric
   light curve and the H${\alpha}$ profile of SN~1987A \citep{Utr_05}.
This and the fact that the outer layers of the pre-SN models have a suitable
   structure permit us to obtain a good reproduction of the initial luminosity
   peak within the first $\sim$7\,days by all binary-merger models listed above
   (Figure~\ref{fig:lcurves}).
During the transition from the adiabatic cooling phase to the CRW phase
   a prominent minimum in the luminosity forms, which is observed at around
   day 7.
All models of the subset compared here, except models M16-4a-1 and M17-8a-4,
   reproduce well the observed minimum (Figure~\ref{fig:lcurves}).
However, only one hydrodynamic model, M15-7b-3, also reproduces
   fairly well both the smooth rising part of the observed light curve and
   the major broad maximum with a duration of $\sim$100\,days.
The smoothness of the bolometric light curve of model M15-7b-3 results from
   the outward mixing of the bulk of radioactive $^{56}$Ni in velocity space
   up to 2980\,km\,s$^{-1}$ (Table~\ref{tab:hydmod}), which is consistent with
   the observed value of about 3000\,km\,s$^{-1}$ \citep{CHELH_94}, while
   all other models fall short of the required mixing.

The wide dome-like light curve maximum, in turn, arises from mixing
   a significant amount of hydrogen-rich matter down to the center.
The analysis of the line profiles of hydrogen emission in the nebular phase
   \citep{Chu_91, KF_98, MJS_12}, and the 3D view of H$\alpha$ emission
   \citep{LFS_16} and molecular hydrogen in SN~1987A \citep{LSF_19} show
   that hydrogen is mixed deeply down to velocities $\le$700\,km\,s$^{-1}$.
It is noteworthy that in all hydrodynamic models hydrogen
   in the innermost layers of the ejecta expands with velocities lower than
   40\,km\,s$^{-1}$ (Table~\ref{tab:hydmod}), in good agreement with
   the observations.
The mass of hydrogen confined to the inner layers ejected with velocities
   less than 2000\,km\,s$^{-1}$ may be considered as a measure of the optical
   depth of these layers, because hydrogen is the dominant contributor to opacity
   and the velocity at the photosphere observed in SN~1987A around the top of
   the dome-like maximum is about 2000\,km\,s$^{-1}$ \citep{PHHN_88}.
It is the high optical depth of the inner ejecta that produces a wide and
   dome-shaped light-curve maximum.
In model M15-7b-3 the hydrogen mass expanding with velocities less than
   2000\,km\,s$^{-1}$ is 3.1\,$M_{\sun}$ (Table~\ref{tab:hydmod}), which
   is in good agreement with the observational constraint of 2.2\,$M_{\sun}$
   \citep{KF_98}.

The bolometric luminosity of SN~1987A in the radioactive tail can be formally
   fitted by the directly observed mass of radioactive $^{56}$Ni
   (see Section~\ref{sec:results-lcmod}).
In our case the required mass of $^{56}$Ni is 0.0722\,$M_{\sun}$, which is
   consistent within the errors with other empirical estimates of
   ($0.069\pm0.003$)\,$M_{\sun}$ \citep{BPS_91, McC_93} and
   ($0.071\pm0.003$)\,$M_{\sun}$ \citep{STM_14}.
Unfortunately, the above observational numbers differ from
   the exact amount of radioactive $^{56}$Ni synthesized and ejected
   during an SN explosion.
Even if there is no fallback of $^{56}$Ni, expansion-work effects
   definitely increase the $^{56}$Ni amount required in the hydrodynamic
   modeling to match the observed bolometric luminosity in the radioactive
   tail (compare the directly observed mass of 0.0722\,$M_{\sun}$ with
   the values of $M^{f}_{\mathrm{Ni}}$ in Table~\ref{tab:hydmod}).
Moreover, both fallback of $^{56}$Ni and expansion-work effects make
   an assessment of the ejected $^{56}$Ni mass model-dependent.

Analyzing the evolution of the intensity and the profile of the oxygen doublet
   [\ion{O}{1}] $\lambda\lambda 6300, 6364\,\AA$ in the nebular phase showed
   that the mass of oxygen in SN~1987A is in the range from 0.7\,$M_{\sun}$
   to 2.0\,$M_{\sun}$ \citep{LM_92, Chu_94, CCKC_97, KF_98, JSS_15}.
Our favorite model M15-7b-3 yields 0.83\,$M_{\sun}$ (Table~\ref{tab:hydmod}),
   which agrees with these observational constraints.

\citet{KLF_10} and \citet{LFK_13} studied the morphology of the ejecta using
   images and spectra in the emission lines of [\ion{Si}{1}]+[\ion{Fe}{2}]
   and \ion{He}{1}, and in the emission lines of H$\alpha$ and
   [\ion{Si}{1}]+[\ion{Fe}{2}], respectively, and reconstructed the 3D shape
   of the inner regions of the ejecta.
\citet{KLF_10} approximated the actual 3D shape of the ejecta by an elongated
   triaxial ellipsoid.
The 3D morphology of the observed emission lines reflects thermal and
   nonthermal emissivity produced, as a final result, by the energy
   deposition of gamma rays with energies of about 1\,MeV from the decay
   chain $^{56}$Ni $\to ^{56}$Co $\to ^{56}$Fe.
These gamma rays undergo Compton scattering by free and bound electrons
   in the ejecta, resulting in high-energy electrons which, in turn,
   deposit their energy by heating free electrons or by ionizing and exciting
   atoms and ions.
The energy deposition and the ionization and excitation of individual elements
   are sensitive to the chemical composition.
This fact and both gamma-ray and photon transport are responsible for hiding
   the details of the density distribution of the radioactive matter, but
   leave the possibility to trace at least its global properties.
In other words, the observed emissivity distribution can provide valuable
   information about the spatial distribution of radioactive $^{56}$Ni.
The 3D morphology of the $^{56}$Ni-rich ejecta in the model sequence M17-8a-4,
   M16-4a-1, M15-7b-3, M17-7a-2, M16-7b-2, and M15-8b-1 long after shock
   breakout gradually changes from a practically one-sided, dipolar
   configuration to an almost basically spherical one with a dominant monopole
   (Figure~\ref{fig:3D_models}, second and fourth columns).
Among these models only our favorite model M15-7b-3 has
   a distribution of $^{56}$Ni-rich matter in velocity space with
   a strong dipole component, which resembles the elongated ellipsoid
   inferred from observations better than any other model.

\section{Constraining the progenitor mass}
\label{sec:prmass}
%
\citet{ALM_19} showed that our favorite model M15-7b-3 having an ejecta mass
   of 19.46\,$M_{\sun}$ fits the early X-ray and gamma-ray emission of SN~1987A
   best compared to other binary-merger explosion models.
\citet{JWJ_20}, in turn, found that a good agreement between 3D models and
   the gamma-ray decay lines and the UVOIR bolometric light curve of SN~1987A
   could be achieved with a lower ejecta mass of about 14\,$M_{\sun}$.
In the light of these results we computed a set of artificial models
   M15-7b-3-m2, M15-7b-3-m4, and M15-7b-3-m6 with ejecta masses of
   17.46\,$M_{\sun}$, 15.46\,$M_{\sun}$, and 13.46\,$M_{\sun}$, respectively
   (Table~\ref{tab:auxmod}).
To realize the reduction of the ejecta mass, we decreased the density by
   the corresponding factor in each mass zone.
This simple approach preserves both the shape of the profiles of density and
   all species in velocity space (i.e. the extent of hydrogen and nickel
   mixing), and the ratio of explosion energy and ejecta mass.
Thereby the structure of the helium core becomes incompatible with
   that of the evolutionary model, but this is the price to pay for the
   simplicity of the approach.
The total $^{56}$Ni mass for the auxiliary hydrodynamic models is fitted
   to the luminosity in the radioactive tail in the same way as in the case
   of the 3D explosion models.

\begin{deluxetable}{ l c c c c }
\tabletypesize{\small}
\tablewidth{0pt}
\tablecaption{Basic properties of auxiliary hydrodynamic models%
\label{tab:auxmod}}
\tablehead{
\colhead{Model} & \colhead{$M_\mathrm{ej}$}
                & \colhead{$E_\mathrm{exp}$}
       & \colhead{$M_\mathrm{Ni}^{\,\mathrm{f}}$}
       & \colhead{$\Delta M_\mathrm{H}^\mathrm{2000}$} \\
\colhead{} & \colhead{$(M_{\sun})$}
       & \colhead{(B)}
       & \colhead{$(10^{-2}\,M_{\sun})$}
       & \colhead{$(M_{\sun})$}
}
\startdata
M15-7b-3    & 19.46 & 1.432 & 7.28 & 3.10 \\
M15-7b-3-m2 & 17.46 & 1.289 & 7.28 & 2.76 \\
M15-7b-3-m4 & 15.46 & 1.146 & 7.28 & 2.41 \\
M15-7b-3-m6 & 13.46 & 1.001 & 7.28 & 2.07 \\
\enddata
\tablecomments{%
The auxiliary models are based on the pre-SN model M15-7b
   (Table~\ref{tab:presnm}) and the explosion model M15-7b-3
   (Table~\ref{tab:hydmod}).
$M_\mathrm{ej}$ is the ejecta mass;
   $E_\mathrm{exp}$ the explosion energy;
   $M^{f}_{\mathrm{Ni}}$ the $^{56}$Ni mass ejected at day 150;
   $\Delta M_\mathrm{H}^\mathrm{2000}$ the mass of hydrogen confined to the
      inner layers ejected with velocities less than 2000\,km\,s$^{-1}$.
}
\end{deluxetable}
\begin{figure*}[t]
\centering
   \includegraphics[width=\hsize, clip, trim=20 379 45 208]{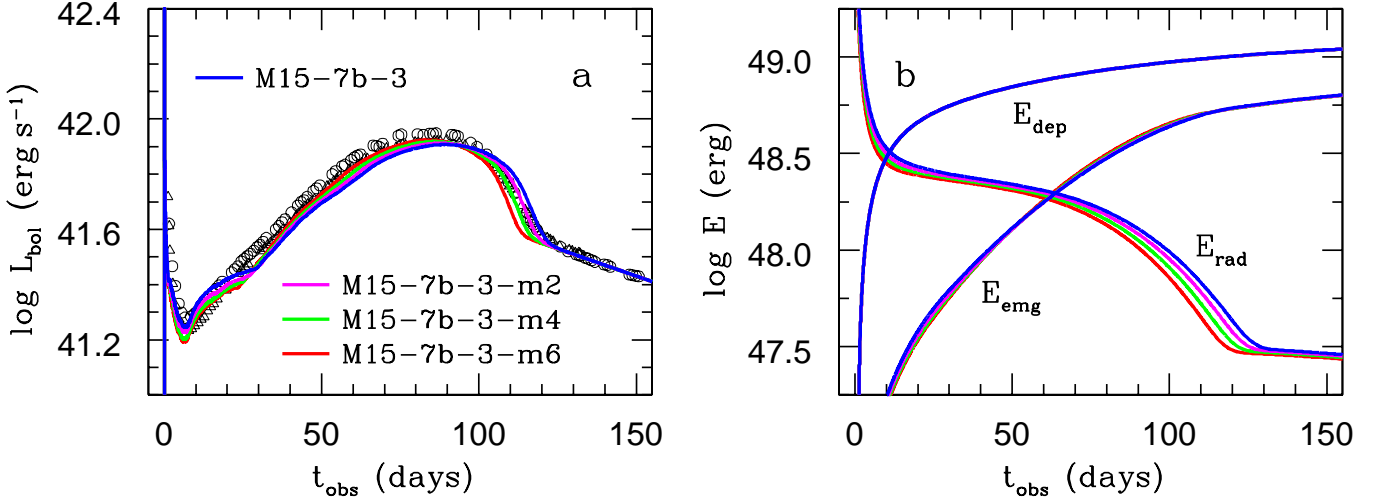}
   \caption{%
   Dependence on the ejecta mass.
   Bolometric light curves (a) and time-integrated energy deposition
      by radioactive decays, $E_\mathrm{dep}$, time-integrated emergent energy,
      $E_\mathrm{emg}$, and total radiation energy of the ejecta,
      $E_\mathrm{rad}$, as a function of time (b) for models M15-7b-3
      ($M_\mathrm{ej}=19.46\,M_{\sun}$), M15-7b-3-m2 (17.46\,$M_{\sun}$),
      M15-7b-3-m4 (15.46\,$M_{\sun}$), and M15-7b-3-m6 (13.46\,$M_{\sun}$).
   The computed light curves are compared with the observed bolometric
      luminosity of SN~1987A obtained by \citet{CMM_87, CWF_88} (open circles)
      and \citet{HSGM_88} (open triangles).
   }
   \label{fig:varmej}
\end{figure*}
The corresponding light curves are depicted in Figure~\ref{fig:varmej}(a).
The original model M15-7b-3 shows a good overall agreement with
   the observed light curve even though the dome of the calculated light
   curve should be slightly shifted as a whole to earlier times.
The latter effect is easily realized by decreasing the mass of the ejected
   envelope \citep{Utr_05}, which is in perfect agreement with the analysis of
   X-ray and gamma-ray observations, and Fe IR line properties of SN~1987A
   by \citet{JWJ_20}.
Decreasing the ejecta mass and, as a consequence, the mass of slow-moving
   hydrogen
   confined to the inner layers (Table~\ref{tab:auxmod}) shifts both the rising
   part of the dome-like maximum of the light curve and the branch declining
   from the maximum to the radioactive tail to earlier times.
The former is due to a shortening of the diffusion time, which is linked to
   the reduction of the total optical depth of the envelope.
The latter effect is due to a decrease of the total radiation energy of
   the ejecta, which is clearly demonstrated in Figure~\ref{fig:varmej}(b).
We ignore the total internal gas energy of the ejecta because it is negligible
   compared to that of radiation \citep{Utr_07}.
Note that the time-integrated energy deposition by radioactive decays
   does not change with varied ejecta mass because the envelope
   remains optically thick for gamma rays.
Also the time-integrated emergent energy is almost independent of the ejecta
   mass (Figure~\ref{fig:varmej}(b)).

Varying the ejecta mass of models M15-7b-3, M15-7b-3-m2, M15-7b-3-m4,
   and M15-7b-3-m6 from 19.46\,$M_{\sun}$ to 13.46\,$M_{\sun}$ reveals that
   the light curves of models M15-7b-3-m2 and M15-7b-3-m4 agree better with
   the dome-like maximum of the observed light curve than the original model
   M15-7b-3, whereas model M15-7b-3-m6 shows a noticeable deficit in the
   luminosity during the transition from the maximum to the radioactive tail
   (Figure~\ref{fig:varmej}(a)).
This luminosity deficit in model M15-7b-3-m6 is connected to a deficit in
   the radiated energy, which, in turn, is related to the lower total radiation
   energy of the ejecta compared to that of other models under consideration
   (Figure~\ref{fig:varmej}(b)).
The seemingly small deficit of the total radiation energy in model M15-7b-3-m6
   during the transition from the maximum to the radioactive tail turns out
   to be fairly significant because of a lack of other sources of energy
   except for a little remaining energy deposition by radioactive decays.
This fact is of great importance and allows us to discard model M15-7b-3-m6
   from our consideration because it is less consistent with the observations
   of SN~1987A and has a severe deficiency in the energy budget during
   the transition from the broad maximum to the radioactive tail.

The experiment undertaken here demonstrates convincingly that a hydrodynamic
   model based on model M15-7b-3 with a lower ejecta mass in the range of
   17.46\,$M_{\sun}$ to 15.46\,$M_{\sun}$ allows for a better agreement with
   the observations, confirming the conclusions of \citet{JWJ_20}.
Of course, an open question arises, whether a lower explosion energy
   (Table~\ref{tab:auxmod}) would allow for sufficiently strong mixing and
   a sufficiently high maximum velocity of the bulk mass of $^{56}$Ni
   consistent with observations of SN~1987A.

\section{Comparison between single-star and binary-merger models}
\label{sec:ssvsbm}
%
\begin{deluxetable*}{ l c c c c c c c c c c c l }
\tabletypesize{\small}
\tablewidth{0pt}
\tablecaption{Subset of the best explosion models of single-star and
              binary-merger progenitors%
\label{tab:subset}}
\tablehead{
\colhead{Model} & \colhead{$M_\mathrm{NS}$}
       & \colhead{$M_\mathrm{CO}^{\,\mathrm{core}}$}
       & \colhead{$M_\mathrm{He}^{\,\mathrm{core}}$}
       & \colhead{$M_\mathrm{ej}$}
       & \colhead{$R_\mathrm{pSN}$}
       & \colhead{$E_\mathrm{exp}$}
       & \colhead{$M_\mathrm{Ni}^{\,\mathrm{f}}$}
       & \colhead{$v_\mathrm{Ni}^{\,\mathrm{bulk}}$}
       & \colhead{$v_\mathrm{H}^\mathrm{min}$}
       & \colhead{$\Delta M_\mathrm{H}^\mathrm{2000}$}
       & \colhead{$M_\mathrm{O}$}
       & \colhead{Remark} \\
\colhead{} & \multicolumn{4}{c}{$(M_{\sun})$}
       & \colhead{$(R_{\sun})$}
       & \colhead{(B)}
       & \colhead{$(10^{-2}\,M_{\sun})$}
       & \multicolumn{2}{c}{(km\,s$^{-1}$)}
       & \multicolumn{2}{c}{$(M_{\sun})$}
       & \colhead{}
}
\startdata
   B15-2 & 1.25 & 1.64 & 4.05 & 14.20 & 56.1 & 1.40 & 7.25 & 3370 &  28 & 0.922 & 0.16 & single-star \\
  W18x-2 & 1.52 & 2.12 & 5.13 & 16.03 & 30.4 & 1.45 & 7.55 & 2460 & 147 & 0.847 & 0.36 & single-star \\
M15-7b-3 & 1.58 & 2.48 & 2.90 & 19.46 & 37.0 & 1.43 & 7.28 & 2980 &  29 & 3.10 & 0.83 & binary-merger \\
M15-8b-1 & 1.32 & 2.50 & 2.95 & 20.73 & 31.8 & 1.57 & 7.40 & 1829 &  27 & 3.14 & 0.96 & binary-merger \\
\hline
Constraint &  &  &  &  & 28.7--57.5 &  & $\approx$7.22 & $\sim$3000 & $\le$700 & $\sim$2.2 & 0.7--2.0 &
\enddata
\end{deluxetable*}
The study undertaken in this paper in the framework of the neutrino-driven
   explosion mechanism provides a unique opportunity to compare single-star
   and binary-merger progenitor models in reproducing the observed properties
   of the BSG Sanduleak\,$-69^{\circ}202$ star and the SN~1987A explosion.
To carry out a comparative analysis of the single-star and binary-merger models,
   we briefly summarize the key observational constraints for SN~1987A:
\begin{enumerate}
\item
The pre-SN star, Sanduleak\,$-69^{\circ}202$, was a blue B3\,Ia supergiant
   \citep{RMP_78, GCC_87, WLL_87, SAK_87} whose location in the HRD is
   determined by an effective temperature of 14\,000\,K to 17\,000\,K
   \citep{HM_84, TLP_04, CLW_06, SPM_08} and a luminosity of
   $(3-6)\times10^{38}$\,erg\,s$^{-1}$ \citep{ABKW_89}.
\item
The directly observed mass of radioactive $^{56}$Ni deduced from a formal fit
   to the actual bolometric luminosity in the radioactive tail is equal to
   0.0722\,$M_{\sun}$ \citep{UWJM_15}, which is consistent within
   the errors with other empirical estimates of
   ($0.069\pm0.003$)\,$M_{\sun}$ \citep{BPS_91, McC_93} and
   ($0.071\pm0.003$)\,$M_{\sun}$ \citep{STM_14}.
\item
The analysis of the infrared emission lines of [\ion{Ni}{2}] and [\ion{Fe}{2}]
   in the nebular phase shows that the bulk of the radioactive $^{56}$Ni was
   moving with a maximum velocity of $\sim$3000\,km\,s$^{-1}$ \citep{CHELH_94}.
\item
The line profiles of the hydrogen emission in the nebular phase
   \citep{Chu_91, KF_98, MJS_12}, and the 3D geometry of the H$\alpha$ emission
   \citep{LFS_16} and of molecular hydrogen \citep{LSF_19} showed
   that hydrogen-rich matter is mixed deeply down to velocities
   $\le$700\,km\,s$^{-1}$.
\item
\citet{KF_98} constrained quantitatively the mass of hydrogen-rich matter
   that expanded with velocities lower than 2000\,km\,s$^{-1}$ to about
   2.2\,$M_{\sun}$.
\item
The observed bolometric light curve has a prominent, broad maximum of
   dome-like shape \citep{CMM_87, CWF_88, HSGM_88}.
\item
The overall shape of the bolometric light curve, which is characterized by
   a narrow initial luminosity peak during the first $\sim$7\,days and
   the broad dome-like maximum with a characteristic width of $\sim$100\,days
   \citep{CMM_87, CWF_88, HSGM_88}, is a specific feature of the peculiar
   Type IIP SN~1987A in contrast to ordinary Type IIP SNe.
\item
The kinematics of the ejected envelope is well deduced from the evolution of
   the photospheric velocity, which was estimated by \citet{PHHN_88} using the
   absorption minimum of the Fe\,II~5169\,\AA\ line.
\item
The mass of oxygen in the SN ejecta is in the range of 0.7\,$M_{\sun}$ to
   2.0\,$M_{\sun}$ \citep{LM_92, Chu_94, CCKC_97, KF_98, JSS_15}.
\item
\citet{KLF_10} and \citet{LFK_13} studied spectra and images in the
   emission lines of [\ion{Si}{1}]+[\ion{Fe}{2}] and \ion{He}{1} and
   in the emission lines of H$\alpha$ and [\ion{Si}{1}]+[\ion{Fe}{2}],
   respectively, and reconstructed the 3D shape of the inner regions of
   the ejecta.
\citeauthor{KLF_10} approximated the actual 3D shape by an elongated
   triaxial ellipsoid.
\item
The hard X-ray and gamma-ray emission is very sensitive to macroscopic
   mixing in the SN ejecta, the ejecta structure, and the ejecta mass.
\citet{ALM_19} analyzed the early hard X-ray and gamma-ray observations
   of SN~1987A to test 3D hydrodynamic simulations of neutrino-driven
   explosions.
\item
Using 3D hydrodynamic simulations of neutrino-driven explosions,
   \citet{JWJ_20} studied the gamma-ray decay lines and
   the UVOIR bolometric light curve of SN~1987A to deduce constraints on
   the asymmetry of radioactive ejecta and the ejected envelope mass.
\end{enumerate}
\begin{figure}[t]
\centering
   \includegraphics[width=\hsize, clip, trim=18 153 67 99]{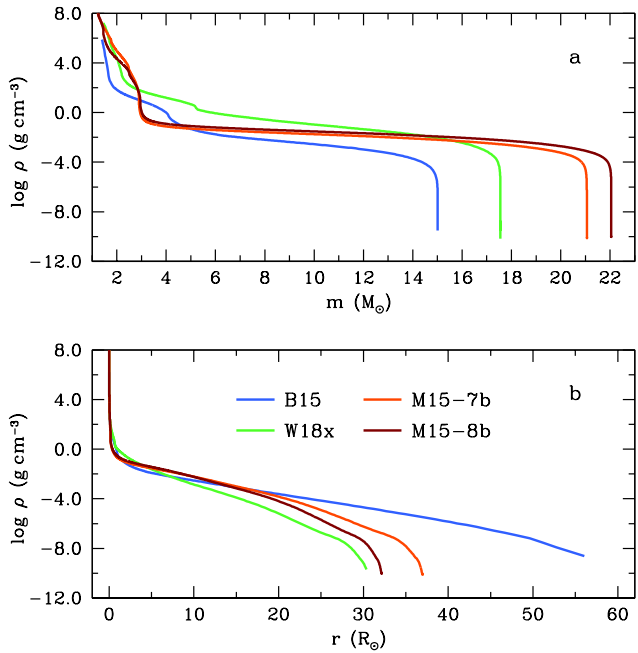}
   \caption{%
   Density profiles as function of interior mass (a) and as function
      of radius for the whole star (b) in the single-star pre-SN models B15
      and W18x \citep{UWJ_19}, and in the binary-merger pre-SN models M15-7b
      and M15-8b (Table~\ref{tab:presnm}).
   }
   \label{fig:denssbm}
\end{figure}
\begin{figure*}[t]
\centering
   \includegraphics[width=\hsize, clip, trim=20 379 47 208]{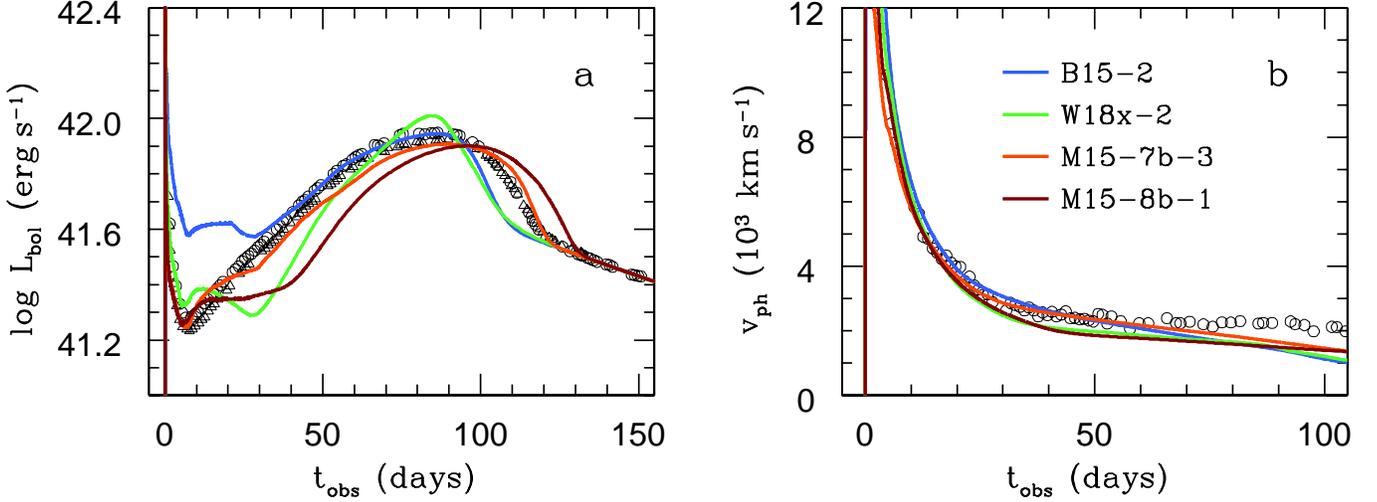}
   \caption{%
   Bolometric light curves (a) and photospheric velocity (b) as function
      of time for models B15-2, W18x-2, M15-7b-3, and M15-8b-1.
   We define the photosphere to be the spherical surface at which the
      Thomson optical depth is $2/3$.
   The computed results are compared with the observed bolometric luminosity of
      SN~1987A obtained by \citet{CMM_87, CWF_88} (open circles) and
      \citet{HSGM_88} (open triangles), and with the velocity at the
      photosphere estimated by \citet{PHHN_88} with the absorption minimum of
      the Fe\,II~5169\,\AA\ line (open circles).
   }
   \label{fig:lcvph}
\end{figure*}
To quantify the quality of single-star and binary-merger explosion models,
   we compare the best and second-best model based on the agreement between
   simulations and observations using wide range of criteria.
Hydrodynamic simulations of single stars resulted in the best and second-best
   model being models B15-2 and W18x-2 \citep{UWJ_19}, while our simulations
   of binary mergers yield M15-7b-3 and M15-8b-1 as the best and second-best
   model, respectively.
These models constitute the subset of our best single-star and binary-merger
   explosion models, whose basic properties are listed in
   Table~\ref{tab:subset}.
The structure of their progenitor models are depicted in
   Figure~\ref{fig:denssbm}, the bolometric light curves in 
   Figure~\ref{fig:lcvph}(a), and the evolution of the photospheric velocity
   in Figure~\ref{fig:lcvph}(b).
This subset of models allows us not only to estimate how well the selected
   models match the observations, but also to demonstrate differences between
   the best and second-best model.

\begin{deluxetable*}{ l c c }
\tabletypesize{\small}
\tablewidth{0pt}
\tablecaption{Comparative analysis of single-star and binary-merger models%
\label{tab:companal}}
\tablehead{
\colhead{Observational constraint} & \colhead{Single-star}
                                   & \colhead{Binary-merger} \\
\colhead{} & \colhead{B15-2 / W18x-2} & \colhead{M15-7b-3 / M15-8b-1}
}
\startdata
\phantom{1}1. location of Sanduleak\,$-69^{\circ}202$ in the HRD & $-$ / $+$ & $+$ / $+$ \\
\phantom{1}2. production of $^{56}$Ni in 3D simulations    & $+$ / $+$ & $+$ / $+$ \\
\phantom{1}3. maximum velocity of the bulk of $^{56}$Ni    & $+$ / $-$ & $+$ / $-$ \\
\phantom{1}4. minimum velocity of hydrogen matter          & $+$ / $+$ & $+$ / $+$ \\
\phantom{1}5. mass of hydrogen with $v<2000$\,km\,s$^{-1}$ & $-$ / $-$ & $+$ / $+$ \\
\phantom{1}6. dome of the light curve                      & $+$ / $-$ & $+$ / $+$ \\
\phantom{1}7. the entire light curve                       & $-$ / $-$ & $+$ / $-$ \\
\phantom{1}8. evolution of the photospheric velocity       & $+$ / $-$ & $+$ / $-$ \\
\phantom{1}9. oxygen mass in the SN ejecta                 & $-$ / $-$ & $+$ / $+$ \\
10. 3D shape of the $^{56}$Ni ejecta                       & $-$ / $+$ & $+$ / $-$ \\
11. X-ray and gamma-ray emission                 & $+$ / \phantom{$-$} & $+$ / $-$ \\
12. gamma-ray decay lines                                  & $+$ / $-$ & $-$ / $-$ \\
\hline
Total score & $7:12$ / $4:11$ & $11:12$ / $6:12$
\enddata
\end{deluxetable*}
Comparing the models in the subset, in the framework of neutrino-driven
   explosions, we obtain the following results:
\begin{enumerate}
\item
\citet{SEWBJ_16} clearly demonstrated that the single-star progenitor model B15
   (W15 in the nomenclature of \citeauthor{SEWBJ_16}) has a helium-core mass
   and a luminosity that are both too small compared to the observational
   constraints imposed by the location of the Sanduleak\,$-69^{\circ}202$ star
   in the HRD, whereas model W18x agrees well with these constraints.
All of the binary-merger progenitor models provided by \citet{MH_17} for this
   study are located in or close to the region of the observed properties of
   Sanduleak\,$-69^{\circ}202$ in the HRD.
\item
The 3D neutrino-driven explosion simulations of both the single-star models
   B15-2 and W18x-2 and the binary-merger models M15-7b-3 and M15-8b-1 are
   capable of producing the initial (at the onset of light-curve modeling)
   $^{56}$Ni masses required to match the observations of the radioactive tail
   of SN~1987A (Table~\ref{tab:subset}).
\item
Both of the two best models, B15-2 and M15-7b-3, yield a maximum velocity
   of the bulk of $^{56}$Ni consistent with the observed value of about
   3000\,km\,s$^{-1}$ in contrast to the second-best models, W18x-2 and
   M15-8b-1, which do not fulfill this constraint of SN~1987A
   (Table~\ref{tab:subset}).
\item
In the single-star and binary-merger models considered
   here the inward mixing of hydrogen, occurring during the 3D neutrino-driven
   simulations, results in minimum velocities of hydrogen-rich matter of less
   than 150\,km\,s$^{-1}$ (Table~\ref{tab:subset}), which is in good
   agreement with the spectral observations of SN~1987A.
\item
Table~\ref{tab:subset} shows that the single-star models B15-2 and W18x-2
   contain masses of hydrogen-rich matter expanding with velocities lower
   than 2000\,km\,s$^{-1}$ that are a factor of two lower than the
   observational constraint of about 2.2\,$M_{\sun}$, whereas the binary-merger
   models M15-7b-3 and M15-8b-1 even exceed it.
\item
For the binary-merger models M15-7b-3 and M15-8b-1 the broad domes
   of the calculated light curves are very similar in shape to the observed
   one, but they are slightly less luminous and shifted to a slightly later
   epoch (Figure~\ref{fig:lcvph}(a)).
In contrast, the dome of the single-star model B15-2 satisfactorily fits the
   observed one, while the dome of model W18x-2 displays a peaked shape and
   disagrees with the observations.
The quality of the dome-shape fits partially correlates with the mass of
   hydrogen-rich matter mixed inward to a velocity of
   $\lesssim$2000\,km\,s$^{-1}$ and the mass of the ejecta
   (Table~\ref{tab:subset}).
\item
The binary-merger model M15-7b-3 matches the overall observed
   bolometric light curve much better than any of the other models
   (Figure~\ref{fig:lcvph}(a)).
The narrow initial luminosity peak during the first $\sim$7\,days results
   from a relatively small radius of the corresponding pre-SN model of
   37.0\,$R_{\sun}$ (Table~\ref{tab:subset}) and a suitable structure of its
   outer layers (Figure~\ref{fig:denssbm}).
A combination of the maximum velocity of the bulk mass of $^{56}$Ni around
   3000\,km\,s$^{-1}$ and the mass of hydrogen-rich matter of 3.10\,$M_{\sun}$
   mixed inward to a velocity of $\lesssim$2000\,km\,s$^{-1}$
   (Table~\ref{tab:subset}) produces the smooth and monotonic rising part of
   the light curve and its broad dome-like maximum with a characteristic width
   of $\sim$100\,days.
\item
The evolution of the photospheric velocity of both best models B15-2 and
   M15-7b-3 is in good agreement with that observed during the first 60\,days
   when the position of the photosphere is more or less correctly determined
   in the light-curve simulations, whereas the agreement is worse for the other
   two models (Figure~\ref{fig:lcvph}(b)).
After the photospheric velocity has decreased below 3000\,km\,s$^{-1}$,
   which is the maximum speed of the bulk mass of the $^{56}$Ni-rich matter,
   around day 30, the asymmetry in the 3D morphology of the inner regions of
   the ejecta studied by \citet{KLF_10} and \citet{LFK_13} can gradually
   affect the shape of the photosphere.
The disparity between the calculated and observed photospheric velocities
   after about 60\,days most likely stems from assuming a spherical $^{56}$Ni
   distribution in our 1D simulations.
\item
The mass of oxygen in the ejecta of the binary-merger models M15-7b-3 and
   M15-8b-1 falls in the observed range of 0.7\,$M_{\sun}$ to 2.0\,$M_{\sun}$,
   but the single-star models B15-2 and W18x-2 have oxygen masses
   significantly below this range (Table~\ref{tab:subset}).
\item
The 3D morphology of the $^{56}$Ni-rich ejecta in the binary-merger model
   M15-7b-3 long after shock breakout has a pronounced dipole component
   (Figure~\ref{fig:3D_models}, upper row, second column), which resembles well
   the elongated-ellipsoid shape that was extracted from spectra and images
   of SN~1987A in the emission lines in the nebular phase.
Except for a single prominent clump moving with a velocity of about
   3000\,km\,s$^{-1}$, the 3D explosion morphology of the single-star model
   W18x-2 is of approximately elongated ellipsoidal shape
   (\citealt[Figure 4, lower row, last column]{UWJ_19}).
The two other models of M15-8b-1 (Figure~\ref{fig:3D_models}, upper row,
   last column) and B15-2 (\citealt[Figure 4, lower row, first column]{UWJ_19})
   have a morphology that is closer to a spherical shape than to an elongated
   ellipsoid.
\item
\citet{ALM_19} computed the early hard X-ray and gamma-ray emission produced
   by 3D neutrino-driven explosion models of single-star and binary-merger
   progenitors, and compared the emergent emission with the
   corresponding spectra, continuum light curves, and line fluxes of SN\,1987A.
They showed that the single-star model B15-2 and the
   binary-merger model M15-7b-3 are basically consistent with the X-ray and
   gamma-ray observations, whereas model M15-8b-1 fails to match them.
Model W18x-2 is not investigated in that study.
\item
Analyzing the gamma-ray decay lines and the UVOIR bolometric light curve of
   SN~1987A up to 600\,days, \citet{JWJ_20} found that the 3D explosion models
   show good agreement with the observations only for an ejecta mass of about
   14\,$M_{\sun}$ and a kinetic energy of around $1.5\times10^{51}$\,erg.
It is important that our auxiliary hydrodynamic models based on model M15-7b-3
   and having an ejecta mass in the range of 17.46\,$M_{\sun}$ to
   15.46\,$M_{\sun}$, i.e., lower than in the reference model M15-7b-3, result
   in a better agreement with the photometric observations of SN\,1987A
   (see Section~\ref{sec:prmass}).
\end{enumerate}

For a concise presentation of the results of the above comparative analysis and
   their adequate assessment, we list all twelve observational constraints
   in Table~\ref{tab:companal}, where plus signs stand for matching the
   constraint, minus signs for not matching, and a blank space for a lack of
   the relevant data.
Assuming equal weights for all constraints of the analysis, we conclude that
   our favorite binary-merger explosion model M15-7b-3 matches all of the
   constraints except for point No.~12, demonstrating an overwhelming
   superiority over the other models.
Only point 12 still poses a problem, because the ejecta mass of the explosion
   model M15-7b-3 is too large.
This points to a shortcoming of the binary-merger progenitor model M15-7b,
   whose pre-collapse mass is too high to avoid this problem.
Among the single-star explosions, model B15-2 has some important favorable
   properties -- namely the maximum velocity of the bulk of $^{56}$Ni,
   the dome of the light curve, and the X-ray and gamma-ray emission -- that
   agree with observations.

Note that \citet{ONF_20} compared 3D single-star and binary-merger
   explosions by their ability to reproduce the observed infrared emission
   lines of [\ion{Fe}{2}] \citep{HCE_90} and to satisfy the constraint imposed
   by the fast $^{56}$Ni clump \citep{UCA_95}.
Their 3D explosion modeling was not self-consistent (because they imposed
   the explosion asymmetry by hand), but in line with our conclusions, they
   independently found that the explosion of the binary-merger progenitor
   model, which was initiated by an asymmetric bipolar-like SN shock,
   reproduces the observations better than the single-star explosion.

\section{Conclusions}
\label{sec:conclsns}
%
In this work we compare the results of 3D neutrino-driven explosion
   simulations and light curve modeling, based on progenitor models of
   Sanduleak\,$-69^{\circ}202$ evolved in a binary-merger scenario with
   the observations of the peculiar Type IIP SN~1987A.
We draw the following conclusions:
\begin{itemize}
\item
The radii of the compact blue binary-merger progenitor models are in between
   31.8\,$R_{\sun}$ and 37.3\,$R_{\sun}$ (Table~\ref{tab:presnm}) and agree
   well with the photometric radius of the BSG Sanduleak\,$-69^{\circ}202$
   star of 28.7\,$R_{\sun}$ to 57.5\,$R_{\sun}$ \citep{ABKW_89}.
In addition, these pre-SN radii are in agreement with those of 30\,$R_{\sun}$
   to 40\,$R_{\sun}$ allowed by hydrodynamic explosion modeling and 
   a time-dependent atmosphere model that explain the photometric and
   spectroscopic observations of SN~1987A \citep{Utr_05}.
Such pre-SN radii and the structure of the outer layers are suitable to
   reproduce the observed initial luminosity peak during the first $\sim$7\,days
   (Figure~\ref{fig:lcurves}) with explosions of all of our binary-merger
   models M15-7b-3, M15-8b-1, M16-4a-1, M16-7b-2, M17-7a-2, and M17-8a-4,
   which have comparable and SN~1987A-like explosion energies
   (Table~\ref{tab:hydmod}).
\item
After the first 20 days, we see some non-monotonic behavior (and partly
   short-timescale variability) in the light curves of all of the explosion
   models instead of the smooth and monotonic behavior of the observed light
   curve (Figure~\ref{fig:lcurves}).
The rising part of the light curve depends both on the structure of the pre-SN
   model \citep{Utr_04} and on the extent of outward mixing of radioactive
   $^{56}$Ni (\citealt{Utr_04}; \citealt[Figure~8]{UWJ_19}).
Strong outward mixing of $^{56}$Ni in models M15-7b-1, M15-7b-2, M15-7b-3,
   and M16-4a-1 (Table~\ref{tab:hydmod}) -- which exceeds the observational
   constraint on the maximum velocity for the bulk mass of $^{56}$Ni of
   about 3000\,km\,s$^{-1}$ -- produces an almost smooth and monotonic rising
   part of the calculated light curves.
In contrast, weaker mixing of $^{56}$Ni in some of the explosion models
   results in non-smooth and non-monotonic behavior.
\item
The domes of all calculated light curves are very similar in shape to the
   observed behavior, but they are less luminous up to about 0.1\,dex
   and shifted to a later epoch up to nearly 30\,days
   (Figure~\ref{fig:lcurves}).
However, for the 3D explosion models based on progenitor M15-7b both of these
   shortcomings are smaller and are about 0.02\,dex and 10\,days,
   respectively.
This effect is explained by a larger optical depth of the inner layers than
   in the single-star explosion model B15-2, whose dome of the light curve
   agrees with the observations of SN~1987A much better than that of all
   other single-star explosion models (\citealt[Figure~11]{UWJM_15}).
The higher optical depth in the binary-merger explosion models is caused by
   a much higher abundance of hydrogen in the inner layers.
The last fact is evident from the mass of hydrogen mixed into the helium
   shell, which is about 0.5\,$M_{\sun}$ in the explosion models of our
   binary-merger progenitors (Table~\ref{tab:hydmod}) compared to
   0.15\,$M_{\sun}$ in model B15-2 (\citealt[Table~2]{UWJ_19}).
\item
The initial $^{56}$Ni mass (at the time of shock breakout) required
   to match the observations of SN~1987A in the radioactive tail
   (Figure~\ref{fig:lcurves}) is below the maximum production of $^{56}$Ni
   estimated for all of our 3D explosion models (Table~\ref{tab:hydmod}).
Thus, the 3D neutrino-driven explosion simulations considered in this study
   are able to synthesize the ejected amount of radioactive $^{56}$Ni.
\item
The sequence of the binary-merger explosion models M15-7b-3, M15-8b-1,
   M16-4a-1, M16-7b-2, M17-7a-2, and M17-8a-4, all of which have nearly
   the same and SN~1987A-like explosion energies, showed, similar to
   the single-star models, that the extent of outward radioactive $^{56}$Ni
   mixing in the framework of the 3D neutrino-driven simulations depends
   mainly on the following three explosion-related properties, which depend
   on the structure of the progenitor models:
   high growth factors of RT instabilities at the (C+O)/He and He/H composition
   interfaces, and a weak interaction of fast RT plumes with the reverse shock
   occurring below the He/H composition interface.
An optimal combination of these three factors turned out to be a sufficient
   condition for efficient outward mixing of $^{56}$Ni into the hydrogen
   envelope.
\item
\citet{KF_98} quantitatively constrained the mass of hydrogen-rich gas mixed
   to velocities of $\lesssim$2000\,km\,s$^{-1}$ to about 2.2\,$M_{\sun}$.
Models M15-7b-4 and M16-4a-1 agree well with this observational constraint,
   whereas other models yield significantly higher masses of
   hydrogen matter in the range of 3.07\,$M_{\sun}$ to 6.41\,$M_{\sun}$
   mixed below 2000\,km\,s$^{-1}$ (Table~\ref{tab:hydmod}).
\item
The masses of oxygen in the ejecta of the binary-merger models fall in
   the range of 0.83\,$M_{\sun}$ to 1.29\,$M_{\sun}$ (Table~\ref{tab:hydmod})
   and are fully consistent with the observed range of 0.7\,$M_{\sun}$ to
   2.0\,$M_{\sun}$ \citep{LM_92, Chu_94, CCKC_97, KF_98, JSS_15}.
\item
Analyzing the gamma-ray decay lines and the UVOIR bolometric light curve of
   SN~1987A, \citet{JWJ_20} find that 3D neutrino-driven explosion simulations
   show good agreement with the observations only for an ejecta mass of about
   14\,$M_{\sun}$ and a kinetic energy of around $1.5\times10^{51}$\,erg.
Being aware of this result, we computed auxiliary explosion models based on
   model M15-7b-3, which had an ejecta mass in the range of 17.46\,$M_{\sun}$
   to 15.46\,$M_{\sun}$, which is lower than in the reference model M15-7b-3.
These models show a better agreement with the photometric
   observations than the reference model (Section~\ref{sec:prmass}).
\item
A comparative analysis of 3D neutrino-driven explosion simulations and
   light curve modeling of SN~1987A (Section~\ref{sec:ssvsbm}) based on
   progenitors evolved in both single-star and binary-merger scenarios
   shows that only binary-merger model M15-7b-3 matches eleven of twelve
   crucial observational constraints, including the observational properties
   of the BSG Sanduleak $-69^{\circ}202$ star, demonstrating a remarkable
   superiority compared to other single-star and binary-merger models
   (Table~\ref{tab:companal}).
Among the single-star models, model B15-2 has
   some favorable properties -- namely the maximum velocity of the bulk of
   $^{56}$Ni, the dome of the light curve, and the X-ray and gamma-ray
   emission -- that agree with observations.
\end{itemize}

As in \citet{UWJM_15, UWJ_19}, we did not fine-tune the modeling approach
   for SN 1987A, except for choosing a suitable explosion energy, to achieve
   good agreement of the 3D neutrino-driven explosion models and the
   corresponding light-curve results with observations of SN~1987A.

Our results are a promising step forward towards fully self-consistent models
   that can explain the observed properties of SN~1987A.
They provide strong support to the binary-merger scenario for the BSG
   progenitor star of this supernova.
Moreover, they permit the expectation that further progress will be achieved
   by improved binary-merger progenitor models, having a smaller pre-collapse
   mass, higher resolution 3D explosion simulations to later evolution stages
   beyond the decay time of $^{56}$Ni and $^{56}$Co \citep{GWJ_21},
   and a treatment of radiative transfer including 3D effects
   \citep[e.g.,][]{DA_19}.

\acknowledgements
%
V.P.U. was supported by the guest program of the Max-Planck-Institut f\"ur
   Astrophysik and by Russian Scientific Foundation grant 19-12-00229
   in his work on radiation-hydrodynamics simulations for supernova
   light-curve modeling with the code CRAB.
A.H. was supported by the Australian Research Council (ARC) Centre of
   Excellence (CoE) for Gravitational Wave Discovery (OzGrav) project
   number CE170100004 and by the ARC CoE for All Sky Astrophysics in
   3 Dimensions (ASTRO 3D) project number CE170100013.
A.H.'s contribution benefited from support, in part, by the National Science
   Foundation under Grant No. PHY-1430152 (JINA Center for the Evolution of
   the Elements).
At Garching, funding by the Deutsche  Forschungsgemeinschaft (DFG, German
   Research Foundation) through Sonderforschungsbereich (Collaborative Research
   Center) SFB-1258 ``Neutrinos and Dark Matter in Astro- and Particle Physics
   (NDM)'' and under Germany’s Excellence Strategy through Cluster of
   Excellence ORIGINS (EXC-2094)--390783311, and by the European Research
   Council through Grant ERC-AdG No. 341157-COCO2CASA is acknowledged.
Computation of the 3D models and postprocessing of the data were done on Hydra
   of the Rechenzentrum Garching.



\end{document}